\documentclass[aps,floatfix,showpacs,superscriptaddress,showkeys]{revtex4}

\usepackage{amssymb,amsmath,amsfonts,latexsym,graphicx,epsfig,here}

\usepackage{mkPhys}

\newcommand{\bea}{\begin{eqnarray}}
\newcommand{\ena}{\end{eqnarray}}

\newcommand{\be}{\begin{equation}}
\newcommand{\en}{\end{equation}}

\newcommand{\nn}{\nonumber\\}

\newcommand{\la}{\langle}
\newcommand{\ra}{\rangle}
\newcommand{\Bla}{\Big\langle}
\newcommand{\Bra}{\Big\rangle}

\newcommand{\Tr}{\mbox{\rm{tr}}}
      
\begin{document}

\hfill DSF-2012-6 (Napoli), MZ-TH/12-50 (Mainz)

\title{Rare baryon decays $\Lambda_b \to  \Lambda \ell^+\ell^-$ 
$(\ell=e,\mu,\tau)$ and $\Lambda_b \to  \Lambda \gamma$ : \\ 
Differential and total rates, lepton-- and hadron--side 
forward--backward asymmetries}

\author{Thomas Gutsche}
\affiliation{Institut f\"ur Theoretische Physik, Universit\"at T\"ubingen,\\
Kepler Center for Astro and Particle Physics,\\
Auf der Morgenstelle 14, D-72076, T\"ubingen, Germany}

\author{Mikhail A. Ivanov}
\affiliation{Bogoliubov Laboratory of Theoretical Physics, \\
Joint Institute for Nuclear Research, 141980 Dubna, Russia}

\author{J\"{u}rgen G. K\"{o}rner}
\affiliation{PRISMA Cluster of Excellence, Institut f\"{u}r Physik, 
Johannes Gutenberg-Universit\"{a}t, \\ 
D-55099 Mainz, Germany}

\author{Valery E. Lyubovitskij
\footnote{On leave of absence
from Department of Physics, Tomsk State University,
634050 Tomsk, Russia}}
\affiliation{Institut f\"ur Theoretische Physik, Universit\"at T\"ubingen,\\
Kepler Center for Astro and Particle Physics,\\
Auf der Morgenstelle 14, D-72076, T\"ubingen, Germany}

\author{Pietro Santorelli}
\affiliation{Dipartimento di Fisica, Universit\`a di Napoli
Federico II, Complesso Universitario di Monte Sant'Angelo,
Via Cintia, Edificio 6, 80126 Napoli, Italy}
\affiliation{Istituto Nazionale di Fisica Nucleare, Sezione di
Napoli, 80126 Napoli, Italy}

\today

\begin{abstract}
Using the covariant constituent quark model previously developed by us  
we calculate the differential rate and the 
forward--backward asymmetries on the lepton and hadron side for the rare
baryon decays $\Lambda_b \to  \Lambda \ell^+\ell^-$ $(\ell=e,\mu,\tau)$ 
and $\Lambda_b \to  \Lambda \gamma$.
We use helicity methods to write down a three--fold joint angular decay 
distribution for the cascade decay
$\Lambda_{b} \to \Lambda (\to p \pi^{-})+ J_{\rm eff}(\to \ell^{+}\ell^{-})$.
Through appropriate angular integrations we obtain expressions for the rates, 
the lepton--side forward--backward (FB) asymmetry and the polarization of the 
daughter baryon $\Lambda$ leading to a hadron--side forward--backward 
asymmetry. We present numerical results on these observables using the
covariant quark model and compare our results to the results of other
calculations that have appeared in the literature.
\end{abstract}

\pacs{12.39.Ki,13.20.He,14.20.Jn,14.20.Mr}
\keywords{relativistic quark model, light and bottom baryons, 
rare decays, angular distributions, asymmetries}

\maketitle

\section{Introduction}

In a recent paper the CDF Collaboration has reported on 24 events of the
rare baryon decay 
$\Lambda_{b} \to \Lambda+\mu^{+}\mu^{-}$~\cite{Aaltonen:2011qs}. 
The collaboration measures the total branching ratio and
gives first results on the $q^{2}$--spectrum. This experimental result
is quite remarkable since, given the measured branching ratio of 
$1.73\times10^{-6}$,
the CDF collaboration must have had a data sample of at least 42 million 
$\Lambda_{b}$'s. 
The physics of heavy baryon decays appears to have entered a new era with
this experimental result. With the LHC running so well,
many more $\Lambda_{b}$'s will be recorded by e.g. LHCb in the near future
which makes the study of rare $\Lambda_{b}$ decays a worthwhile 
enterprise. The decay $\Lambda_{b} \to \Lambda\,\ell^{+}\ell^{-}$
can be considered to be a welcome complement to the well--analyzed
rare meson decays $B \to K^{(\ast)}\,\ell^{+}\ell^{-}$, 
$B_{s} \to \phi\,\ell^{+}\ell^{-}$ etc. to study the short-- and 
long--distance dynamics of
rare decays induced by the transition $b \to s\,\ell^{+}\ell^{-}$.
     
There have been a number of theoretical papers on the rare 
$\Lambda_b \to  \Lambda$ baryon decays involving the one-photon mode
$\Lambda_{b} \to \Lambda \gamma$ and the dilepton modes
$\Lambda_b \to  \Lambda\, \ell^+\ell^-$
$(\ell=e,\mu,\tau)$.
They use the same set of (penguin) operators or their non--Standard Model 
extensions to describe the short distance
dynamics but differ in their use of theoretical models to calculate the 
nonperturbative transition matrix element 
$\la \Lambda| \, O_{i} \, |\Lambda_{b} \ra$.
Among the phenomenological models used are the bag model~\cite{Cheng:1994kp},
a pole model \cite{Mannel:1997xy,Colangelo:2007jy}, the covariant oscillator 
quark model~\cite{Mohanta:1999id}, nonrelativistic quark 
models~\cite{Cheng:1995fe,Mott:2011cx}, a perturbative QCD 
approach using light--cone distribution amplitudes calculated from QCD
sum rules~\cite{He:2006ud}.  
The authors of~\cite{Mannel:1997xy,Colangelo:2007jy,Huang:1998ek,
Chen:2001ki,Chen:2001sj,
Chen:2001zc,Zolfagharpour:2007eh} have made use of the heavy quark mass 
limit to write their form factors in terms of two independent heavy quark 
effective theory (HQET) form factors~\cite{Hussain:1990uu,%
Hussain:1992rb,Mannel:1990vg} for which they provide some
estimates. Mannel and Recksiegel~\cite{Mannel:1997xy} and 
Colangelo et al.~\cite{Colangelo:2007jy} 
use experimental input for this 
estimate, while~\cite{Huang:1998ek,Chen:2001ki,Chen:2001sj,%
Chen:2001zc,Zolfagharpour:2007eh} 
use QCD sum rules to determine the two HQET form factors. 
Recently the $\Lambda_b$ rare decays were studied in
lattice QCD~\cite{Lb_lattice} and used an improved version of the
QCD sum rules~\cite{Lb_QCDSR}.

As has been emphasized by~\cite{Huang:1998ek,Korner:1991zx,Korner:1992uw} 
heavy quark symmetry should
not be relied on at small $q^{2}$, i.e. close to maximal recoil, in 
particular for heavy--to--light transitions. Heavy quark 
symmetry is expected to be reliable close to zero recoil where not much
momentum is transferred to the spectator system. As one moves away from the
zero recoil point and more momentum gets transferred to the spectator system, 
hard gluon exchange including spin flip interactions become more important
and the heavy quark symmetry can be expected to break 
down~\cite{Korner:1991zx,Korner:1992uw}. 
One should,   
therefore, not rely on a form factor parametrization in terms of only two
heavy quark symmetry form factors for the whole $q^{2}$ range as has been done
in Refs.~\cite{Mannel:1997xy,Huang:1998ek,Chen:2001ki,Chen:2001sj,Chen:2001zc,%
Zolfagharpour:2007eh}.

Furthermore, the results of QCD sum rules for heavy--to--light transitions 
have been shown to be unreliable close to maximum 
recoil~\cite{Ball:1997rj}. One should rather rely on light--cone 
sum rules for the near maximum recoil region for the evaluation of
the hadronic form factors. 
It is for these reasons that most recent calculations have employed 
light--cone sum rules to calculate the whole set of hadronic form 
factors~\cite{Wang:2008sm,Aslam:2008hp,Wang:2008ni,Aliev:2010uy,%
Mannel:2011xg,Wang:2011uv,Feldmann:2011xf,Aliev:2005np,Azizi:2010qk,%
Azizi:2011ri,Aliev:2012sx}. The form factors are then  extrapolated to the
low recoil region by using some pole--type parametrizations. The authors
of~\cite{Mannel:2011xg,Wang:2011uv,Feldmann:2011xf} have derived some very
interesting form factor relations in the large recoil region using soft
collinear effective theory (SCET).
 
Apart from rates, $q^{2}$--spectra, lepton--side forward-backward 
asymmetries treated in most of the papers some of the authors have also 
discussed polarization effects of the final state particles. For example,
in Refs.~\cite{Mannel:1997xy,Chen:2001ki,Chen:2001sj,Mannel:2011xg} 
the polarization of the daughter baryon $\Lambda$ was calculated  
which can be measured experimentally by analyzing the decay
$\Lambda \to p\,\pi^{-}$. The 
polarization components of a single lepton have been considered 
in~\cite{Chen:2001sj,Chen:2001zc} while the authors 
of Ref.~\cite{Zolfagharpour:2007eh} 
have even studied double--lepton polarization
asymmetries. Polarization effects of the decaying $\Lambda_{b}$ have been
investigated in~\cite{Aliev:2005np}. Non-Standard Model effects 
for various observables have been examined 
in Refs.~\cite{Colangelo:2007jy,Huang:1998ek,Chen:2001zc,Zolfagharpour:2007eh,
Aslam:2008hp,Wang:2008ni,Azizi:2010qk,Azizi:2011ri,Aliev:2012sx}.

There are large discrepancies in the predictions of the different models
for the various observables, in particular for the photonic decay 
$\Lambda_{b} \to \Lambda\,\gamma$ and for the near--zero--recoil  
$\tau$--mode $\Lambda_b \to  \Lambda\, \tau^+\tau^-$. It will be interesting
to compare the results of future experiments on rare baryon decays with
the various model predictions, in particular for the above two decay modes. 

In the present paper we use the covariant constituent quark model (for 
short: covariant quark model) as dynamical input to calculate the 
nonperturbative transition matrix elements.
In the covariant quark model the current--induced transitions between 
baryons are calculated from two--loop Feynman diagrams with free quark 
propagators in which the divergent high energy behavior of the loop 
integrations is tempered by Gaussian vertex 
functions~\cite{Ivanov:1996pz}-\cite{Branz:2010pq}. 
An attractive new feature has recently
been added to the covariant quark model in as much as quark confinement has
now been incorporated in an effective way, i.e. there are no quark thresholds 
and thus no free quarks in the relevant Feynman diagrams~\cite{Branz:2009cd,%
Ivanov:2011aa,Gutsche:2012ze}. We emphasize that the
covariant quark model described here is a truly frame--independent
field--theoretic quark model in contrast to other constituent quark models 
which are basically quantum mechanical with built--in relativistic elements. 
One of the advantages of the covariant quark model is that it allows one to
calculate the transition form factors in the full accessible range of 
$q^{2}$--values.

We somehow deviate from the traditional order when presenting our results. 
We first discuss the model independent aspects of the problem involving spin 
physics. This is done by use of the helicity amplitude method which leads to
comprehensive, compact and clearly organized formulae for the joint angular 
decay distributions of the decay products and the spin density matrix 
elements of the final state particles. Lepton mass effects are automatically 
included. At a later stage we present the details of the dynamics for which 
we give numerical results towards the end of the paper. We believe that, in
this paper, and in a forthcoming paper, we provide the first complete 
and comprehensive discussion of all the spin physics aspects of the decay 
$\Lambda_b \to  \Lambda\, \ell^+\ell^-$.

We mention that our helicity formulae are ideally suited as input in an event
generator. This has been previously demonstrated for the charged current decay
$\Xi^{0} \to \Sigma^{+}\ell^{-}\bar{\nu}_{\ell}$ $(\ell=e,\mu)$ followed by 
the decay $\Sigma^{+} \to p\pi^{0}$ including even polarization effects for 
the $\Xi^{0}$~\cite{Kadeer:2005aq}. We have written a Monte Carlo (MC) event 
generator for the above process based on decay distribution formulae 
derived from a corresponding helicity analysis. The event generator is 
based on the {\it genbod} routine from the CERNLIB library and has been used 
by the NA48 experiment to analyze their data on the above process.

Our paper is structured as follows. 
In Sec.~II, we present a detailed discussion of the helicity formalism
that allows one to write down the 
joint angular distribution of the cascade decay 
$\Lambda_{b} \to \Lambda (\to p \pi^{-})+ j_{\rm eff}(\to \ell^{+}\ell^{-})$. 
In Sec.~III we review the basic notions of our dynamical approach --- 
{\it the covariant quark model for baryons}. In particular, we
i) derive the phenomenological Lagrangians describing the interaction 
of baryons with their constituent quarks 
ii) introduce the corresponding 
interpolating 3--quark currents with the quantum numbers of the respective 
baryon iii) discuss the idea and implementation of quark confinement
iv) present the calculational loop integration techniques. 
In Sec.~IV, we consider the application of our approach to the rare 
one-photon decay 
$\Lambda_b \to  \Lambda \gamma$ and the dilepton decay
$\Lambda_{b} \to \Lambda (\to p \pi^{-})+ j_{\rm eff}(\to \ell^{+}\ell^{-})$. 
Finally, in Sec.~V, we summarize our results. Some technical material has 
been relegated to the Appendices A--E. 

\section{Joint angular decay distribution}

As in the case of the rare meson decays $B \to K^{(\ast)} \ell^+\ell^-$
 $(\ell=e,\mu,\tau)$ treated in~\cite{Faessler:2002ut} one 
can exploit the cascade nature of the decay
$\Lambda_{b} \to \Lambda (\to p \pi^{-})+ j_{\rm eff}(\to \ell^{+}\ell^{-})$ to
write down a joint angular decay distribution involving the polar angles 
$\theta,\, \theta_{B}$ and the azimuthal angles $\chi$ defined by the
decay products in their respective (center of mass) CM systems as shown 
in Fig.1. The angular decay distribution involves
the helicity amplitudes $h^{m}_{\lambda_1\lambda_2}$ for the decay
$j_{\rm eff} \to \ell^{+}\ell^{-}$,
$H^{m}_{\lambda_{\Lambda}\lambda_{j}}$ for the decay 
$\Lambda_{b} \to \Lambda + j_{\rm eff}$ and $h^{B}_{\lambda_{p}0}$ 
for the decay $\Lambda \to p + \pi^{-}$. The joint angular decay
distribution for the decay of an unpolarized $\Lambda_{b}$ reads
\begin{eqnarray}
\label{bjoint3}
W(\theta,\theta_{B},\chi)  &\propto& 
\sum_{\lambda_1,\lambda_{2},\lambda_j,\lambda'_j,J,J',m,m',\lambda_{\Lambda},
\lambda'_{\Lambda},\lambda_{p}} 
h^{m}_{\lambda_1\lambda_2}(J)h^{m'}_{\lambda_1\lambda_2}(J')
e^{i(\lambda_{j}-\lambda'_{j})\chi}
\nonumber\\ 
&\times&
\delta_{\lambda_{j}-\lambda_{\Lambda},\lambda'_{j}-\lambda'_{\Lambda}}
\delta_{JJ'}
d^J_{\lambda_j,\lambda_1-\lambda_{2}}(\theta)
d^{J'}_{\lambda'_j,\lambda_1-\lambda_{2}}(\theta)
H^{m}_{\lambda_{\Lambda}\lambda_{j}}(J)
H^{m'\dagger}_{\lambda'_{\Lambda}\lambda'_{j}}(J')
\nonumber \\
&\times& 
d^{1/2}_{\lambda_{\Lambda}\lambda_{p}}(\theta_{B})
d^{1/2}_{\lambda'_{\Lambda}\lambda_{p}}(\theta_{B})
h^{B}_{\lambda_{p}0}h^{B\,\dagger}_{\lambda_{p}0}\,. 
\end{eqnarray}
The Kronecker delta in 
$\delta_{\lambda_{j}-\lambda_{\Lambda},\lambda'_{j}-\lambda'_{\Lambda}}$ in
(\ref{bjoint3}) expresses the fact that we are considering the decay of an
unpolarized $\Lambda_{b}$. In Eq.~(\ref{bjoint3}) one has to observe that
$|\lambda_{j}-\lambda_{\Lambda}|=|\lambda'_{j}-\lambda'_{\Lambda}|=1/2$ due
to the spin $1/2$ nature of the decaying $\Lambda_{b}$. The corresponding
Kronecker delta has not been included in~(\ref{bjoint3}) and 
will also be omitted in the subsequent
formulas. The $d^{j}_{mm'}$ in Eq.~(\ref{bjoint3}) with $(j=0,1/2,1)$ are 
Wigner's $d$--functions where $d^{0}_{00}=1$. In Appendix~\ref{app:jadd}
we provide an explicit expression for 
the three-fold angular decay distribution by expanding out the r.h.s.
of Eq.~(\ref{bjoint3}). 
We mention that it is not 
difficult to generalize Eq.~(\ref{bjoint3}) to the case of a decaying 
polarized $\Lambda_{b}$ as has been done in~\cite{Aliev:2005np} by 
transcribing the results of the corresponding 
semileptonic charged current decays~\cite{Bialas:1992ny,Kadeer:2005aq}. 

Let us discuss the helicity amplitudes appearing in Eq.~(\ref{bjoint3}) in
turn. The lepton--side helicity amplitudes $h^{m}_{\lambda_1\lambda_2}$ for 
the process 
$j_{\rm eff} \to \ell^{+}\ell^{-}$ are defined by 
$(\hat{\lambda}_{j}=\lambda_{1}-\lambda_{2})$ 
\begin{eqnarray}
\label{defhm}
m=1 \quad(V)\qquad\qquad  h^{1}_{\hat{\lambda}_{j};\lambda_{1}\lambda_{2}}(J)
&=&
\bar{u}_{2}(\lambda_{2})\,\gamma_{\mu}\,v_{1}(\lambda_{1})
\,\epsilon(\hat{\lambda}_{j})\,,\nonumber\\
m=2\quad(A)\qquad\qquad h^{2}_{\hat{\lambda}_{j};\lambda_{1}\lambda_{2}}(J)&=&
\bar{u}_{2}(\lambda_{2})\,\gamma_{\mu}\gamma_{5}\,v_{1}(\lambda_{1})
\,\epsilon(\hat{\lambda}_{j}) \,. 
\end{eqnarray}
We have put a hat on the helicity label $\hat{\lambda}_{j}$ to emphasize 
that $\hat{\lambda}_{j}$
is not the $\lambda_{j}$ appearing in Eq.~(\ref{bjoint3}).
We have also included the label 
$(\hat{\lambda}_{j}=\lambda_{1}-\lambda_{2})$ in Eq.~(\ref{defhm}) for clarity
even if the notation is redundant.
We evaluate the helicity amplitudes in the $(\ell^{+}\ell^{-})$ CM system 
with $\ell^{+}$ defining the $z$--direction. The label $(J)$ takes the values 
$(J=0)$ with $\lambda_{j}=0$ and $(J=1)$ with $\lambda_{j}=\pm1,0$ for the 
scalar and vector parts of the effective current 
$j_{\rm eff}$, respectively. In order to
distinguish between the two $\lambda_{j}=0$ cases we write 
$\lambda_{j}=t$ for the $(J=0)$ scalar case and $\lambda_{j}=0$ for the 
$(J=1)$ vector case. 
The lepton--side helicity amplitudes can be calculated to be 
\begin{eqnarray}
\label{lhelamp}
h^{1}_{\,t\,;\frac{1}{2}\,\frac{1}{2}}(J=0)&=& 0\,, \nonumber \\
h^{1}_{\,0\,;\frac{1}{2}\,\frac{1}{2}}(J=1)&=& 2m_{\ell}\,, \nonumber \\
h^{1}_{+1\,;\frac{1}{2}\,-\frac{1}{2}}(J=1)&=& -\sqrt{2q^{2}}\,, 
\\
h^{2}_{\,t;\frac{1}{2}\,\frac{1}{2}}(J=0)&=& 2m_{\ell}\,, \nonumber \\
h^{2}_{\,0;\frac{1}{2}\,\frac{1}{2}}(J=1)&=&  0\,, \nonumber \\
h^{2}_{+1;\frac{1}{2}\,-\frac{1}{2}}(J=1)&=& \sqrt{2q^{2}}\,\,\,v 
\,.  
\nonumber
\end{eqnarray}
where $v=\sqrt{1-4m_{\ell}^{2}/q^{2}}$ 
is the lepton velocity in the $(\ell^+\ell^-)$ CM frame and 
$m_\ell$ is the leptonic mass.  

From parity one has
\begin{eqnarray}
\label{leppar}
h^{1}_{-\lambda_{j};-\lambda_{1}-\lambda_{2}}&=&
h^{1}_{\lambda_{j};\lambda_{1}\lambda_{2}}\,,\\
h^{2}_{-\lambda_{j};-\lambda_{1}-\lambda_{2}}&=&-\,
h^{2}_{\lambda_{j};\lambda_{1}\lambda_{2}} \nonumber \,. 
\end{eqnarray}

In Eq.~(\ref{bjoint3}) we are summing over the lepton helicities. 
A closer look at the relations (\ref{lhelamp},\ref{leppar}) shows 
that there are no $(J=0;J=1)$ interference effects in the 
joint angular decay distribution (\ref{bjoint3}). 
This has been annotated by the Kronecker delta $\delta_{JJ'}$ 
in~(\ref{bjoint3}). $(J=0;J=1)$ interference effects come into play when 
one leaves the lepton helicities unsummed, i.e. when one considers lepton 
polarization effects. In this case one has to replace $\delta_{JJ'}$ by
$(-1)^{J+J'}$ in Eq.~(\ref{bjoint3}) where the extra minus sign for the 
$(J=0;J=1)$ interference 
contribution results from the Minkowskian form of the metric 
tensor~\cite{KS,Bialas:1992ny,Kadeer:2005aq}. We mention that $(J=0;J=1)$
interference effects occur in charged current transitions already in the 
unpolarized lepton case~\cite{KS,Bialas:1992ny,Kadeer:2005aq}.

From Eq.~(\ref{lhelamp}) it is clear that the scalar contributions enter the 
game only for nonzero lepton masses $m_\ell \neq 0$. The scalar contributions 
will thus only be important for the $\tau$--mode. 

The hadronic helicity amplitudes $H^{m}_{\lambda_{\Lambda}\lambda_{j}}(J)$ 
in Eq.~(\ref{bjoint3}) describe the full dynamics of the 
current--induced transitions $\Lambda_{b} \to \Lambda + j_{\rm eff}$ including
the structure and the values of the short distance coefficients of the 
pertinent penguin operators. The helicity labels on the helicity amplitudes 
$H^{m}_{\lambda_{\Lambda}\lambda_{j}}(J)$  take the values 
$\lambda_{\Lambda}=\pm\frac{1}{2}$, $\lambda_{j}=t$ for $(J=0)$ and
$\lambda_{j}=\pm1,0$ for $(J=1)$ as described above. 
The superscript $m$ on $H^{m}_{\lambda_{\Lambda}\lambda_{j}}(J)$ defines
whether the hadronic helicity amplitude multiplies the
lepton vector current $\bar{u}(\ell^{-})\gamma_{\mu}\,v(\ell^{+})$ ($m=1$) 
or the axial vector
current $\bar{u}(\ell^{-})\gamma_{\mu}\gamma_{5}\,v(\ell^{+})$ ($m=2$). 
More details on the definitions of the hadronic helicity amplitudes and their 
calculation in the covariant quark model can be found in Sec.~III.

If desired one can switch from the helicity amplitudes used here to 
the transversity amplitudes used e.g. in~\cite{Kruger:2005ep} by use of the
relations
\begin{equation}
A^{m}_{\lambda_{\Lambda}\perp,\parallel}=(H^{m}_{\lambda_{\Lambda}+1}
\mp H^{m}_{\lambda_{\Lambda}-1})/\sqrt{2},
\quad A^{m}_{\lambda_{\Lambda}0}=H^{m}_{\lambda_{\Lambda}0}\,,
\quad A^{m}_{\lambda_{\Lambda}t}=H^{m}_{\lambda_{\Lambda}t}\,. 
\end{equation} 
The advantage of the transversity amplitudes is that they have definite 
transformation properties under parity.

The helicity amplitudes $h^{B}_{\lambda_{p}\lambda_{\pi}=0}$, finally,
describe the decay $\Lambda \to p + \pi^{-}$.  We shall use experimental
input for the relevant bilinear forms of the helicity amplitudes. The 
helicity labels on the helicity amplitudes 
$h^{B}_{\lambda_{p}\lambda_{\pi}=0}$ are 
self--explanatory.  

As mentioned in the introduction the joint angular decay distribution 
Eq.~(\ref{bjoint3}) is ideally suited as input to an event generator 
using sequential boosts to the respective rest systems of the secondary 
particles as exemplified by the cascade--type decay 
distribution~(\ref{bjoint3}).

We mention that the decay distribution Eq.~(\ref{bjoint3}) reproduces the
results in Ref.~\cite{Faessler:2002ut} for 
$B \to K^{(\ast)}(\to K \pi) \ell^+\ell^-$ when one replaces the helicity
amplitudes for $\Lambda \to p + \pi^{-}$ by the corresponding helicity
amplitude for $K^{\ast} \to K \pi$. The necessary replacement is
\begin{equation}
\label{replace}
H^{m}_{\lambda_{\Lambda}\lambda_{j}}
H^{m'\dagger}_{\lambda'_{\Lambda}\lambda'_{j}}
d^{1/2}_{\lambda_{\Lambda}\lambda_{p}}(\theta_{B})
d^{1/2}_{\lambda'_{\Lambda}\lambda_{p}}(\theta_{B})
h^{B}_{\lambda_{p}0}h^{B\,\dagger}_{\lambda_{p}0}
\quad \to \quad H^{m}_{\lambda_{K^{\ast}}\lambda_j}
H^{m'\dagger}_{\lambda'_{K^{\ast}}\lambda'_j} 
d^{1}_{0\,\lambda_{j}}(\theta^{\ast})
d^{1}_{0 \,\lambda'_{j}}(\theta^{\ast})
h^{K^{\ast}}_{00}h^{K^{\ast}\dagger}_{00}
\end{equation}
where, in the mesonic case, $\lambda_{K^{\ast}}=-\lambda_j$ and
$\lambda'_{K^{\ast}}=-\lambda'_j$ since the decaying $B$ has spin zero.
In Eq.~(\ref{replace}) we have omitted the label $(J)$ on the hadronic
helicity amplitudes for brevity.
Note that there is only one helicity amplitude for $K^{\ast} \to K \pi$
compared to the two helicity amplitudes for $\Lambda \to p + \pi^{-}$. A new 
feature of the baryonic case
is that the decay $\Lambda \to p + \pi^{-}$ is parity nonconserving which leads
to a forward--backward asymmetry on the hadron side as is the case on the 
lepton side.

In this paper we will not further consider the azimuthal $\chi$--dependence of
the joint angular decay distribution Eq.~(\ref{bjoint3}) but we shall rather 
integrate (\ref{bjoint3}) over the azimuthal angle $\chi$. This leads to 
$\lambda_{j}=\lambda'_{j}$ via 
$\int_{0}^{2\pi} \, d\chi \, \exp[\,i(\lambda_{j}-\lambda'_{j})\,\chi] 
= 2\pi \,\delta_{\lambda_{j}\lambda'_{j}}.$  
Since we are considering the decay 
of an unpolarized $\Lambda_{b}$ such that 
$\lambda_{\Lambda_{b}}=\lambda'_{\Lambda_{b}}$ one also has the equality 
$\lambda_{\Lambda}=\lambda'_{\Lambda}$ after azimuthal integration due to 
the fact that
$\lambda_{\Lambda_{b}}=\lambda_{j}-\lambda_{\Lambda}$ and
$\lambda'_{\Lambda_{b}}=\lambda_{j}-\lambda'_{\Lambda}$. One then obtains the 
two--fold angular decay distribution
\begin{eqnarray}
\label{bjoint2}
W(\theta,\theta_{B})  &\propto& 2\pi 
\sum_{\lambda_1,\lambda_{2},\lambda_j,J,m,m',\lambda_{\Lambda},
\lambda_{p}} 
h^{m}_{\lambda_1\lambda_2}(J)h^{m'}_{\lambda_1\lambda_2}(J)
\nonumber\\ 
&\times&
d^J_{\lambda_j,\lambda_1-\lambda_{2}}(\theta)
d^{J}_{\lambda_j,\lambda_1-\lambda_{2}}(\theta)
H^{m}_{\lambda_{\Lambda}\lambda_{j}}(J)
H^{m'\dagger}_{\lambda_{\Lambda}\lambda_{j}}(J)
\nonumber \\
&\times& 
d^{1/2}_{\lambda_{\Lambda}\lambda_{p}}(\theta_{B})
d^{1/2}_{\lambda_{\Lambda}\lambda_{p}}(\theta_{B})
h^{B}_{\lambda_{p}0}h^{B\,\dagger}_{\lambda_{p}0}\,. 
\end{eqnarray}
The fact that $\lambda_{\Lambda}=\lambda'_{\Lambda}$ in~(\ref{bjoint2}) 
implies that the spin density matrix of the daughter baryon $\Lambda$ 
appearing in~(\ref{bjoint2})  is purely diagonal
implying that in a polar angle analysis such as the one in 
Eq.~(\ref{bjoint2}) one can only probe the longitudinal polarization
$P^{\Lambda}_{z}$ of the $\Lambda$.

In Eq.~(\ref{bjoint3}) we have summed over the helicity labels of the leptons,
i.e. we have taken the trace of the respective spin density matrices.
By leaving the respective helicity labels unsummed one can then obtain
single lepton and double lepton spin asymmetries as have been discussed 
in Refs.~\cite{Chen:2001sj} and~\cite{Zolfagharpour:2007eh}, respectively. As
mentioned before one can treat the decay of a polarized $\Lambda_{b}$ in a
similar vein.

In the following subsections we will consider various integrated forms of
Eq.~(\ref{bjoint2}).

\subsection{Differential rate}

The differential rate is obtained from the two--fold angular decay distribution
Eq.~(\ref{bjoint2}) by integrations over $(\cos\theta_{B},\cos\theta)$.
For the $\cos\theta_{B}$--
integration one uses $\int_{-1}^{1}d\cos\theta_{B}
d^{1/2}_{\lambda_{\Lambda}\lambda_{p}}(\theta_{B})
d^{1/2}_{\lambda_{\Lambda}\lambda_{p}}(\theta_{B})=1$. The $\cos\theta$--
integration, finally, can be done by using
$\int_{-1}^{1}d\cos\theta\,d^J_{\lambda_j,\lambda_1-\lambda_{2}}(\theta)
d^{J}_{\lambda_j,\lambda_1-\lambda_{2}}(\theta)=2/3$ for $J=1$, and 2 for
$J=0$. One obtains
\begin{eqnarray}
\label{bjoint0}
W  &\propto&\tfrac{2}{3}\cdot2\pi\cdot 
\sum_{\lambda_1,\lambda_{2},\lambda_j,J,m,\lambda_{\Lambda}} 
h^{m}_{\lambda_1\lambda_2}(J)h^{m}_{\lambda_1\lambda_2}(J)
 \nonumber \\ 
&\times&
(3\,\delta_{J\,0}+\delta_{J\,1})
H^{m}_{\lambda_{\Lambda}\lambda_{j}}(J)
H^{\dagger m}_{\lambda_{\Lambda}\lambda_{j}}(J) 
\sum_{\lambda_{p}}
h^{B}_{\lambda_{p}0}h^{B\,\dagger}_{\lambda_{p}0} \,. 
\end{eqnarray} 
Note that the differential rate obtains only parity conserving diagonal 
contributions such that $m=m'$. 

Let us define rate functions $\Gamma_{X}^{mm'}$ ($X=U,L,S$) through
\begin{equation}
\label{hel_rateb}
\frac{d\Gamma_{X}^{mm'}}{dq^2} =\frac{1}{2} \frac{G^2_F}{(2\pi)^3} 
\left(\frac{\alpha|\lambda_t|}{2\pi}\right)^2
\frac{|{\bf p_2}|\,q^2\,v}{12\,M_1^2}\,H_X^{mm'} \,, 
\end{equation} 
where $\alpha = 1/137.036$ is the fine structure constant, 
$G_F = 1.16637 \times 10^{-5}$ GeV$^{-2}$ is the Fermi coupling constant, 
$\lambda_t = V^\dagger_{ts} \, V_{tb} = 0.041$ 
is the product of corresponding Kabayashi-Maskawa matrix elements and 
$|{\bf p_2}|=\lambda^{1/2}(M_1^2,M_2^2,q^2)/2M_1$ is
the momentum of the $\Lambda$-hyperon in the $\Lambda_b$-rest frame. Note 
that we have included the statistical factor $1/(2S_{\Lambda_b}+1)=1/2$
in the definition of the rate functions.

The bilinear expressions $H^{mm'}_{X}$ ($X=U,L,S$) are defined by
\begin{equation}
\label{helcom2}
\qquad
\begin{array}{lr}
\mbox{$ H^{mm'}_U = 
{\rm Re}(H^{m}_{\frac{1}{2}1} H^{\dagger m'}_{\frac{1}{2}1}) + 
{\rm Re}(H^{m}_{-\frac{1}{2}-1} H^{\dagger m'}_{-\frac{1}{2}-1}) $}  & 
\hfill\mbox{ \rm unpolarized-transverse}\,, 
\\
\mbox{$ H^{mm'}_L = 
{\rm Re}(H^{m}_{\frac{1}{2}0} H^{\dagger m'}_{\frac{1}{2}0}) + 
{\rm Re}(H^{m}_{-\frac{1}{2}0} H^{\dagger m'}_{-\frac{1}{2}0}) $}     & 
\hfill\mbox{ \rm longitudinal}\,, 
\\
\mbox{$ H^{mm'}_S =  
{\rm Re}(H^{m}_{\frac{1}{2}t}H^{\dagger m'}_{\frac{1}{2}t}) +  
{\rm Re}(H^{m}_{-\frac{1}{2}t}H^{\dagger m'}_{-\frac{1}{2}t})$} &  
\hfill\mbox{ \rm scalar}\,. 
\end{array}\\
\end{equation}
Note that, compared to~\cite{Faessler:2002ut}, 
we have redefined the scalar structure function $H^{mm'}_S$ by omitting a 
factor of 3 in the definition of the scalar structure function.

Putting in the correct normalization factors one 
obtains the differential rate
$d\Gamma/dq^{2}$ which reads
\begin{eqnarray}
\label{bjoint00}
\frac{d\Gamma(\Lambda_b \to \Lambda \,\ell^{+}\ell^{-})}{d q^2}=
\frac{v^{2}}{2}\cdot\bigg( U^{11+22} + L^{11+22} \bigg)
+\frac{2m_\ell^{2}}{q^{2}}\cdot\frac{3}{2}\cdot
\bigg( U^{11} + L^{11} + S^{22} \bigg)\,, 
\end{eqnarray}
where we have adopted the notations 
$d\Gamma_X^{mm'}/d q^2=X^{mm'}$ and $X^{11+22}=X^{11}+X^{22}$. 
Here, and in the following, we do an importance 
sampling of our rate expressions by sorting the contributions according to 
powers of the threshold factor $v$. When one wants to compare our results to 
the corresponding results for the mesonic case written down 
in Ref.~\cite{Faessler:2002ut} one has to rearrange the contributions 
in Ref.~\cite{Faessler:2002ut} accordingly. And, one has to take 
into account the
factor of 3 difference in the definition of the scalar structure function.
We mention that the authors of~\cite{Kruger:2005ep} have also written their 
mesonic decay distributions in terms of powers of the threshold factor $v$. 
The second term proportional to
$2m_\ell^{2}/q^{2}$ in~(\ref{bjoint00}) can be seen to arise from the 
$s$--wave contributions
to $\ell^{+}\ell^{-}$ production: $(J=1)$ in $U^{11}$ and $L^{11}$ associated 
with the vector current $m=1$,
and $(J=0)$ in $S^{22}$ associated with the axial vector current $m=2$ 
(see Eq.~(\ref{lhelamp})). 
  
The total rate, finally, is obtained by $q^{2}$--integration in the range
\begin{equation}
4m_{\ell}^{2}\,\,\le\,\, q^{2} \,\, \le\,\, (M_{1}-M_{2})^{2} \,. 
\end{equation}
For the lower $q^{2}$ limit one has $4m_{\ell}^{2}=(1.04\times 10^{-6},0.045,
12.6284)$ GeV$^2$ for $\ell=(e,\mu,\tau)$. The upper limit of the 
$q^{2}$--integration is given by
$(M_{\Lambda_b}-M_{\Lambda})^{2}=20.29$ GeV$^2$. For $\ell=(e,\mu)$ one 
is practically probing the whole $q^{2}$ region while for $\ell=\tau$ the 
$q^{2}$--range is restricted to the low recoil half of phase-space starting 
at $\sqrt{q^{2}}=3.55$ GeV just below the position of the $\Psi(2S)$ 
vector meson resonance.

\subsection{Lepton--side decay distribution}

Integrating the two--fold decay distribution (\ref{bjoint2}) over 
$\cos\theta_{B}$ one obtains
\begin{eqnarray}
\label{bjoint1l}
W(\theta)  &\propto&2\pi  
\sum_{\lambda_1,\lambda_{2},\lambda_j,J,m,m',\lambda_{\Lambda}
} 
h^{m}_{\lambda_1\lambda_2}(J)h^{m'}_{\lambda_1\lambda_2}(J)
 \nonumber \\ 
&&
d^J_{\lambda_j,\lambda_1-\lambda_{2}}(\theta)
d^{J}_{\lambda_j,\lambda_1-\lambda_{2}}(\theta)
H^{m}_{\lambda_{\Lambda}\lambda_{j}}(J)
H^{\dagger m'}_{\lambda_{\Lambda}\lambda_{j}}(J) 
\,\,\sum_{\lambda_{p}}
h^{B}_{\lambda_{p}0}h^{B\,\dagger}_{\lambda_{p}0} \,. 
\end{eqnarray}
The lepton--side decay distribution involves one more parity--odd
structure function which is defined by
\begin{equation}
\label{helcom2a}
\qquad\begin{array}{lr}
\mbox{$H^{mm'}_P =  
{\rm Re}(H^{m}_{\frac{1}{2}1}H^{\dagger m'}_{\frac{1}{2}1})- 
{\rm Re}(H^{m}_{-\frac{1}{2}-1}H^{\dagger m'}_{-\frac{1}{2}-1})$} & 
\hfill\mbox{ \rm parity--odd}
\\
\end{array}\\
\end{equation}

Putting in the correct normalization one obtains
\begin{eqnarray}
\label{costheta2}
\frac{d\Gamma(\Lambda_{b}\to \Lambda \,\ell^{+}\ell^{-})}{dq^2d\cos\theta} 
&=&\,
v^{2}\cdot\bigg[\frac{3}{8}\,(1+\cos^2\theta)\cdot
\frac{1}{2} U^{11+22}  
\ + \ \frac{3}{4}\,\sin^2\theta\cdot
\frac{1}{2} L^{11+22} \bigg]\label{distr2}\nonumber\\[2mm]
&-&\,v \cdot\frac{3}{4}\cos\theta\cdot P^{12} 
\ + \ \frac{2m_{\ell}^{2}}{q^{2}}\cdot \frac{3}{4}\cdot
\bigg[ U^{11}+ L^{11} + S^{22} \bigg]\,.
\end{eqnarray}

One can define a lepton--side forward--backward asymmetry $A_{FB}^{\ell}$
by $A_{FB}^{\ell}=(F-B)/(F+B)$ where $F$ and $B$ denote the rates in the
forward and backward hemispheres 

\begin{equation}
A_{FB}^{\ell}(q^{2}) 
= - \frac{3}{2}\,\frac{v\cdot P^{12}}
{v^{2}\cdot\big(\,U^{11+22}
+L^{11+22}\big)+\frac{2m_{\ell}^{2}}{q^{2}}\cdot 3 \cdot
\big(U^{11}+L^{11}
+S^{22}\,\big) }\,.
\end{equation}
Note that the lepton--side forward--backward asymmetry vanishes at threshold
$q^{2} \to 4m^{2}_{\ell}$. 
The integrated forward--backward asymmetry is defined as 
\begin{eqnarray} 
\bar A^{\ell}_{FB} = 
- \frac{3}{2}\,
\frac{\int\limits_{4m_{\ell}^2}^{(M_1-M_2)^2} \, dq^2 \ 
\Big( v\cdot P^{12} \Big)} 
{\int\limits_{4m_{\ell}^2}^{(M_1-M_2)^2} \, dq^2 \, \biggl(
v^{2}\cdot\big(\,U^{11+22}
+L^{11+22}\big)+\frac{2m_{\ell}^{2}}{q^{2}}\cdot 3 \cdot
\big(U^{11}+L^{11}
+S^{22}\,\big) \biggr) }\,.
\end{eqnarray} 

\subsection{$\Lambda$--polarization and hadron--side decay 
distribution}

Integrating the two--fold decay distribution~(\ref{bjoint2}) over 
$\cos\theta$ one obtains
\begin{eqnarray}
\label{bjoint1h}
W(\theta_{B})  &\propto&\tfrac{2}{3}\cdot2\pi\cdot 
\sum_{\lambda_1,\lambda_{2},\lambda_j,J,m,m',\lambda_{\Lambda},
\lambda_{p}} 
h^{m}_{\lambda_1\lambda_2}(J)h^{m'}_{\lambda_1\lambda_2}(J)
 \nonumber \\ 
&\times&
(\delta_{J\,1}+3\,\delta_{J\,0})
H^{m}_{\lambda_{\Lambda}\lambda_{j}}(J)
H^{\dagger m'}_{\lambda_{\Lambda}\lambda_{j}}(J) 
\,\,d^{1/2}_{\lambda_{\Lambda}\lambda_{p}}(\theta_{B})
d^{1/2}_{\lambda_{\Lambda}\lambda_{p}}(\theta_{B})
h^{B}_{\lambda_{p}0}h^{B\,\dagger}_{\lambda_{p}0} \,. 
\end{eqnarray} 
In fact, one finds from the structure of the lepton helicity amplitudes
Eq.~(\ref{lhelamp}) and~(\ref{leppar}) that only $(m=m')$ contributions
survive in Eq.~(\ref{bjoint1h}).
 
By chopping off the $\Lambda \to p\pi$ decay structure
$d^{1/2}_{\lambda_{\Lambda}\lambda_{p}}(\theta_{B})
d^{1/2}_{\lambda_{\Lambda}\lambda_{p}}(\theta_{B})
h^{B}_{\lambda_{p}0}h^{B\,\dagger}_{\lambda_{p}0}$ and leaving the helicity
$\lambda_{\Lambda}$ of the $\Lambda$ unsummed one obtains the diagonal 
spin density matrix of the $\Lambda$ given by
\begin{equation}
W_{\lambda_{\Lambda}\lambda_{\Lambda}}  \propto\tfrac{2}{3}\cdot2\pi\cdot 
\sum_{\lambda_1,\lambda_{2},\lambda_j,J,m} 
h^{m}_{\lambda_1\lambda_2}(J)h^{m}_{\lambda_1\lambda_2}(J)
(\delta_{J\,1}+3\,\delta_{J\,0})
H^{m}_{\lambda_{\Lambda}\lambda_{j}}(J)
H^{\dagger m}_{\lambda_{\Lambda}\lambda_{j}}(J) \,. 
\end{equation}
The $z$--component of the polarization of the daughter baryon $\Lambda$
can then be calculated from the diagonal spin density matrix elements
according to 
\begin{equation}
P_{z}^{\Lambda} 
=\frac{W_{\frac{1}{2}\,\frac{1}{2}}-W_{-\frac{1}{2}\,-\frac{1}{2}}}
{W_{\frac{1}{2}\,\frac{1}{2}}+W_{-\frac{1}{2}\,-\frac{1}{2}}}\,, 
\end{equation}
which gives
\begin{equation}
\label{pz1}
P_{z}^{\Lambda}=\frac{
v^{2}\cdot\big(\,P^{11+22}
+L_{P}^{11+22}\big)+\frac{2m_{\ell}^{2}}{q^{2}}\cdot 3 \cdot 
\big(P^{11}+L_{P}^{11}
+S_{P}^{22}\,\big)
}
{v^{2}\cdot\big(\,U^{11+22}
+L^{11+22}\big)+\frac{2m_{\ell}^{2}}{q^{2}}\cdot 3 \cdot
\big(U^{11}+L^{11}
+S^{22}\,\big) 
} \,. 
\end{equation}
One has to keep in mind that $X^{mm'}$ stands for the differential expression
$d\Gamma_X^{mm'}/d q^2$ ($X=U,L,S,P,L_{P},S_{P}$). When averaging the 
polarization $P_{z}^{\Lambda}$ over $q^{2}$ one has to remember to separately
average the numerator and denominator in~(\ref{pz1}). 

In (\ref{pz1}) we have defined two new parity--violating structure functions
according to
\begin{equation}
\label{helcom3}
\qquad\begin{array}{lr}
\mbox{$H^{mm'}_{L_{P}}=
{\rm Re}
(H^{m}_{\frac{1}{2}\,0}H^{\dagger m'}_{\frac{1}{2}\,0}-H^{m}_{-\frac{1}{2}\,0}
H^{\dagger m'}_{-\frac{1}{2}\,0})$} & 
\hfill\mbox{ \rm longitudinal--polarized}\,, 
\\
\mbox{$H^{mm'}_{S_{P}}=
{\rm Re}
(H^{m}_{\frac{1}{2}\,t}H^{\dagger m'}_{\frac{1}{2}\,t}-H^{m}_{-\frac{1}{2}\,t}
H^{\dagger m'}_{-\frac{1}{2}\,t})$}
 & 
\hfill\mbox{ \rm scalar--polarized}\,. 
\end{array}\\
\end{equation}

Returning to Eq.~(\ref{bjoint1h}) we write down 
the correctly normalized single--angle decay distribution which reads

\begin{eqnarray}
\label{bjointh1}
\frac{d\Gamma(\Lambda_{b}\to \Lambda(\to p \pi^{-})\ell^{+}\ell^{-})}
     {dq^2\,d\cos\theta_{B}} &=&
{\rm Br}(\Lambda \to p\pi^{-}) 
\nonumber\\ 
&\times& \frac{1}{2}
\Bigg\{ 
\,\, \frac{v^{2}}{2}\cdot\bigg[\,U^{11+22} + L^{11+22} \, + \,  
                               \Big( P^{11+22} + L_{P}^{11+22} \Big) \, 
                    \alpha_{B}\cos\theta_{B}\,\bigg]
\nonumber\\
&+&
\frac{2m_\ell^{2}}{q^{2}}\cdot\frac{3}{2}\cdot
\bigg[
   U^{11} + L^{11} + S^{22} \, + \, 
 \Big( P^{11} + L_{P}^{22}  + S_{P}^{22} \Big) \, \alpha_{B}\cos\theta_{B}
\,\bigg]
\,\,\Bigg\}\,\,, 
\end{eqnarray}
which, using the rate Eq.~(\ref{costheta2}) and the polarization (\ref{pz1}),
can be rewritten as
\begin{equation}
\label{pz2}
\frac{d\Gamma(\Lambda_{b}\to \Lambda(\to p \pi^{-})\ell^{+}\ell^{-})}
     {dq^2\,d\cos\theta_{B}} 
={\rm Br}(\Lambda \to p\pi^{-})\frac{1}{2}
 \frac{d\Gamma(\Lambda_b \to \Lambda\, \ell^{+}\ell^{-})}{d q^2}
\Big(1+\alpha_{B}P_{z}^{\Lambda}\cos\theta_{B}\Big) \,. 
\end{equation}  
In Eqs.~(\ref{bjointh1}) and~(\ref{pz2}) we have made use of the asymmetry 
parameter $\alpha_{B}$ in the decay $\Lambda \to p+\pi^{-}$ which is defined 
by 
\begin{equation}
\alpha_{B}\,=\,\frac{|h^{B}_{{\frac{1}{2}}0}|^{2}-
|h^{B}_{-{\frac{1}{2}}0}|^{2}}
{|h^{B}_{{\frac{1}{2}}0}|^{2}
+|h^{B}_{-{\frac{1}{2}}0}|^{2}}\,.
\end{equation}
The asymmetry parameter has been measured to be 
$\alpha_{B}=0.642\pm0.013$~\cite{Beringer:1900zz}. 
As a check on~(\ref{bjointh1}) one recovers the differential rate expression 
Eq.~(\ref{bjoint0}) by integrating over $\cos\theta_{B}$ and removing the 
factor ${\rm Br}(\Lambda \to p\pi^{-})$.

Corresponding to the lepton--side forward--backward asymmetry $A_{FB}^{\ell}$
one can define a hadron--side forward--backward asymmetry $A_{FB}^{h}$
defined by $A_{FB}^{h}=(F-B)/(F+B)$. Contrary to the mesonic case 
$B \to K^{\ast}(\to K\pi)+\ell^{+}\ell^{-}$ the 
hadron--side forward--backward asymmetry in 
$\Lambda_b \to  \Lambda \ell^+\ell^-$ is nonzero since the decay
$\Lambda \to p + \pi^{-}$ is parity nonconserving, i.e.
one has $h^{B}_{\frac{1}{2}0}\neq h^{B}_{-\frac{1}{2}0}$\, with $64\%$ 
analyzing power. Using the form Eq.~(\ref{pz2}) one finds that the 
forward--backward asymmetry is simply related to the
polarization $P_{z}^{\Lambda}$ via    
\begin{equation}
A^{h}_{FB}(q^2)=\frac{\alpha_{B}}{2}P_{z}^{\Lambda}(q^2) \,. 
\end{equation} 
The integrated forward--backward asymmetry is defined as 
\begin{eqnarray} 
\bar A^{h}_{FB} = \frac{\alpha_{B}}{2} \, 
\frac{
 \int\limits_{4m_{\ell}^2}^{(M_1-M_2)^2} \, dq^2 \, 
\biggl(v^{2}\cdot\big(\,P^{11+22}
+L_{P}^{11+22}\big)+\frac{2m_{\ell}^{2}}{q^{2}}\cdot 3 \cdot 
\big(P^{11}+L_{P}^{11}+S_{P}^{22}\,\big)\biggr)}
{\int\limits_{4m_{\ell}^2}^{(M_1-M_2)^2} \, dq^2 \, 
\biggl(v^{2}\cdot\big(\,U^{11+22}
+L^{11+22}\big)+\frac{2m_{\ell}^{2}}{q^{2}}\cdot 3 \cdot
\big(U^{11}+L^{11}
+S^{22}\,\big)
\biggr)} 
\,. 
\end{eqnarray}  
For the sake of completeness let us also list the $\theta_B$--dependence 
of the polarization $P_{z}^{\Lambda}(\theta_B)$ which is obtained by chopping 
off the $\Lambda \to p\pi$ decay structure in Eq.~(\ref{bjoint2}) and 
proceeding as before when calculating the $\cos\theta_B$--independent 
polarization~(\ref{pz1}). One obtains
\begin{equation}
P_{z}^{\Lambda}(q^2,\theta_B) = \frac{ 
v^{2}\,\Big({\textstyle \frac{3}{8}}
(1+\cos^2\theta_B)\cdot
P^{11+22}
+{\textstyle \frac{3}{2}}\,\sin^2\theta_B\cdot
L_{P}^{11+22}\Big)
-\,v \,{\textstyle \frac{3}{2}}\cos\theta_B\cdot U^{12}
+\frac{2m_{\ell}^{2}}{q^{2}}\cdot {\textstyle \frac{3}{2}}\cdot
\Big(P^{11} 
+L_{P}^{11}+
S_{P}^{22}\Big)
}
{v^{2}\,\Big({\textstyle \frac{3}{8}}
(1+\cos^2\theta_B)\cdot
U^{11+22}
+{\textstyle \frac{3}{2}}\,\sin^2\theta_B\cdot
L^{11+22}\Big)
-\,v \,{\textstyle \frac{3}{2}}\cos\theta_B\cdot P^{12}
+\frac{2m_{\ell}^{2}}{q^{2}}\cdot {\textstyle \frac{3}{2}}\cdot
\Big(U^{11} 
+L^{11}+
S^{22}\Big)} \,. 
\end{equation}

In this section we have only discussed the $z$--component of the polarization
of the daughter baryon $\Lambda$. As mentioned before
the contribution of the transverse polarization component $P_{x}^{\Lambda}$ 
(as e.g. calculated in~\cite{Huang:1998ek}) averages out after 
$\chi$--integration since it enters the angular decay distribution with an 
angular factor 
$\propto P_{x}^{\Lambda}\alpha_{B}\cos\chi\,\sin\theta_{B}$ (see 
Eq.~(\ref{bjoint3}) or Ref.~\cite{Kadeer:2005aq}). 

\section{$\Lambda$-type baryons in the the covariant quark model}

In this section we discuss the basic ingredients of the covariant quark model 
which will be used for the calculation of the rare decays of $\Lambda_b$ 
baryon. A detailed description of baryons as bound states of three 
quarks can be found in Refs.~\cite{Faessler:2001mr,Faessler:2008ix,%
Faessler:2009xn,Branz:2010pq,Gutsche:2012ze}. 
This includes a description 
of the structure of the Gaussian vertex factor, the choice of interpolating
baryon currents as well as the compositeness condition for baryons.
   
The new features introduced to the meson sector 
in Refs.~\cite{Branz:2009cd,Ivanov:2011aa} and applied to the baryon sector 
in Ref.~\cite{Gutsche:2012ze} are both technical and conceptual. 
Instead of using Feynman
parameters for the evaluation of the two--loop baryonic quark model
Feynman diagram we now use Schwinger parameters. The technical advantage
is that this leads to a simplification of the tensor loop integrations in
as much as the loop momenta occurring in the quark propagators can be 
written as derivative operators. Furthermore, the use of Schwinger parameters 
allows one to incorporate quark confinement in an effective way.
Details of these two new features of the covariant quark model have been 
described in~\cite{Branz:2009cd,Ivanov:2011aa,Gutsche:2012ze}. 

In the following we consider $\Lambda = (Q[ud])$-type baryons needed 
in the present application. They consist of a heavy quark 
and two light quarks in a $^{1}S_{0}$ spin 0 configuration. 
The coupling of a $\Lambda$-type baryon to its constituent quarks 
is described by the Lagrangian 
\be\label{eq:Lagr_str}
{\cal L}^{\,\Lambda}_{\rm int}(x) 
                           = g_\Lambda \,\bar\Lambda(x)\cdot J_\Lambda(x) 
                           + g_\Lambda \,\bar J_\Lambda(x)\cdot \Lambda(x)\,,  
\en
where we make use of the same 
$J_\Lambda(\bar J_\Lambda)$ 
interpolating three quark current
for all three $\Lambda$-type baryons.

In general, for the 
$\Lambda$-type baryons one can construct three types of currents 
without derivatives --- pseudoscalar $J^P$, scalar $J^S$ and 
axial-vector $J^A$ (see, 
Refs.~\cite{Ivanov:1996fj,Ivanov:1999bu,Faessler:2001mr,Branz:2010pq}): 
\bea
J^P_{\Lambda_{Q[ud]}}&=& \epsilon^{a_1a_2a_3} \, 
Q^{a_1} \, u^{a_2} C \gamma_5 d^{a_3} \,, 
\nn
J^S_{\Lambda_{Q[ud]}}&=& \epsilon^{a_1a_2a_3} \, \gamma^5 \, 
Q^{a_1} \, u^{a_2} C  d^{a_3} \,,
\nn
J^A_{\Lambda_{Q[ud]}}&=& \epsilon^{a_1a_2a_3} \, 
\gamma^\mu \, Q^{a_1} \, u^{a_2}  
C\gamma_5\gamma_\mu d^{a_3} \,. 
\label{eq:Lambda-cur1}
\ena 
The symbol $[ud]$ in the suffixes of the currents denote antisymmetrization
of flavor and spin indices with respect to the light quarks $u$
and $d$. We will consider the three flavor types of the  
$\Lambda$-baryons: 
$\Lambda^0_s(sud)$, $\Lambda^+_c(cud)$ and $\Lambda^0_b(bud)$. 
In Ref.~\cite{Faessler:2009xn} we have shown 
that, in the nonrelativistic limit, the $J^P$ and $J^A$ interpolating currents
of the $\Lambda_{Q[ud]}$ baryons become degenerate and attain 
the correct nonrelativistic limit  
(in the case of single-heavy baryons this limit coincides with the heavy 
quark limit), while the 
$J^S$ current vanishes in the nonrelativistic limit. 
On the other hand, the $J^P$ and $J^A$ interpolating currents 
of the $\Lambda$-type baryons become degenerate
with SU(N$_{\mathrm f}$)-symmetric currents. 
In Ref.~\cite{Ivanov:1996fj} we have shown that, in the case of 
the heavy-to-light baryon transition 
$\Lambda_c^+ \to \Lambda^0 e^+ \nu_e$, the use of a SU(3) symmetric 
$\Lambda^0$ hyperon is essential in order to describe data on 
$\Gamma(\Lambda_c^+ \to \Lambda^0 e^+ \nu_e)$ 
(see also discussion in Refs.~\cite{Korner:1994nh,Cheng:1995fe}). 
In Appendix~\ref{app:Lambda_currents} we explicitly demonstrate 
that our $J^P$ and $J^A$ are degenerate with the SU(N$_{\mathrm f}$)-symmetric 
currents. 
Therefore, in the following we restrict ourselves 
to the simplest pseudoscalar $J^P$  current. 
The nonlocal interpolating  three--quark current is written as 
\bea
J_\Lambda(x) &=& \int\!\! dx_1 \!\! \int\!\! dx_2 \!\! \int\!\! dx_3 \, 
F_\Lambda(x;x_1,x_2,x_3) \, J^{(\Lambda)}_{3q}(x_1,x_2,x_3)\,,
\label{eq:Lambda-cur2}\\
J^{(\Lambda)}_{3q}(x_1,x_2,x_3) &=& 
\tfrac12 \epsilon^{a_1a_2a_3} \, Q^{a_1}(x_1)\,
\left( u^{a_2}(x_2) \,C \, \gamma^5 \, d^{a_3}(x_3)
      -d^{a_2}(x_3) \,C \, \gamma^5 \, u^{a_3}(x_2) \right) 
\nn
&=&  \epsilon^{a_1a_2a_3} \, Q^{a_1}(x_1)\,
 u^{a_2}(x_2) \,C \, \gamma^5 \, d^{a_3}(x_3)\,,
\nn
\nn
\bar J_\Lambda(x) &=& \int\!\! dx_1 \!\! \int\!\! dx_2 \!\! \int\!\! dx_3 \, 
F_\Lambda(x;x_1,x_2,x_3) \, \bar J^{(\Lambda)}_{3q}(x_1,x_2,x_3)\,,
\nn
\bar J^{(\Lambda)}_{3q}(x_1,x_2,x_3) &=& 
\epsilon^{a_1a_2a_3} \, \bar d^{a_3}(x_3)\, \gamma^5 \,C\, \bar u^{a_2}(x_2)  
\cdot \bar Q^{a_1}(x_1)\,, 
\nonumber
\ena
where $Q=s,c,b$. Here the matrix $C=\gamma^{0}\gamma^{2}$ is
the usual charge conjugation matrix and the $a_i$ $(i=1,2,3)$ are color 
indices. 

The vertex function $F_\Lambda$ characterizes the finite size 
of the $\Lambda$-type baryon. 
We assume that the vertex function is real. 
To satisfy translational invariance 
the function $F_N$ has to fulfill the identity 
\be
F_\Lambda(x+a;x_1+a,x_2+a,x_3+a) \, = \, 
F_\Lambda(x;x_1,x_2,x_3) 
\label{eq:trans_inv}
\en
for any given four-vector $a\,$. 
In the following we use a particular form for the vertex function 
\be
F_\Lambda(x;x_1,x_2,x_3) \, = \, 
\delta^{(4)}(x - \sum\limits_{i=1}^3 w_i x_i) \;  
\Phi_\Lambda\biggl(\sum_{i<j}( x_i - x_j )^2 \biggr) 
\label{eq:vertex}
\en 
where $\Phi_\Lambda$ is the correlation function of the three constituent 
quarks with the coordinates $x_1$, $x_2$, $x_3$ and masses $m_1$, $m_2$, $m_3$,
respectively. The variable $w_i$ is defined by 
$w_i=m_i/(m_1+m_2+m_3)$ such that $\sum_{i=1}^3 w_i=1.$ 

We shall make use of the Jacobi coordinates $\rho_{1,2}$ and the CM 
coordinate $x$ which are defined by

\bea
x_1 & = & x \, + \, \tfrac{1}{\sqrt{2}}\,w_3\, \rho_1 
            \, - \, \tfrac{1}{\sqrt{6}} \, (2w_2 + w_3)\,\rho_2 \,,
\nn
x_2 & = & x \, + \, \tfrac{1}{\sqrt{2}}\,w_3\, \rho_1
            \, + \, \tfrac{1}{\sqrt{6}}\, (2w_1 + w_3)\,\rho_2\,, 
\nn
x_3 & = & x \, - \, \tfrac{1}{\sqrt{2}}\,(w_1+w_2)\,\rho_1
            \, + \, \tfrac{1}{\sqrt{6}}\,(w_1 - w_2)\,\rho_2\,.
\label{eq:Jacobi}
\ena 
The CM coordinate is given by $x \,= \, \sum_{i=1}^3 w_i x_i$. 
In terms of the Jacobi coordinates one obtains
\be
\sum\limits_{i<j}( x_i - x_j )^2 = \rho_1^2 + \rho_2^2\,.
\label{eq:relative} 
\en 
Note that the choice of Jacobi coordinates is not unique.
By using the particular choice of Jacobi coordinates given by 
Eq.~(\ref{eq:Jacobi})
one obtains the following representation for the
correlation function $\Phi_\Lambda$ in Eq.~(\ref{eq:vertex})  
\bea
\Phi_\Lambda\biggl(\sum_{i<j}( x_i - x_j )^2 \biggr) &=&
\int\!\frac{d^4p_1}{(2\pi)^4}\!\int\!\frac{d^4p_2}{(2\pi)^4}\,
e^{-ip_1(x_1-x_3) - ip_2(x_2-x_3)}\,\bar\Phi_\Lambda(-P^2_1-P^2_2)\,,
\label{eq:Fourier}\\
\bar\Phi_\Lambda(-P^2_1-P^2_2) &=&
\tfrac{1}{9}\int\!d^4\rho_1\int\!d^4\rho_2
\,e^{iP_1\rho_1 + iP_2\rho_2}\,\Phi_\Lambda(\rho^2_1+\rho^2_2)\,,
\nn
&&P_1=\tfrac{1}{\sqrt{2}}(p_1+p_2)\,,\qquad 
  P_2=-\tfrac{1}{\sqrt{6}}(p_1-p_2)\,.
\nonumber
\ena 
This representation is valid for any choice of the set of Jacobi coordinates.
The particular choice~(\ref{eq:Jacobi}) is a preferred choice since it
leads to the specific form of the argument 
$-P^2_1-P^2_2=-\tfrac23(p_1^2+p_2^2+p_1p_2)$.
Since this expression is invariant under the transformations:
$p_1\leftrightarrow p_2$, $p_2\to -p_2-p_1$ and $p_1\to -p_1-p_2$,  
the r.h.s. in Eq.~(\ref{eq:Fourier}) is invariant under permutations 
of all $x_i$ as it should be.

In the next step we have to specify the function 
$\bar\Phi_\Lambda(-P^2_1-P^2_2)\equiv\bar\Phi_\Lambda(-P^2)$ which 
characterizes the finite size of the baryons.  
We will choose a simple Gaussian form for the function $\bar\Phi_\Lambda$: 
\be
\bar\Phi_\Lambda(-P^2) = \exp(P^{\,2}/\Lambda_\Lambda^2) \,,
\label{eq:Gauss}
\en  
where $\Lambda_\Lambda$ is a size parameter parametrizing the distribution 
of quarks inside a $\Lambda$-type baryon. We use different values 
of the $\Lambda_\Lambda$ parameter for different types of the 
$\Lambda$-type baryon: 
$\Lambda_{\Lambda_s}$, $\Lambda_{\Lambda_c}$ and $\Lambda_{\Lambda_b}$ 
for the $\Lambda$, $\Lambda_c$ and $\Lambda_b$ baryons, respectively. 
One has to note that we have used another
definition of the $\Lambda_\Lambda$ in our previous papers:
 $\Lambda_\Lambda = \Lambda_\Lambda^{\rm old}/(3\sqrt{2}).$   

Since $P^{\,2}$ turns into $-\,P^{\,2}_E$ in Euclidean space 
the form~(\ref{eq:Gauss}) has the appropriate falloff behavior in 
the Euclidean region.
We emphasize that any choice for $\Phi_\Lambda$ is appropriate
as long as it falls off sufficiently fast in the ultraviolet region of
Euclidean space to render the corresponding Feynman diagrams ultraviolet 
finite. The choice of a Gaussian form for $\Phi_\Lambda$ has obvious
calculational advantages.

The coupling constants $g_\Lambda$ are determined by 
the compositeness condition suggested by Weinberg~\cite{Weinberg:1962hj}
and Salam~\cite{Salam:1962ap} (for review, see Ref.~\cite{Hayashi:1967hk})
and extensively used in our approach (for details, 
see Ref.~\cite{Efimov:1993ei}).  
The compositeness condition in the case of baryons 
implies that the renormalization constant of 
the baryon wave function is set equal to zero: 
\be 
Z_\Lambda = 1 - \Sigma^\prime_\Lambda(m_\Lambda)  = 0 \, 
\label{eq:Z=0}
\en 
where $\Sigma^\prime_\Lambda$ is the on-shell derivative of the 
$\Lambda$-type baryon mass function $\Sigma_\Lambda$, i.e. 
$\Sigma^\prime_\Lambda=\partial\,\Sigma_\Lambda/\partial\!\!\not\!\!p$, at 
$\not\!\! p=m_\Lambda$. 
The compositeness condition is the
central equation of our covariant quark model.  
The physical meaning, the implications and corollaries of the compositeness 
condition have been discussed in some detail in our previous papers 
(see e.g.~\cite{Branz:2009cd}).

The calculation of the $\Lambda$-type mass function (Fig.2) and 
the electromagnetic vertex (Fig.3) proceed in the same way as 
shown in the nucleon case in Ref.~\cite{Gutsche:2012ze}. 
The matrix elements in momentum space read
\be
\Sigma_\Lambda(p) = 6 g_\Lambda^2
\Bla\Bla\bar\Phi_\Lambda^2(-z_0)
S_Q(k_1+w_1 p)
\Tr\left[S_u(k_2-w_2 p)\gamma^5 S_d(k_2-k_1+w_3 p)\gamma^5\right]\Bra\Bra\,,
\label{eq:mass-lam}
\en
where we use the same shorthand notation $<<...>>$ for the double 
loop-momentum integration. The variable $z_{0}$ is defined as 
\be
z_0 =\tfrac12(k_1-k_2)^2+\tfrac16(k_1+k_2)^2\,.
\label{eq:short2}
\en

The various contributions to the electromagnetic vertex are given by 

\bea
\Lambda^\mu_{\Lambda\, Q}(p,p')
&=& 6\, e_Q\, g_\Lambda^2
\Bla\Bla\bar\Phi_\Lambda(-z_0)
\bar\Phi_\Lambda\Big(-\tfrac12(k_1-k_2+w_3 q)^2
                   -\tfrac16(k_1+k_2+(2w_2+w_3) q)^2\Big)
\nn
&\times&
S_Q(k_1+w_1 p')\gamma^\mu S_Q(k_1+w_1 p'+q)
\Tr\left[S_u(k_2-w_2 p')\gamma^5 S_d(k_2-k_1+w_3 p')\gamma^5\right]\Bra\Bra\,,
\nn\nn
\Lambda^\mu_{\Lambda\, u}(p,p')
&=& -\,6\, e_u\, g_\Lambda^2
\Bla\Bla\bar\Phi_\Lambda(-z_0)
  \bar\Phi_\Lambda\Big(-\tfrac12(k_1-k_2+w_3 q)^2
                     -\tfrac16(k_1+k_2-(2w_1+w_3) q)^2\Big)
\nn
&\times&
S_Q(k_1+w_1 p')
\Tr\left[S_u(k_2-w_2 p'-q)\gamma^\mu S_u(k_2-w_2 p')
\gamma^5 S_d(k_2-k_1+w_3 p')\gamma^5\right]\Bra\Bra\,,
\nn\nn
\Lambda^\mu_{\Lambda\, d}(p,p')
&=& 6\, e_d\, g_\Lambda^2
\Bla\Bla\bar\Phi_\Lambda(-z_0)
  \bar\Phi_\Lambda\Big(-\tfrac12(k_1-k_2-(w_1+w_2) q)^2
                     -\tfrac16(k_1+k_2-(w_1-w_2) q)^2\Big)
\nn
&\times&
S_Q(k_1+w_1 p')
\Tr\left[S_u(k_2-w_2 p')
\gamma^5 S_d(k_2-k_1+w_3 p')\gamma^\mu S_d(k_2-k_1+w_3 p'+q) 
\gamma^5\right]\Bra\Bra\,,
\nn\nn
\Lambda^\mu_{\Lambda\,(a)}(p,p')
&=& 6\, g_\Lambda^2
\Bla\Bla\bar\Phi_\Lambda(-z_0)
\tilde E_\Lambda^\mu(k_1+w_1p',-k_2+w_2p',k_2-k_1+w_3p';q)
\nn
&\times&
S_Q(k_1+w_1 p')
\Tr\left[S_u(k_2-w_2 p')\gamma^5 S_d(k_2-k_1+w_3 p')\gamma^5\right]\Bra\Bra\,,
\nn\nn
\Lambda^\mu_{\Lambda\, (b)}(p,p')
&=& 6\, g_\Lambda^2
\la\la\bar\Phi_\Lambda(-z_0)
\tilde E_\Lambda^\mu(k_1+w_1p,-k_2+w_2p,k_2-k_1+w_3p;-q)
\nn
&\times&
S_Q(k_1+w_1 p)
\Tr\left[S_u(k_2-w_2 p)\gamma^5 S_d(k_2-k_1+w_3 p)\gamma^5\right]\Bra\Bra\,.
\label{eq:em-vertex-lam}
\ena 
The free quark propagator in momentum space is given by 
\bea
S_f(k)  = \frac{1}{m_{f}-\not\! k}  
\label{eq:Green-quark2}
\ena 
where $f=u,d,s,c,b$ denotes the flavor of the freely propagating quark. We 
restrict ourselves to 
isospin invariance $m_u=m_d$. 
The function $\tilde E_\Lambda^\mu(r_1,r_2.r_3;r)$ is defined as 
\bea 
\widetilde E^\alpha_\Lambda(p_1,p_2,p_3;r) =
\sum\limits_{i=1}^3e_{q_i}\int\limits_0^1 d\tau
\Big\{- w_{i1}(w_{i1}r^\alpha + 2 q_1^\alpha) \bar\Phi'_\Lambda(-z_1)
      - w_{i2}(w_{i2}r^\alpha + 2 q_2^\alpha) \bar\Phi'_\Lambda(-z_2) 
\Big\} \,. 
\label{eq:cur-em-nonloc}
\ena
The variables $q_1=\sum_{i=1}^3w_{i1}r_i$ and $q_2=\sum_{i=1}^3w_{i2}r_i$
in $\tilde E_\Lambda^\mu(r_1,r_2.r_3;r)$ can be seen to be related to the
loop momenta by
\be
q_1 = \tfrac{1}{\sqrt{2}}(k_1-k_2)\,, \qquad
q_2 = -\,\tfrac{1}{\sqrt{6}}(k_1+k_2)
\label{eq:q1q2}
\en
for both bubble diagrams. By using Eq.~(\ref{eq:q1q2})
one finds the $q=0$ relations 
\bea
\Lambda^\mu_{\Lambda\,(a)}(p,p)+\Lambda^\mu_{\Lambda\,(b)}(p,p)
&=& - \,8\,g_\Lambda^2
\Bla\Bla
(Q_1 k_1^\mu + Q_2 k_2^\mu)  \bar\Phi'_\Lambda(-z_0)\bar\Phi_\Lambda(-z_0)
\nn
&\times&
S_Q(k_1+w_1 p)
\Tr\left[S_u(k_2-w_2 p)\gamma^5 S_d(k_2-k_1+w_3 p)\gamma^5\right]\Bra\Bra\,,
\nn\nn
Q_1 &=& e_1 (w_2+2w_3) - e_2(w_1-w_3)  - e_3 (2w_1+w_2)\,, 
\nn
Q_2 &=&   e_1 (w_2-w_3)  - e_2(w_1+2w_3) + e_3 (w_1+2w_2)\,,
\label{Q1Q2}
\ena
where the subscripts on the charges $e_{i}$ refer to the flavors
of the three quarks:
$"i=1"\to "s,c,b"$, $"i=2"\to "u"$ and $"i=3"\to "d"$.  
Next we will use an integration-by-parts identity to write  
\be
\Bla\Bla\frac{\partial}{\partial k_i^\mu}
\Big\{\bar\Phi^2_\Lambda(-z_0)
S_Q(k_1+w_1 p)
\Tr\left[S_u(k_2-w_2 p)\gamma^5 S_d(k_2-k_1+w_3 p)\gamma^5\right]
\Big\}\Bra\Bra\, \equiv\, 0\,, \qquad (i=1,2)\,.
\label{eq:IP-lam}
\en
One finds
\bea
\Bla\Bla 
k_1^\mu\,A_0\Bra\Bra &=& \frac14 \Bla\Bla(2\,A^\mu_1 + A^\mu_2 - A^\mu_3 )
\Bra\Bra\,,
\nn
\Bla\Bla k_2^\mu\,A_0 \Bra\Bra &=& \frac14 
\Bla\Bla ( A^\mu_1 + 2\,A^\mu_2 + A^\mu_3 ) \Bra\Bra\,,
\label{eq:identity-lam}
\ena
where
\bea
A_0 &=& \bar\Phi'_\Lambda(-z_0)\,\bar\Phi_\Lambda(-z_0)
S_Q(k_1+w_1 p)
\Tr\left[S_u(k_2-w_2 p)\gamma^5 S_d(k_2-k_1+w_3 p)\gamma^5\right]\,,
\nn
A^\mu_1&=& \bar\Phi^2_\Lambda(-z_0)
S_Q(k_1+w_1 p)\gamma^\mu S_Q(k_1+w_1 p)
\Tr\left[S_u(k_2-w_2 p)\gamma^5 S_d(k_2-k_1+w_3 p)\gamma^5\right]
\nn
A^\mu_2 &=& \bar\Phi^2_\Lambda(-z_0)
S_Q(k_1+w_1 p)
\Tr\left[S_u(k_2-w_2 p)\gamma^\mu S_u(k_2-w_2 p) 
\gamma^5 S_d(k_2-k_1+w_3 p)\gamma^5\right]\,,
\nn
A^\mu_3 &=& \bar\Phi^2_\Lambda(-z_0)
S_Q(k_1+w_1 p)
\Tr\left[S_u(k_2-w_2 p)
\gamma^5 S_d(k_2-k_1+w_3 p)\gamma^\mu S_d(k_2-k_1+w_3 p)\gamma^5\right]\,.
\label{eq:bricks}
\ena
Using these identities and collecting all pieces together, one obtains
\be
\Lambda^\mu_{\Lambda}(p,p)
=(e_Q+e_u+e_d)\,\frac{\partial\Sigma_\Lambda(p)}{\partial p^\mu}\,,
\qquad \not\! p=m_\Lambda\,.
\label{eq:WI-lam}
\en
As was discussed above, this Ward identity allows one to use
the compositeness condition $Z_\Lambda=0$ written in the form
\be
\Lambda^\mu_{\Lambda}(p,p) = \gamma^\mu\,,\qquad \not\! p=m_\Lambda\,,
\label{eq:Z=0-lam}
\en
where we take $e_Q=e_c$ for the present discussion.  
Again we have checked analytically that,
on the $\Lambda$-type baryon mass shell, the triangle diagrams are gauge 
invariant by themselves
and the non-gauge invariant parts coming from the bubble diagrams
Fig.3(c) and 3(d) cancel each other. 
The standard definition of the electromagnetic form factors is
\be 
\Lambda^{\mu}_{\Lambda}(p,p^{\prime}) = 
\gamma_{\mu}F_{1}(q^2)-\frac{i\sigma^{\mu q}}{2m_{\Lambda}} F_2(q^2)\,,
\label{eq:em-ff-lam}
\en 
where 
$\sigma^{\mu q}=\tfrac{i}{2}(\gamma^\mu\gamma^\nu-\gamma^\nu\gamma^\mu)q_\nu.$ 
The magnetic moment of the $\Lambda$-type baryon is defined by 
\be 
\mu_\Lambda \, = \, 
\left( \, F_{1}(0) + F_{2}(0) \, \right) \,\, \frac{e}{2 m_\Lambda} \,. 
\en 
In terms of the nuclear magneton (n.m.) 
$\frac{e}{2 m_p}$  
the $\Lambda$--hyperon magnetic moment is given by 
\be 
\mu_\Lambda  = 
\ ( \, F_1(0) + F_2(0) \, ) \,\, \frac{m_p}{m_\Lambda}  
\en 
where $m_p$ is the proton mass. 

The magnetic moment has been measured for the $\Lambda_s$ only and
is given by~\cite{Beringer:1900zz}  
\be\label{L_mm}
\mu_{\Lambda_s} = -\,0.613 \pm 0.004 \,. 
\en 

Since we want to fit the size
parameters $\Lambda_{\Lambda_s}$,
$\Lambda_{\Lambda_c}$ and $\Lambda_{\Lambda_b}$ also to semileptonic
$b \to c$ and $c \to s$ charged current transitions we need to briefly
set up the formalism for the description of these transitions,
i.e. for the transitions
\bea\label{Matr_sl}
M(\Lambda_{Q[ud]} \to \Lambda_{Q'[ud]} \ell^- \bar\nu_\ell) =
\frac{G_F}{\sqrt{2}} \, V_{QQ'} \,
\la\Lambda_{Q'[ud]}\,|\,\bar{Q'}\,O^\mu\, Q\,|\,\Lambda_{Q[ud]}\ra
\ (\ell^- \, O_\mu \, \bar\nu_\ell) \,,
\ena
where $O^\mu=\gamma^\mu(1-\gamma_5)$.
These processes are described in our model by the triangle diagram 
shown in Fig.4. 
The hadronic matrix element in~(\ref{Matr_sl}) is expanded in terms
of the dimensionless
form factors $f_i^{J}$ ($i=1, 2, 3$ and $J = V, A$), viz.
\bea\label{FF_sl}
\la \Lambda_{Q'[ud]}\,|\,\bar Q'\, \gamma^\mu\, Q\,| \Lambda_{Q[ud]} \ra
&=&
\bar u_2(p_2)
\Big[ f^V_1(q^2) \gamma^\mu - f^V_2(q^2) i\sigma^{\mu q}/M_1
     + f^V_3(q^2) q^\mu/M_1 \Big] u_1(p_1)\,,
\nn
\la \Lambda_{Q'[ud]}\,|\,\bar Q'\,\gamma^\mu\,\gamma^5 Q\,|
\Lambda_{Q[ud]}\ra
&=& \bar u_2(p_2)
\Big[ f^A_1(q^2) \gamma^\mu - f^A_2(q^2) i\sigma^{\mu q}/M_1
     + f^A_3(q^2) q^\mu/M_1 \Big]\gamma^5 u_1(p_1)\,.
\ena
The calculation of the form factors in our approach is automated by the
use
of FORM~\cite{Vermaseren:2000nd} and FORTRAN packages written 
for this purpose. To be able to compare with our earlier 
calculations which did not contain confinement the packages exist 
for the confined and the unconfined versions of the covariant quark model.

The results of our numerical calculations are well represented
by the double--pole parametrization
\be\label{DPP}
f(s)=\frac{f(0)}{1 - a s + b s^2}\,,
\en
where $s=q^2/M_{\Lambda_i}^2$ and $M_{\Lambda_i}$ is the
mass of the initial baryon.
Using such a parametrization facilitates further integrations.
The values of $f(0)$, $a$ and $b$ are listed
in Tables~\ref{tab:fflcs}-\ref{tab:fflbs}.
We plot the form factors in the full kinematical regions
($0 \le s \le s_{\rm max}$) in Figs.~\ref{fig:ff_cs} ($c \to s$)
and \ref{fig:ff_bc} ($b \to c$):
solid and dotted lines correspond to approximated and exact results,
respectively. The agreement between the approximate and numerically
calculated form factors is excellent except for the form factors
$A_{2}(s)$ and $T_{A1}(s)$ for which the agreement is not so good.
This is due to the steep ascent of $A_{2}$ and descent of $T_{A1}$ 
at the high end of the $q^{2}$ spectrum. A better fit would require 
the addition of a linear $s$ term in the numerator of Eq.~(\ref{DPP}) 
for these two form factors.

As in the case of the rare meson decays $B \to K(K^\ast) \bar{\ell}
\ell$
and $B_c \to D(D^\ast) \bar{\ell} \ell$ treated in 
Ref.~\cite{Faessler:2002ut}
all physical observables
(the rate $\Gamma(\Lambda_{Q[ud]} \to \Lambda_{Q'[ud]}
+ \ell^- \bar\nu_\ell)$ and asymmetry parameter
$\alpha$ etc.) are conveniently written down
in terms of helicity
amplitudes $H_{\lambda_2,\lambda_j}$. Note that the corresponding
helicity amplitudes do not carry any superscripts as they are needed
in the description of the corresponding rare decays.
The relations of these helicity amplitudes to the
invariant form factors $f^J_i$ is given in Appendix~\ref{app:Hel_amp}.
The rate for the charged current transitions can be written as
\bea
\Gamma(\Lambda_{Q[ud]} \to \Lambda_{Q'[ud]}
+ \ell^- \bar\nu_\ell) &=& \frac{G_F^2 |V_{QQ'}|^2}{192 \pi^3 M_1^2} \,
\int\limits_{m_\ell^2}^{(M_1-M_2)^2} \, \frac{dq^2}{q^2} \,
(q^2 - m_\ell^2)^2 \, |{\bf p_2}|  \ {\cal H}\,.
\ena
For the asymmetry parameter $\alpha$ in these decays  
(see Ref.~\cite{Ivanov:1996fj} for the definition of the 
asymmetry parameter) one obtains
\bea
\alpha  &=&
\frac{\int\limits_{m_\ell^2}^{(M_1-M_2)^2} \, \frac{dq^2}{q^2} \,
(q^2 - m_\ell^2)^2 \, |{\bf p_2}|  \ {\cal G}}
{\int\limits_{m_\ell^2}^{(M_1-M_2)^2} \, \frac{dq^2}{q^2} \,
(q^2 - m_\ell^2)^2 \, |{\bf p_2}|  \ {\cal H}} \,,
\ena
where
\bea
{\cal H} &=& H_U + H_L + \frac{m_\ell^2}{2q^2}
\biggl( 3H_S + H_U + H_L \biggr)\,, \nonumber\\
{\cal G} &=& H_P + H_{L_P} + \frac{m_\ell^2}{2q^2}
\biggl( 3H_{S_P} + H_P + H_{L_P} \biggr)\,,
\ena
and the
$H_X$ are the following combinations of the helicity amplitudes:
\bea
H_U &=& |H_{\frac{1}{2}1}|^2 + |H_{-\frac{1}{2}-1}|^2 \,, \nonumber\\
H_L &=& |H_{\frac{1}{2}0}|^2 + |H_{-\frac{1}{2}0}|^2  \,, \nonumber\\
H_P &=& |H_{\frac{1}{2}1}|^2 - |H_{-\frac{1}{2}-1}|^2 \,, \nonumber\\
H_{L_P} &=& |H_{\frac{1}{2}0}|^2 - |H_{-\frac{1}{2}0}|^2  \,, \nonumber\\
H_S &=& |H_{\frac{1}{2}t}|^2 + |H_{-\frac{1}{2}t}|^2 \,, \nonumber\\
H_{S_P} &=& |H_{\frac{1}{2}t}|^2 - |H_{-\frac{1}{2}t}|^2 \,.
\ena

We determine the set of size parameters $\Lambda_{\Lambda_s}$, 
$\Lambda_{\Lambda_c}$ and $\Lambda_{\Lambda_b}$ by fitting data on the 
magnetic moment of the $\Lambda$-hyperon~(\ref{L_mm}) and the nominal 
branching ratios of the semileptonic 
decays $\Lambda_c \to \Lambda \ell^+ \nu_\ell$ and 
$\Lambda_b \to \Lambda_c \ell^- \bar\nu_\ell$ by a one-parameter fit to these 
values. Using the results of Table~\ref{tab:fflbc} for the 
$\Lambda_b \to \Lambda_c$ case one finds the zero recoil values 
$f_{1}^{V}=0.87$ and $f_{1}^{A}=0.86$. These form factor values are somewhat
lower than the values  $f_{1}^{V}=f_{1}^{A}=1$ predicted by HQET. This can
be interpreted as an indication that the nominal value for the 
$\Lambda_b \to \Lambda_c \ell^- \bar\nu_\ell$ 
branching ratio listed in the Particle Data Group~\cite{Beringer:1900zz} 
and used by us in our fit is underestimated. 
With the choice of dimensional parameters 
$\Lambda_{\Lambda_s} = 0.490$ GeV, 
$\Lambda_{\Lambda_c} = 0.864$ GeV 
and $\Lambda_{\Lambda_b} = 0.569$ GeV we get a reasonable agreement 
with current data on exclusive Cabibbo-allowed decays of 
$\Lambda_c$ and $\Lambda_b$ (see Tables~\ref{tab:rates} 
and~\ref{tab:asym}). For the magnetic moments we get the following results: 
\bea
\quad \mu_{\Lambda_s} = -0.73\,,
\quad \mu_{\Lambda_c} =  0.39\,,
\quad \mu_{\Lambda_b} = -0.06\,, 
\label{eq:mag-mom-lam}
\ena 
which compares well with data for the $\mu_{\Lambda_s}$ and 
theoretical estimates for the 
$\mu_{\Lambda_c}$ and $\mu_{\Lambda_b}$ (see the detailed discussion 
in Ref.~\cite{Faessler:2006ft}). In particular, 
our present results for the magnetic moments of heavy $\Lambda$-hyperons 
are very close to our predictions done before in the model 
without taking account of the mechanism of quark confinement: 
$\mu_{\Lambda_c} =  0.42$ and 
$\mu_{\Lambda_b} = -0.06$~\cite{Faessler:2006ft}. 
Note, the other model parameters $m_q$ and $\lambda$ 
are taken from the fit done in the Ref.~\cite{Ivanov:2011aa}:
\be
\def\arraystretch{2}
\begin{array}{ccccccc}
     m_u        &      m_s        &      m_c       &     m_b & \lambda  &   
\\\hline
 \ \ 0.235\ \   &  \ \ 0.424\ \   &  \ \ 2.16\ \   &  \ \ 5.09\ \   & 
\ \ 0.181\ \   & \ {\rm GeV} 
\end{array}
\label{eq: fitmas}
\en

\section{
The rare baryon decays $\Lambda_b\to \Lambda + \ell^+\ell^-$ 
and $\Lambda_b\to \Lambda + \gamma$
}  

The effective Hamiltonian \cite{Buchalla:1995vs}
leads to the quark decay amplitudes $b\to s l^+l^-$ and 
$b\to s \gamma$: 
\bea
M(b\to s\ell^+\ell^-) & = & 
\frac{G_F}{\sqrt{2}}\frac{\alpha \lambda_t}{2\pi} \, \, 
\biggl\{
 C_9^{\rm eff}\,
       \left( \bar{s} O^\mu b \right) \,\left(  \bar\ell\gamma_\mu \ell \right)
+C_{10}\left( \bar{s} O^\mu  b \right) \,
\left(  \bar \ell\gamma_\mu\gamma_5 \ell \right)
\nn
&-&\frac{2}{q^2}\,C_7^{\rm eff}\, 
\left[
  m_b \left(\bar{s}\,i\sigma^{\mu q}\,(1+\gamma^5)\, \,b \right)
+ m_s \left(\bar{s}\,i\sigma^{\mu q}\,(1-\gamma^5)\, \,b \right)
\right]
\left( \bar \ell \gamma_\mu \ell \right) 
\biggr\} \,. \label{eq:free} 
\ena 
and 
\bea
M(b\to s\gamma) \ = \ 
- \frac{G_F}{\sqrt{2}} \frac{e \lambda_t}{4\pi^2} \,   
C_7^{\rm eff}\, 
\left[
  m_b \left(\bar{s}\,i\sigma^{\mu q}\,(1+\gamma^5)\, \,b \right)
+ m_s \left(\bar{s}\,i\sigma^{\mu q}\,(1-\gamma^5)\, \,b \right)
\right]
\epsilon_\mu \,, 
\ena 
where $\sigma^{\mu q}=\tfrac{i}{2}(\gamma^\mu\gamma^\nu 
-\gamma^\nu\gamma^\mu)q_\nu$, $O^\mu = \gamma^\mu (1 - \gamma^5)$ 
and $\lambda_t\equiv V^\dagger_{ts} V_{tb}$.
The Wilson coefficient $C_9^{\rm eff}$ effectively takes 
into account, first, the contributions from the four-quark
operators $Q_i (i = 1, \cdots, 6)$ and, second, the nonperturbative effects 
(long--distance contributions) coming from the $c\bar c$-resonance 
contributions what are, as usual, parametrized by a Breit-Wigner 
ansatz~\cite{Ali:1991is} (see details in Appendix~\ref{app:Wilson}). 

The Feynman diagrams contributing to the exclusive transitions 
$\Lambda_b\to \Lambda \bar \ell \ell$ and $\Lambda_b\to \Lambda \gamma$ 
are shown in Fig.6. 
The corresponding matrix elements of the exclusive transitions 
$\Lambda_b\to \Lambda \bar \ell \ell$ and $\Lambda_b\to \Lambda \gamma$ 
are defined by 
\bea\label{Matr_bsll}
M(\Lambda_b\to \Lambda \bar \ell \ell) & = & 
\frac{G_F}{\sqrt{2}} \frac{ \alpha\lambda_t}{2\,\pi}  
\biggl\{
 C_9^{\rm eff}\,
  \la\Lambda\,|\,\bar{s}\,O^\mu\, b\,|\,\Lambda_b\ra \,\bar\ell\gamma_\mu \ell
\nonumber\\
&+&
C_{10}\, \la\Lambda\,|\,\bar{s}\,O^\mu\, b\,|\,\Lambda_b\ra
\, \bar\ell\gamma_\mu \gamma_5 \ell
\nn
&-& 
\frac{2m_b}{q^2}\,C_7^{\rm eff}\, 
\la\Lambda\,|\,  \bar{s}\,i\sigma^{\mu q}\,(1+\gamma^5)\,\,b\,|\,\Lambda_b\ra
\,  \bar\ell\gamma_\mu \ell 
\biggr\} \, 
\label{eq:exclus}
\ena
and 
\bea\label{Matr_bsg}
M(\Lambda_b \to \Lambda \gamma)  =  
- \frac{G_F}{\sqrt{2}} \frac{e \lambda_t}{4\,\pi^2}  \, m_b 
\,C_7^{\rm eff}\, \la\Lambda\,|\,  
\bar{s}\,i\sigma^{\mu q}\,(1+\gamma^5)\,\,b\,|\,\Lambda_b\ra
\,  \epsilon_\mu \,. 
\ena

The hadronic matrix elements in~(\ref{Matr_bsll}) 
and~(\ref{Matr_bsg}) are expanded in terms of dimensionless 
form factors $f_i^{J}$ ($i=1, 2, 3$ and $J = V, A, TV, TA$), viz. 
\bea\label{FF_def}
\la B_2\,|\,\bar s\, \gamma^\mu\, b\,| B_1 \ra &=&
\bar u_2(p_2)
\Big[ f^V_1(q^2) \gamma^\mu - f^V_2(q^2) i\sigma^{\mu q}/M_1
     + f^V_3(q^2) q^\mu/M_1 \Big] u_1(p_1)\,,
\nn
\la B_2\,|\,\bar s\, \gamma^\mu\gamma^5\, b\,| B_1 \ra &=&
\bar u_2(p_2)
\Big[ f^A_1(q^2) \gamma^\mu - f^A_2(q^2) i\sigma^{\mu q}/M_1
     + f^A_3(q^2) q^\mu/M_1 \Big]\gamma^5 u_1(p_1)\,,
\nn
\la B_2\,|\,\bar s\, i\sigma^{\mu q}/M_1\, b\,| B_1 \ra &=&
\bar u_2(p_2)
\Big[ f^{TV}_1(q^2) (\gamma^\mu q^2 - q^\mu \not\! q)/M_1^2
- f^{TV}_2(q^2) i\sigma^{\mu q}/M_1\Big] u_1(p_1)\,,
\nn
\la B_2\,|\,\bar s\, i\sigma^{\mu q}\gamma^5/M_1\, b\,| B_1 \ra &=&
\bar u_2(p_2)
\Big[ f^{TA}_1(q^2) (\gamma^\mu q^2 - q^\mu \not\! q)/M_1^2
- f^{TA}_2(q^2) i\sigma^{\mu q}/M_1\Big]\gamma^5 u_1(p_1)\,.
\label{eq:rare-ff}
\ena
One can see that, in comparison with the Cabibbo-allowed $b \to c$ and 
$c \to s$ transitions, one has four more form factors $f_{1,2}^{TV,TA}$. 
As was mentioned before, the numerical results for the invariant form factors 
are well represented by the double--pole parametrization~(\ref{DPP}).  
The values of $f(0)$, $a$ and $b$ for the approximated form factors 
describing the $b \to s$ flavor transitions are listed  
in Table~\ref{tab:fflbs}. 
The plots of the form factors in the full kinematical regions 
($0 \le s \le s_{\rm max}$) are shown in Fig.~\ref{fig:ff_bs}: 
the solid and dotted lines correspond to approximated and exact results, 
respectively.  
One can see that both curves are in close agreement with each other. 
There is only a small disagreement 
for the suppressed form factors $f^A_2$ and $f^{TA}_2$. 

Note, that  a form factor approximation similar to the
form~(\ref{DPP}) was successfully used by us in Ref.~\cite{Faessler:2002ut} 
in the analysis of rare decays of bottom mesons. The relations of 
the helicity amplitudes and invariant form factors are given in 
Appendix~\ref{app:Hel_amp}. 

Similar to Eq.~(\ref{pz2}) the angular decay distribution for the cascade 
decay $\Lambda_{b}\to\Lambda(\to p\pi^{-}) \gamma$ can be written as
\bea
\frac{d\Gamma(\Lambda_{b}\to \Lambda(\to p \pi^{-})\gamma)}
{d\cos\theta_{B}} 
={\rm Br}(\Lambda \to p\pi^{-})
\,\frac{1}{2}\,\Gamma(\Lambda_{b}\to\Lambda \gamma)(1+\alpha_{B}P_{z}^{\Lambda}
\cos\theta_{B}) \,, 
\label{onephoton}
\ena 
where $\alpha_{B}$ is the asymmetry parameter in the decay 
$\Lambda \to p + \pi^-$ for which we take the experimental value 
$\alpha_{B}=0.642\pm0.013$~\cite{Beringer:1900zz}. 
The $\Lambda_{b}\to\Lambda \gamma$ decay rate is calculated according to
\bea 
\Gamma(\Lambda_b \to \Lambda \gamma) &=& 
\frac{\alpha}{2} \,  
\left(\frac{G_F \, M_1^2 \, |\lambda_t|}{4\pi^2 \sqrt{2}}\right)^2 
\, \frac{|{\bf p_2}|}{2 M_1^2} 
\ \biggl[ |H_{\frac{1}{2}1}^V|^2 + |H_{-\frac{1}{2}-1}^V|^2 
        + |H_{\frac{1}{2}1}^A|^2 + |H_{-\frac{1}{2}-1}^A|^2 
\biggr] \nonumber\\ 
&=& \frac{\alpha}{2} \, \left(\frac{G_F  m_b \, |\lambda_t| \, 
C_7^{\rm eff}}{4\pi^2 \sqrt{2}}\right)^2 
\, \frac{(M_1^2-M_2^2)^3}{M_1^3} \,  
\ \biggl[ \Big(f_2^{TV}(0)\Big)^2\ 
+ \Big(f_2^{TA}(0)\Big)^2 \biggr] \,.   
\ena 
As before the expressions of helicity amplitudes in terms of invariant 
form factors are given in Appendix~\ref{app:Hel_amp}. 
The $z$--component of the polarization of the $\Lambda$ appearing 
in Eq.~(\ref{onephoton}) is given by
\bea 
\tilde P_{z}^{\Lambda}=\frac{\tilde W_{\frac{1}{2}\,\frac{1}{2}} 
- \tilde W_{-\frac{1}{2}\,-\frac{1}{2}}}
{\tilde W_{\frac{1}{2}\,\frac{1}{2}} 
+ \tilde W_{-\frac{1}{2}\,-\frac{1}{2}}}
\ena 
where 
\bea 
\tilde W_{ \lambda_{\Lambda} \lambda_{\Lambda} }  \propto 
H_{ \lambda_{\Lambda},\lambda_{j}=2\lambda_{\Lambda} }
H^{\dagger}_{ \lambda_{\Lambda}, \lambda_{j}=2\lambda_{\Lambda} } \,. 
\ena 
We have used a tilde notation in $\tilde P_{z}^{\Lambda}$ and
$\tilde W_{ \lambda_{\Lambda} \lambda_{\Lambda} }$ in order to distinguish 
these quantities from the corresponding quantities in the dilepton modes. 
One has, 
\bea 
\tilde P_{z}^{\Lambda}
= - 2 \frac{f_2^{TV}(0)f_2^{TA}(0)}{(f_2^{TV}(0))^2 + (f_2^{TA}(0))^2} \,. 
\ena 
Note, that $f_2^{TV}(0) \equiv f_2^{TA}(0)$ (see proof in 
Appendix~\ref{app:TV2_TA2}), which is in agreement with 
statement of Ref.~\cite{Hiller:2001zj}. 
Therefore, $\tilde P_{z}^{\Lambda} \equiv -1$ and finally 
\bea
\label{raddecay}
\frac{1}{\Gamma_{\rm tot}} \, 
\frac{d\Gamma(\Lambda_{b}\to \Lambda(\to p \pi^{-})\gamma)}
{d\cos\theta_{B}} 
={\rm Br}(\Lambda \to p\pi^{-})
\,\frac{1}{2}\,{\rm Br}(\Lambda_{b}\to\Lambda \gamma) 
(1-\alpha_{B}\cos\theta_{B}) \,. 
\ena

\section{Numerical results}

In this section we present a detailed numerical analysis of the 
rare decays $\Lambda_{b} \to \Lambda \ell^{+}\ell^{-}$ and 
$\Lambda_{b} \to \Lambda \gamma$. 
In Figs.~\ref{fig:Distee}-\ref{fig:ang_radiative} we present 
two-dimensional and three-dimensional polar angle and polarization 
distributions. 
Our predictions for differential rates are shown in the two-dimensional plots 
Figs.~\ref{fig:Distee}-\ref{fig:Disttt}, 
lepton-side and hadron-side forward-backward asymmetries are displayed 
in Figs.~\ref{fig:AeFB}-\ref{fig:AtFB} and 
in Figs.~\ref{fig:AheFB}-\ref{fig:AhtFB}, 
respectively. In all the three cases we plot two respective results what are
labelled by ``LD''
(including long-distance contributions) and ``NLD'' (no long-distance
contributions). In Figs.~\ref{fig:LSDee}-\ref{fig:LSDtt} we provide
three-dimensional plots of the $s$-dependence of the lepton-side 
polar angle decay distributions for each of the $e,\mu$ and $\tau$-cases. 
In Figs.~\ref{fig:HSDee}-\ref{fig:HSDtt} we do the same for the hadron-side 
decay distribution. One clearly sees the long-distance contributions of the  
charmonium resonances. 
In Figs.~\ref{fig:HpSDee}-\ref{fig:HpSDtt} we show plots of of the $\cos\theta$
and $s$ dependence of the longitudinal polarization $P^{\Lambda}_{z}$ 
of the daughter baryon $\Lambda$, again for the $e,\mu$ and $\tau$-cases. 
The polarization is large and negative in all cases. Finally, in
Fig.~\ref{fig:ang_radiative} we show the (hadron-side) polar angle
distribution of the radiative decay 
$\Lambda_{b}\to \Lambda(\to p \pi^{-})\gamma$. As expected from 
Eq.~(\ref{raddecay}) and from the discussion in Sec.~IV the $\cos\theta_{B}$
dependence is given by a straight-line plot with a slope proportional 
to the asymmetry parameter $\alpha_{B}$. 

In Table~\ref{tab:sl_leptons} 
we present our results for the branching ratios of the rare dileptonic 
decay $\Lambda_{b} \to \Lambda \ell^{+}\ell^{-}$.  
The results without long-distance effects are shown in brackets. 
Our predictions for the radiative decay $\Lambda_{b} \to \Lambda \gamma$ 
are shown in Table~\ref{tab:sl_gamma}. Here we also present the results 
of other approaches using the compilation of Ref.~\cite{Lb_QCDSR}. 
The results for the integrated lepton--side and hadron--side 
forward--backward asymmetries are shown in Tables~\ref{tab:FB_int}. 

In our calculations 
we do not include the regions around the two charmonium resonances 
$R_{c\bar c} = J/\psi, \Psi(2S)$. We exclude the regions
$M_{J/\Psi} - 0.20$ GeV
to $M_{J/\Psi} + 0.04$ GeV and  $M_{\Psi(2S)} - 0.10$ GeV
to $M_{\Psi(2S)} + 0.02$ GeV.
As stressed in Ref.~\cite{Mott:2011cx} these regions 
are experimentally vetoed, 
because the rates of nonleptonic decays $\Lambda_b \to \Lambda + R_{c\bar c}$, 
followed by the dileptonic decays of the charmonium, are much larger than 
rates 
of the $b \to s$-induced rare decays $\Lambda_b \to \Lambda \ell^+ \ell^-$. 
Vetoing the regions near the charmonium resonances leads 
to physically acceptable results --- 
the predictions with and without the inclusion of long--distance effects are 
comparable with each other. Otherwise (without such a vetoing) the 
results with long--distance effects are dramatically enhanced (as shown 
in different theoretical calculations, see also results in 
Table~\ref{tab:sl_leptons}). 

\section{Summary and conclusions}

i) We have used the helicity formalism to express a number of 
observables in the rare baryon decay
$\Lambda_{b} \to \Lambda (\to p \pi^{-})\, \ell^{+}\ell^{-}$ in
terms of a basic set of hadronic helicity structure functions. 
In the helicity method one provides complete information on the spin density
matrix of each particle in the cascade decay chain which can be
conveniently read out by considering angular decay distributions in the
rest frame of that particular particle.
We hope that we have demonstrated the advantages of the helicity method over
a traditional covariant calculation. Every conceivable observable can be 
written in terms of bilinear forms of the basic hadronic 
helicity amplitudes calculated in this paper while a covariant evaluation 
requires an {\it ab initio} calculation for every new observable. We have 
provided some examples of such observables in this paper.

ii) There is a multitude of observables to be explored experimentally
and theoretically. These include the polarization of the decaying baryon and
single--lepton and double--lepton polarization asymmetries what have not
been discussed in this paper. The advantage of
the helicity method is that it is straightforward to define any of the
observables of the problem and to express them in terms of bilinear forms
of the hadronic helicity matrix elements defined and calculated in this
paper. There is no need to restart a covariant calculation for every new
spin observable. We mention that it is well--known that hadronically
produced hyperons are found to be
partially polarized perpendicular to the production plane. Similar
polarization effects are expected to occur for hadronically produced
$\Lambda_{b}$'s. Also
$\Lambda_{b}$'s from $Z \to \Lambda_{b} \bar{\Lambda}_{b}$ are expected to
be highly polarized. It would be important to take into account such
polarization effects in the angular decay distribution of the
$\Lambda_{b}$.

iii) We have provided results with and without taking the so--called
long distance effects into account, for which the long distance effects
are calculated by the contributions of the $J/\Psi$ and $\Psi(2S)$ resonances. 

iv) We have described from a unified point of view exclusive Cabibbo-allowed 
semileptonic decays $\Lambda_b \to \Lambda_c \ell^- \bar\nu_\ell$,
$\Lambda_c \to \Lambda \ell^+ \nu_\ell$ and rare decays 
$\Lambda_b \to \Lambda \ell^+ \ell^-$, 
$\Lambda_b \to \Lambda \gamma$ with the use of only three model parameters: 
the size parameters $\Lambda_{\Lambda_{s}}$, $\Lambda_{\Lambda_{c}}$ and
$\Lambda_{\Lambda_{b}}$ defining the distribution of quarks 
in the $\Lambda$, $\Lambda_c$ and $\Lambda_b$ baryons. 

v) The helicity formulas introduced in this paper can be used as input
in a MC event generator patterned after the existing event generator for
$\Xi^{0}(\uparrow) \to \Sigma^{+}(\to p\pi^{0})\ell^{-}\bar{\nu}_{\ell}$
$\ell=(e,\mu)$  which is described and put to use
in~\cite{Kadeer:2005aq} and which has been used by the NA48 Collaboration
to analyze its data on the above decay~\cite{Batley:2012mi}.
Such a MC event generator would require a viable
parametrization of the hadronic transition helicity amplitudes for the
whole range of $q^{2}$ which we provide in this paper.

vi) In a future work, we plan to discuss further rare baryonic ($b\to s$)
and $(b \to d)$ decays such as 
$\Omega_{b}^{-} \to \Omega^{-}\,\ell^{+}\ell^{-}$,
$\Xi_{b}^{-} \to \Sigma^{-}\,\ell^{+}\ell^{-}$, etc., and
$\Lambda_{b} \to n\,\ell^{+}\ell^{-}$. We shall then also compare our form 
factor results with the results of other model calculations.

vii) In this future work we shall also discuss the full 
three--fold joint angular decay distribution including a treatment of 
$\Lambda_{b}$ polarization effects as well as single lepton polarization 
effects.

\begin{acknowledgments}

This work was supported by the DFG under Contract No. LY 114/2-1, 
by Federal Targeted Program ``Scientific and scientific-pedagogical 
personnel of innovative Russia'' Contract No.02.740.11.0238. 
The work is done partially under
the project 2.3684.2011 of Tomsk State University. 
M.A.I.\ acknowledges the support of the
Forschungszentrum of the Johannes Gutenberg--Universit\"at Mainz
``Elementarkr\"afte und Mathematische Grundlagen (EMG)'' and 
the Heisenberg-Landau Grant.  
M.A.I. and V.E.L. would like to thank Dipartimento di Fisica, 
Universit\`a di Napoli Federico II and Istituto Nazionale 
di Fisica Nucleare, Sezione di Napoli for warm hospitality. 

\end{acknowledgments}

\appendix
\section{Joint four-fold angular decay distribution for the decay of 
an unpolarized $\Lambda_b$}
\label{app:jadd}
We write out the three-fold angular decay distribution Eq.~(\ref{bjoint3})
where we collect together terms with the threshold behavior $v^{0}$, $v^{1}$
and $v^{2}$. Including the $q^{2}$ dependence one obtains 
a four-fold joint angular decay distribution for the decay of
an unpolarized $\Lambda_b$. One has  
\be
W(\theta,\theta_{B},\chi)  \propto \frac{32\,q^2}{9}\,
\Big( |h^{B}_{\frac12 0}|^2 + |h^{B}_{-\frac12 0}|^2 \Big)\,
\Big( A\, v^2 + B\, v + C\, \frac{2 m_\ell^2}{q^2} \Big)\,,
\label{eq:3-fold}
\en
where the coefficients $A$,$B$ and $C$ are given by
\bea
A &=& \frac{9}{64}\,(1+\cos^2\theta)\,\left( U^{11}+U^{22} \right)
     + \frac{9}{32}\,\sin^2\theta\, \left( L^{11}+L^{22} \right)
\nn[1.2ex]
&+& \frac{9}{32}\,\alpha_B\,\cos\theta_B\,
\Big[\sin^2\theta\,\left(L_P^{11}+L_P^{22} \right)
  + \frac12\,(1+\cos^2\theta)\, \left( P^{11}+P^{22} \right) \Big]
\nn[1.2ex]
&+& \frac{9}{16\sqrt{2}}\,\alpha_B\,\sin 2\theta\,\sin\theta_B\,
    \Big[   \cos\chi\,\left( I1_P^{11}+I1_P^{22} \right)
          - \sin\chi\,\left( I2_P^{11}+I2_P^{22} \right) \Big] \,,
\nn[2ex]
B &=& -\, \frac{9}{16}\,\cos\theta\,
           \Big[ P^{12} + \alpha_B\,\cos\theta_B\,U^{12} \Big]
\nn[1.2ex]
        &-& \frac{9}{4\sqrt{2}}\,\alpha_B\,\sin\theta\,\sin\theta_B\,
            \Big[\cos\chi\,I3_P^{12} - \sin\chi\,I4_P^{12} \Big]\,,
\nn[2ex]
C &=& \frac{9}{16}\,\left( U^{11} + L^{11} + S^{22} \right)
      + \frac{9}{16}\,\alpha_B\,\cos\theta_B\,
        \left( P^{11} + L_P^{11} + S_P^{22} \right)\,,
\nonumber
\ena
The bilinear expressions $H_X^{mm'}$ $(X=U, L, S, P, L_P, S_P, 
I1_P, I2_P, I3_P, I4_P)$ are defined by 
\bea 
\qquad
\begin{array}{lr}
\hspace*{-.5cm}
\mbox{$ H^{mm'}_U =
{\rm Re}(H^{m}_{\frac{1}{2}1} H^{\dagger m'}_{\frac{1}{2}1}) +
{\rm Re}(H^{m}_{-\frac{1}{2}-1} H^{\dagger m'}_{-\frac{1}{2}-1}) $}  &
\hfill\mbox{ \rm transverse unpolarized}\,, 
\\
\hspace*{-.5cm}
\mbox{$ H^{mm'}_L =
{\rm Re}(H^{m}_{\frac{1}{2}0} H^{\dagger m'}_{\frac{1}{2}0}) +
{\rm Re}(H^{m}_{-\frac{1}{2}0} H^{\dagger m'}_{-\frac{1}{2}0}) $}     &
\hfill\mbox{ \rm longitudinal unpolarized}\,, 
\\
\hspace*{-.5cm}
\mbox{$ H^{mm'}_S =
{\rm Re}(H^{m}_{\frac{1}{2}t}H^{\dagger m'}_{\frac{1}{2}t}) +
{\rm Re}(H^{m}_{-\frac{1}{2}t}H^{\dagger m'}_{-\frac{1}{2}t})$} &
\hfill\mbox{ \rm scalar unpolarized}\,, 
\\
\hspace*{-.5cm}
\mbox{$H^{mm'}_{P} =
{\rm Re}(H^{m}_{\frac{1}{2}1}H^{\dagger m'}_{\frac{1}{2}1})-
{\rm Re}(H^{m}_{-\frac{1}{2}-1}H^{\dagger m'}_{-\frac{1}{2}-1})$} &
\hfill\mbox{ \rm transverse parity--odd polarized}\,, 
\\
\hspace*{-.5cm}
\mbox{$H^{mm'}_{L_{P}}=
{\rm Re}(H^{m}_{1/2\,0}H^{\dagger m'}_{1/2\,0}-H^{m}_{-1/2\,0}
H^{\dagger m'}_{-1/2\,0})$} &
\hfill\mbox{ \rm longitudinal polarized}\,, 
\\
\hspace*{-.5cm}
\mbox{$H^{mm'}_{S_{P}}=
{\rm Re}(H^{m}_{1/2\,t}H^{\dagger m'}_{1/2\,t}-H^{m}_{-1/2\,t}
H^{\dagger m'}_{-1/2\,t})$} & 
\hfill\mbox{ \rm scalar polarized}\,, 
\\
\hspace*{-.5cm}
\mbox{$H^{mm'}_{I1_{P}}= \frac{1}{4}
{\rm Re}(H^{m}_{1/2\,1}H^{\dagger m'}_{-1/2\,0}
+H^{m}_{-1/2\,0}H^{\dagger m'}_{1/2\,1}$} & \\
\hspace*{.5cm}
\mbox{$-H^{m}_{-1/2\,-1}H^{\dagger m'}_{1/2\,0}
-H^{m}_{1/2\,0}H^{\dagger m'}_{-1/2\,-1})
$} &
\hfill\mbox{ \rm longitudinal--transverse interference (1)}\,, 
\\
\hspace*{-.5cm}
\mbox{$H^{mm'}_{I2_{P}}= \frac{1}{4}
{\rm Im}(H^{m}_{1/2\,1}H^{\dagger m'}_{-1/2\,0}
-H^{m}_{-1/2\,0}H^{\dagger m'}_{1/2\,1}$} & \\
\hspace*{.5cm}
\mbox{$+H^{m}_{-1/2\,-1}H^{\dagger m'}_{1/2\,0} 
-H^{m}_{1/2\,0}H^{\dagger m'}_{-1/2\,-1}) 
$} &
\hfill\mbox{ \rm longitudinal--transverse interference (2)}\,, 
\\
\hspace*{-.5cm}
\mbox{$H^{mm'}_{I3_P}= \frac{1}{4}
{\rm Re}(H^{m}_{1/2\,1}H^{\dagger m'}_{-1/2\,0}
+H^{m}_{-1/2\,0}H^{\dagger m'}_{1/2\,1}$} & \\
\hspace*{.5cm}
\mbox{$+H^{m}_{-1/2\,-1}H^{\dagger m'}_{1/2\,0}
+H^{m}_{1/2\,0}H^{\dagger m'}_{-1/2\,-1})
$} &
\hfill\mbox{ \rm longitudinal--transverse interference (3)}\,, 
\\
\hspace*{-.5cm}
\mbox{$H^{mm'}_{I4_P}= \frac{1}{4}
{\rm Im}(H^{m}_{1/2\,1}H^{\dagger m'}_{-1/2\,0}
-H^{m}_{-1/2\,0}H^{\dagger m'}_{1/2\,1}$} & \\
\hspace*{.5cm}
\mbox{$-H^{m}_{-1/2\,-1}H^{\dagger m'}_{1/2\,0}
+H^{m}_{1/2\,0}H^{\dagger m'}_{-1/2\,-1})
$} &
\hfill\mbox{ \rm longitudinal--transverse interference (4)}\,. 
\\
\end{array}
\ena   
Note the three-fold joint angular decay distribution for the decay of
an unpolarized $\Lambda_b$ is factorized in terms of fully 
transverse (unpolarized and parity--odd polarized), 
longitudinal (upolarized and polarized), 
scalar (unpolarized and polarized) bilinear helicity combinations 
and four combinations of longitudinal-transverse interference.  

Another important property of  the three-fold joint angular decay
distribution is its invariance w.r.t. the choice of coordinate systems.
For example,  using the completeness relation for the polarization vectors
of the effective current one can explicitly show that the angular
decay distribution Eq.~(\ref{eq:3-fold})
is the same for two specific choices of coordinate systems:
system (i) $\vec{J}_{\rm eff}$ is directed along the $z$-axis as in this
paper and
system (ii) $\vec{J}_{\rm eff}$ is antiparallel to the direction of the
$z$-axis
as used e.g. in Refs.~\cite{Kadeer:2005aq,Faessler:2009xn,Branz:2010pq}).
In particular, only the transverse helicity
amplitudes $H^{m}_{\pm 1/2\,\pm 1}$ change sign when going from
system (i) to system (ii) while the other helicity amplitudes
remain invariant. The change of sign for the transverse amplitudes
$H^{m}_{\pm 1/2\,\pm 1}$ can be seen to be compensated by the effects 
of rotating the coordinate system (i) by $180^{\circ}$ around the 
$x$-axis when going from system (i) to system (ii).

\section{Interpolating currents of $\Lambda$-hyperons}
\label{app:Lambda_currents}

When constructing interpolating baryon currents it is convenient to use Fierz
transformations and corresponding identities in order to interchange
the quark fields. First we specify five possible spin structures
$J^{\alpha\beta,\rho\sigma} = \Gamma_1^{\alpha\beta} \otimes (C
\Gamma_2)^{\rho\sigma}$ defining the Fierz transformation of the
baryon currents: 
\bea\label{Fierz_trans}
P &=& I \otimes C \gamma_5 \,, \nonumber\\
S &=& \gamma_5 \otimes C  \,, \nonumber\\
A &=& \gamma^\mu \otimes C \gamma_5 \gamma_\mu \,,  \\
V &=& \gamma^\mu \gamma^5 \otimes C \gamma_\mu \,, \nonumber\\
T &=& \frac{1}{2} \sigma^{\mu\nu} \gamma^5 \otimes C \sigma_{\mu\nu} \,. 
\nonumber
\ena
The Fierz transformation of the structures
$J = \{P,S,A,V,T\}$ read
\bea\label{Fierz_trans2}
P &=& \frac{1}{4} \biggl( \tilde P + \tilde S + \tilde A
+ \tilde V + \tilde T  \biggr) \,, \nonumber\\
S &=& \frac{1}{4} \biggl( \tilde P + \tilde S - \tilde A
- \tilde V + \tilde T  \biggr) \,, \nonumber\\
A &=& \tilde P - \tilde S - \frac{1}{2} \biggl(
\tilde A - \tilde V \biggr) \,, \\
V &=& \tilde P - \tilde S + \frac{1}{2} \biggl(
\tilde A - \tilde V \biggr) \,, \nonumber\\
T &=& \frac{3}{2} ( \tilde P + \tilde S)  - \frac{1}{2} \tilde T \,. \nonumber
\ena 
The symbol $\ \tilde{}\ $ is used to denote Fierz-transformed matrices
according to $\tilde J^{\alpha\sigma,\rho\beta} = \Gamma_1^{\alpha\sigma}
\otimes (C \Gamma_2)^{\rho\beta}$ where $\alpha,\beta,\rho$ and $\sigma$ are
Dirac indices. 
Using Eqs.~(\ref{Fierz_trans2}) one can derive useful identities
\bea\label{Fierz_id}
2 ( P - S ) + A + V &=& 2 ( \tilde P - \tilde S ) + \tilde A + \tilde V \,,
\nonumber\\
3 ( P + S ) + T &=& 3 ( \tilde P + \tilde S ) + \tilde T \,. 
\ena
Let us consider hyperons containing two light nonstrange $u$ or $d$ 
quarks and a third quark  $Q = s, c$ or $b$, which contain  
antisymmetrized combination of $u$ and $d$ quarks over spin and flavor. 
There are two possible SU(N$_{\rm f}$)-symmetric interpolating currents of 
$\Lambda$-hyperons without 
derivatives --- the so-called vector $J^V_{\Lambda_{Q[ud]}}$ 
and tensor  $J^T_{\Lambda_{Q[ud]}}$ current: 
\bea 
J^V_{\Lambda_{Q[ud]}}&=& \frac{1}{3} \, \epsilon^{a_1a_2a_3} \, 
\left( \gamma^\mu\gamma^5 d^{a_1} u^{a_2} C \gamma_\mu Q^{a_3} 
    -  \gamma^\mu\gamma^5 u^{a_1} d^{a_2} C \gamma_\mu Q^{a_3} \right) \,, 
\nonumber\\ 
J^T_{\Lambda_{Q[ud]}}&=& \frac{1}{3} \, \epsilon^{a_1a_2a_3} \, 
\left( \sigma^{\mu\nu} \gamma^5 d^{a_1} u^{a_2} C \sigma_{\mu\nu} Q^{a_3} 
    -  \sigma^{\mu\nu} \gamma^5 u^{a_1} d^{a_2} C \sigma_{\mu\nu} Q^{a_3} 
\right) \,.
\ena   
Using Fierz transformations one can rewrite  
$J^V_{\Lambda_{Q[ud]}}$ and $J^T_{\Lambda_{Q[ud]}}$ currents 
as a linear combination of more convenient currents ---  
pseudoscalar $J^P_{\Lambda_{Q[ud]}}$, scalar $J^S_{\Lambda_{Q[ud]}}$ 
and axial $J^A_{\Lambda_{Q[ud]}}$, which manifestly contain the 
spin-0 $[ud]$-diquark: 
\bea
J^P_{\Lambda_{Q[ud]}}&=& \epsilon^{a_1a_2a_3} \, 
Q^{a_1} \, u^{a_2} C \gamma_5 d^{a_3} \,, 
\nn
J^S_{\Lambda_{Q[ud]}}&=& \epsilon^{a_1a_2a_3} \, \gamma^5 \, 
Q^{a_1} \, u^{a_2} C  d^{a_3} \,,
\nn
J^A_{\Lambda_{Q[ud] }}&=& \epsilon^{a_1a_2a_3} \, 
\gamma^\mu \, u^{a_1} \, d^{a_2}  
C\gamma_5\gamma_\mu q_{3}^{a_3} \,. 
\label{eq:Lambda-cur3}
\ena 
The result after the Fierz transformation reads: 
\bea 
J^V_{\Lambda_{Q[ud]}} &=& \frac{2}{3} J^P_{\Lambda_{Q[ud]}} 
- \frac{2}{3} J^S_{\Lambda_{Q[ud]}} 
+ \frac{1}{3} J^A_{\Lambda_{Q[ud]}}\,, 
\nonumber\\ 
J^T_{\Lambda_{Q[ud]}} &=& J^P_{\Lambda_{Q[ud]}} 
+ J^S_{\Lambda_{Q[ud]}}\,.  
\ena 
It is clear that in the nonrelativistic limit the 
$J^V$ and $J^T$ currents become degenerate and coincide with the 
$J^P$ and $J^A$ currents. Therefore, the $J^P$ and $J^A$ currents 
differ from SU(N$_{\mathrm f}$) currents up to relativistic 
corrections. 

\section{Helicity amplitudes} 
\label{app:Hel_amp}

In Sec.II we have shown how to write out the angular distributions
of the rare $\Lambda_{b}$ decays in terms of
hadron--side helicity amplitudes 
$H^m_{\lambda_2,\lambda_j}$, which in turn can be 
related to invariant form factors $f_i^J$ (see details 
Refs.~\cite{Kadeer:2005aq,Faessler:2009xn,Branz:2010pq}). 
The pertinent relation is 
\bea
H^m_{\lambda_2,\lambda_j} = M_\mu^m(\lambda_2)
\, \epsilon^{\,\ast\mu}(\lambda_j) \,.
\ena 
As before the labels $\lambda_2$ and $\lambda_j$ denote the
helicities of the daughter baryon and the effective current,
corresponding to the lepton pair and the photon, respectively. 
We shall work in the rest frame of 
the parent baryon $B_1$ with the daughter baryon $B_2$ moving in the
negative $z$-direction (see Fig.1) such that
$p_1^\mu = (M_1, {\bf 0})$, $p_2^\mu = (E_2, 0, 0,- |{\bf p}_2|)$ and
$q^\mu = (q_0, 0, 0,  |{\bf p}_2|)$,
where $q_0 = (M_1^2 - M_2^2 + q^2)/(2 M_1)$ 
and 
$E_2 = M_1 - q_0 =
(M_1^2 + M_2^2 - q^2)/(2 M_1)$.
Angular momentum conservation fixes the helicity $\lambda_1$
of the parent baryon according to $\lambda_1 = - \lambda_2 + \lambda_j$. 

The $J=\frac{1}{2}$ baryon spinors are 
given by 
\bea 
\bar u_2\Big(p_2, \pm \frac{1}{2}\Big) &=& \sqrt{E_2 + M_2} \,  
\Big( \chi_\pm^\dagger, \frac{\pm |{\bf p}_2|}{E_2 + M_2}  
\chi_\pm^\dagger \Big)\,, \nonumber\\
u_1\Big(p_1, \pm \frac{1}{2}\Big) &=& \sqrt{2M_1} \, 
\left(
\begin{array}{l}
\chi_\pm \\
0 \\
\end{array}
\right)\,,
\ena 
where $\chi_+ = \left(
\begin{array}{l}
1 \\
0 \\
\end{array} \right)$ 
and $\chi_- = \left(
\begin{array}{l}
0 \\
1 \\
\end{array} \right)$ are two--component Pauli spinors. 

The polarization vectors of the effective
current $J_{\rm eff}$ read 
\bea 
\epsilon^\mu(t)&=&
\frac{1}{\sqrt{q^2}}(\,q_0\,,\,\,0\,,\,\,0\,,\,\,|{\bf p_2}|\,)
\,,\nonumber\\
\epsilon^\mu(\pm) &=&
\frac{1}{\sqrt{2}}(\,0\,,\,\,\mp 1\,,\,\,-i\,,\,\,0\,)\,,
\label{hel_basis}\\
\epsilon^\mu(0) &=&
\frac{1}{\sqrt{q^2}}(\,|{\bf p_2}|\,,\,\,0\,,\,\,0\,,\,\,q_0\,)\,.
\nonumber
\ena
Using this basis one can express the components of the hadronic
tensors through the invariant form factors. It is convenient 
to split the helicity amplitudes on 
vector $(H^{Vm}_{\lambda_2,\lambda_j})$ 
and axial--vector $(H^{Am}_{\lambda_2,\lambda_j})$ parts: 
\bea 
H^m_{\lambda_2,\lambda_j} = H^{Vm}_{\lambda_2,\lambda_j}
                          - H^{Am}_{\lambda_2,\lambda_j}\,.
\ena 
From parity or from an explicit calculation  
one has 
\bea 
& &H^{Vm}_{-\lambda_2,-\lambda_j} = H^{Vm}_{\lambda_2,\lambda_j}\,, 
\nonumber\\
& &H^{Am}_{-\lambda_2,-\lambda_j} = - H^{Am}_{\lambda_2,\lambda_j}\,. 
\ena 
In the case of the transitions $\Lambda_{q_1[q2q3]} \to \Lambda_{q_1'[q_2q_3]} 
+ j_{\rm eff}$ the helicity amplitudes $H^{Vm}_{\lambda_2,\lambda_j},
H^{Am}_{\lambda_2,\lambda_j}$ are given by 
\bea
H^{Vm}_{\frac{1}{2} t} &=& 
\sqrt{\frac{Q_+}{q^2}} \, 
\biggl( M_- \, F_1^{Vm} + \frac{q^2}{M_1} \, F_3^{Vm} \biggr)\,, \nonumber\\
H^{Vm}_{\frac{1}{2} 1} &=& \sqrt{2 Q_-} \, 
\biggl( F_1^{Vm} + \frac{M_+}{M_1} \, F_2^{Vm} \biggr)\,, \nonumber\\
H^{Vm}_{\frac{1}{2} 0} &=& \sqrt{\frac{Q_-}{q^2}} \,  
\biggl( M_+ \, F_1^{Vm} + \frac{q^2}{M_1} \, F_2^{Vm} \biggr)\,, \nonumber\\
&&\\
H^{Am}_{\frac{1}{2} t} &=& \sqrt{\frac{Q_-}{q^2}} \, 
\biggl( M_+ \, F_1^{Am} - \frac{q^2}{M_1} \, F_3^{Am} \biggr)\,, \nonumber\\
H^{Am}_{\frac{1}{2} 1} &=& \sqrt{2 Q_+} \, 
\biggl( F_1^{Am} - \frac{M_-}{M_1} \, F_2^{Am} \biggr)\,, \nonumber\\
H^{Am}_{\frac{1}{2} 0} &=& \sqrt{\frac{Q_+}{q^2}} \,  
\biggl( M_- \, F_1^{Am}  - \frac{q^2}{M_1} \, F_2^{Am} \biggr)\,, \nonumber  
\ena 
where $M_\pm = M_1 \pm M_2$, $Q_\pm = M_\pm^2 - q^2$.  
The form factors $F^{Vm}_i,F^{Am}_i$ are linear combinations of 
the form factors $f_i^J$. In case of the rare decays they involve also 
the  Wilson coefficients. In particular, 
the sets of form factors $F^{Vm}_i,F^{Am}_i$ for the semileptonic charged 
current decays with the Cabibbo-allowed $b \to c$ and $c \to s$ transitions 
read  
\bea 
F_1^{V1} &=& f_1^V 
\,, 
\nonumber\\
F_2^{V1} &=& f_2^V 
\,, 
\nonumber\\
F_3^{V1} &=& f_3^V 
\,, 
\nonumber\\
&&\\
F_1^{A1} &=& f_1^A 
\,, 
\nonumber\\
F_2^{A1} &=& f_2^A 
\,, 
\nonumber\\
F_3^{A1} &=& f_3^A 
\,. 
\nonumber 
\ena 
In the case of the $\Lambda_b \to \Lambda + \ell^+ \ell^-$ transitions 
the corresponding form factors are 
\bea 
F_1^{V1} &=& 
C_9^{\rm eff} \, f_1^V 
       - \frac{2 m_b}{M_1} \, C_7^{\rm eff} \, f_1^{TV} 
\,, 
\nonumber\\
F_2^{V1} &=& 
C_9^{\rm eff} \, f_2^V 
       - \frac{2 m_b M_1}{q^2} \, C_7^{\rm eff} \, f_2^{TV} 
\,, 
\nonumber\\
F_3^{V1} &=& 
C_9^{\rm eff} \, f_3^V 
       + \frac{2 m_b M_-}{q^2} \, C_7^{\rm eff} \, f_1^{TV} 
\,, 
\nonumber\\
&&\\
F_1^{A1} &=& 
C_9^{\rm eff} \, f_1^A 
      + \frac{2 m_b}{M_1} \, C_7^{\rm eff} \, f_1^{TA} 
\,, 
\nonumber\\
F_2^{A1} &=& 
C_9^{\rm eff} \, f_2^A 
      + \frac{2 m_b M_1}{q^2} \, C_7^{\rm eff} \, f_2^{TA} 
\,, 
\nonumber\\
F_3^{A1} &=& 
C_9^{\rm eff} \, f_3^A 
       + \frac{2 m_b M_+}{q^2} \, C_7^{\rm eff} \, f_1^{TA} 
\,, 
\nonumber 
\ena 
and 
\bea 
F_i^{V2} &=& C_{10} \, f_i^{V}\,, \nonumber\\
F_i^{A2} &=& C_{10} \, f_i^{A}\,.  
\ena 
Finally, in the case of the one-photon transitions 
$\Lambda_b \to \Lambda + \gamma$ one 
needs the helicity amplitudes 
$H^{V}_{\pm\frac{1}{2},\pm 1} \equiv H^{V1}_{\pm\frac{1}{2},\pm 1}$ and 
$H^{A}_{\pm\frac{1}{2},\pm 1} \equiv H^{A1}_{\pm\frac{1}{2},\pm 1}$ 
They are related to the $f_2^{TV}$ and $f_2^{TA}$ form factors by
\bea 
H^{V}_{\pm\frac{1}{2},\pm 1} &=& 
 \sqrt{2} \, \frac{M_+M_-}{M_1} \, F_2^V  \,, \nonumber\\
H^{A}_{\pm\frac{1}{2},\pm 1} &=& 
\mp \sqrt{2} \, \frac{M_+M_-}{M_1} \, F_2^A  \,, 
\ena 
where 
\bea 
F_2^V &=& - C_7^{\rm eff} \, \frac{m_b}{M_1} f_2^{TV} \,, \nonumber\\
F_2^A &=& - C_7^{\rm eff} \, \frac{m_b}{M_1} f_2^{TA} \,. 
\ena 

\section{Wilson coefficients}
\label{app:Wilson}

In this paper we use the set of Wilson coefficients 
(see Table~\ref{tab:Wilson}) fixed in 
Ref.~\cite{Faessler:2002ut}. 
The Wilson coefficient $ C_9^{\rm eff}$ effectively takes 
into account, first, the contributions from the four-quark
operators $Q_i$ ($i=1,...,6$) and, second, the nonperturbative 
effects coming from the $c\bar c$-resonance contributions
which are as usual parametrized by a Breit-Wigner ansatz 
\cite{Ali:1991is}:
\begin{eqnarray}
C_9^{\rm eff} & = & C_9 + 
C_0 \left\{
h(\hat m_c,  s)+ \frac{3 \pi}{\alpha^2}\,  \kappa\,
         \sum\limits_{V_i = \psi(1s),\psi(2s)}
      \frac{\Gamma(V_i \rightarrow l^+ l^-)\, m_{V_i}}
{  {m_{V_i}}^2 - q^2  - i m_{V_i} \Gamma_{V_i}}
\right\} 
 \nonumber \\
&-& \frac{1}{2} h(1,  s) \left( 4 C_3 + 4 C_4 +3 C_5 + C_6\right)  
 \\
&-& \frac{1}{2} h(0,  s) \left( C_3 + 3 C_4 \right) +
\frac{2}{9} \left( 3 C_3 + C_4 + 3 C_5 + C_6 \right)\,.
\nonumber
\end{eqnarray}
where $C_0\equiv 3 C_1 + C_2 + 3 C_3 + C_4+ 3 C_5 + C_6$.
Here
\begin{eqnarray*} 
h(\hat m_c,  s) & = & - \frac{8}{9}\ln\frac{m_b}{\mu} 
- \frac{8}{9}\ln\hat m_c +
\frac{8}{27} + \frac{4}{9} x \nonumber\\
& - & \frac{2}{9} (2+x) |1-x|^{1/2} \left\{
\begin{array}{ll}
\left( \ln\left| \frac{\sqrt{1-x} + 1}{\sqrt{1-x} - 1}\right| - i\pi 
\right), &
\mbox{for } x \equiv \frac{4 \hat m_c^2}{ s} < 1 \nonumber \\
& \\
2 \arctan \frac{1}{\sqrt{x-1}}, & \mbox{for } x \equiv \frac
{4 \hat m_c^2}{ s} > 1,
\end{array}
\right. \nonumber\\
h(0,  s) & = & \frac{8}{27} -\frac{8}{9} \ln\frac{m_b}{\mu} - 
\frac{4}{9} \ln\ s + \frac{4}{9} i\pi \,,\nonumber 
\end{eqnarray*}
where $\hat m_c=m_c/M_{\Lambda_b}$, $s=q^2/m_{\Lambda_b}^2$ and 
$\kappa=1/C_0$. In our numerical calculations we use $\mu = m_b = 4.19$ GeV, 
$m_c = 1.27$ GeV, $M_{J/\psi} = 3.096916$ GeV, $M_{\Psi(2S)} = 3.68609$ GeV, 
$\Gamma_{J/\psi} = 92.9$ keV, $\Gamma_{\Psi(2S)} = 304$ keV, 
$\Gamma(J/\psi \to l^+ l^-) = 5.55$ keV and 
$\Gamma(\Psi(2S) \to l^+ l^-) = 2.35$ keV. 

\section{Identity $f^{TV}_2(0) = f^{TA}_2(0)$}
\label{app:TV2_TA2}

Here we demonstrate that $f^{TV}_2(0) = f^{TA}_2(0)$ in the case of a
$P$ interpolating current for the $\Lambda$-type baryons.
The same statement is also true for $S$- and $A$-currents. 
After integration over the loop momenta $k_1$ and $k_2$ 
the general structure of the matrix elements involving 
the Dirac matrix $\Gamma^\mu = i\sigma^{\mu q}$ or 
$i\sigma^{\mu q}\gamma^5$ is written as 
\bea 
M^\mu(p_1,p_2) \, = \, F(p^2_1,p^2_2,q^2) \, \bar u_2(p_2) 
(1 + \alpha_1 \not\! p_1 + \alpha_2 \not\! p_2) \, \Gamma^\mu \, 
(1 + \beta_1 \not\! p_1 + \beta_2 \not\! p_2) \, u_1(p_1) \,, 
\ena 
where $F(p^2_1,p^2_2,q^2)$ depends on the invariants 
$p^2_1$, $p^2_2$ and $q^2$. The $\alpha_i$ and $\beta_i$ are 
coefficients whose explicit values are not needed for the proof. 
Using the package FORM one can show that the $f_2^{TV}(q^2)$ and 
$f_2^{TA}(q^2)$ form factors are given by: 
\bea 
f_2^{TV}(q^2) &=& 
f_2^{TA}(q^2) + 2 \alpha_1 \beta_2 q^2 \, F(p^2_1,p^2_2,q^2) 
\nonumber\\
&=& F(p^2_1,p^2_2,q^2) \, \biggl[ 
1 + M_1 (\beta_1 + \beta_2) 
  + M_2 (\alpha_1 + \alpha_2) 
  + M_1 M_2 (\alpha_1 + \alpha_2) (\beta_1 + \beta_2)
  + \alpha_1 \beta_2 q^2
\biggr] \,. 
\ena 
Therefore, the form factors $f_2^{TV}(q^2)$ and $f_2^{TA}(q^2)$ 
differ only by a term linear in $q^2$ and therefore one has 
$f_2^{TV}(0) = f_2^{TA}(0)$.

\newpage 

\begin{figure}[ht]
\begin{center}
\vspace*{.25cm} 
\epsfig{figure=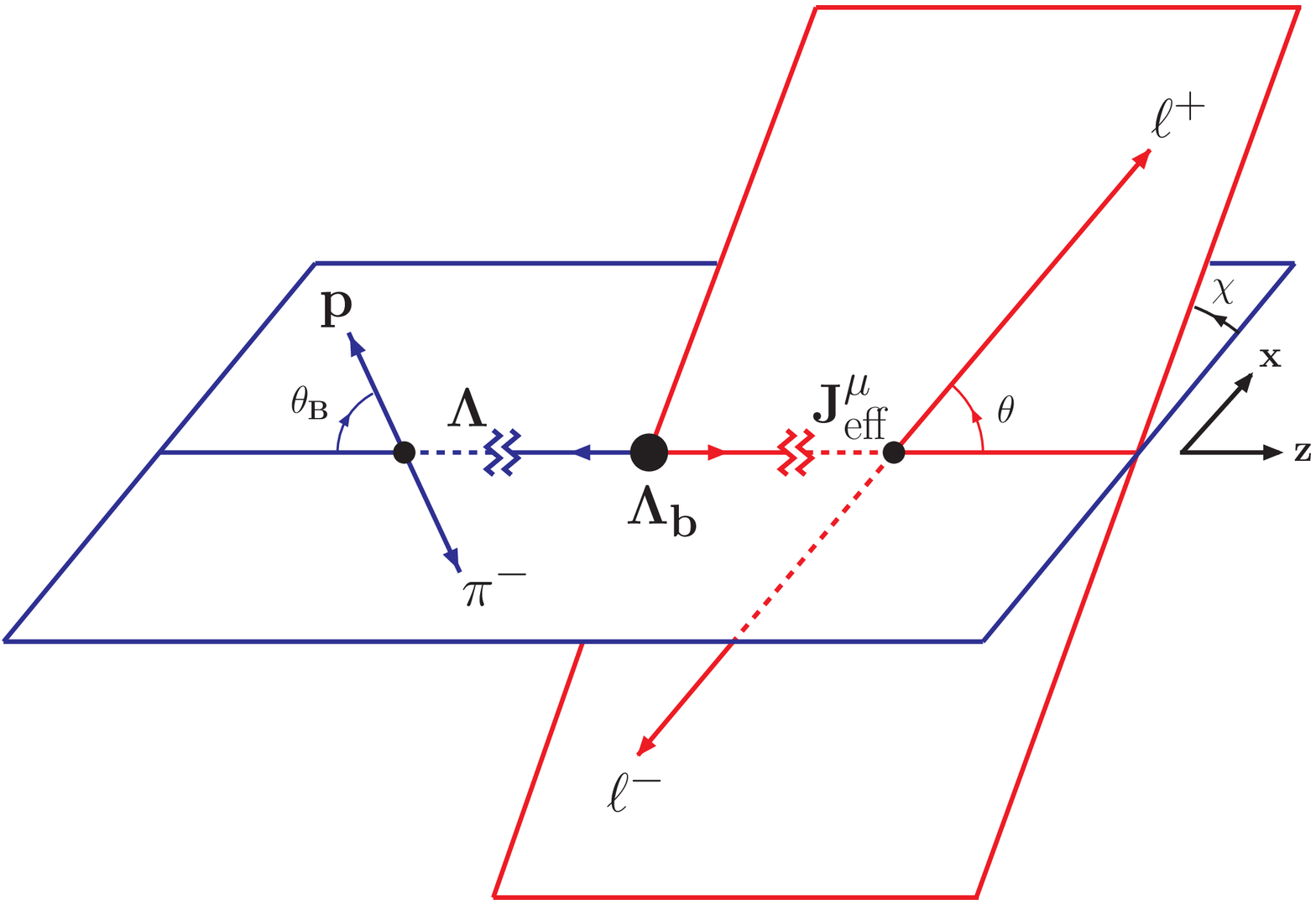,scale=.6} 

\caption{Definition of angles $\theta$, $\theta_B$ and $\chi$ in
the cascade decay 
$\Lambda_{b} \to \Lambda (\to p \pi^{-})
+ J_{\rm eff}(\to \ell^{+}\ell^{-})$\,.}

\end{center}
\end{figure}

\newpage 

\begin{figure}[ht] 
\begin{center}
\epsfig{figure=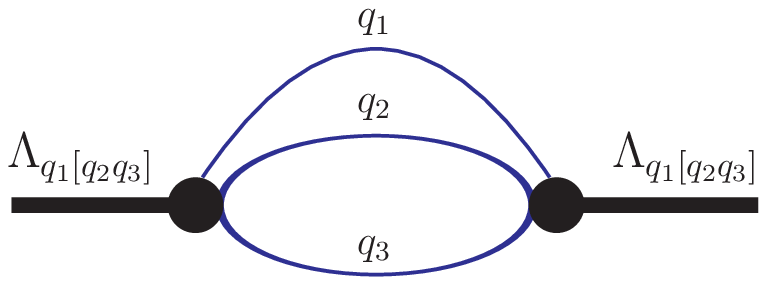,scale=.8} 
\vspace*{-.75cm}
\caption{$\Lambda_{q_1[q_2q_3]}$ baryon mass operator.} 
\end{center}

\begin{center}
\epsfig{figure=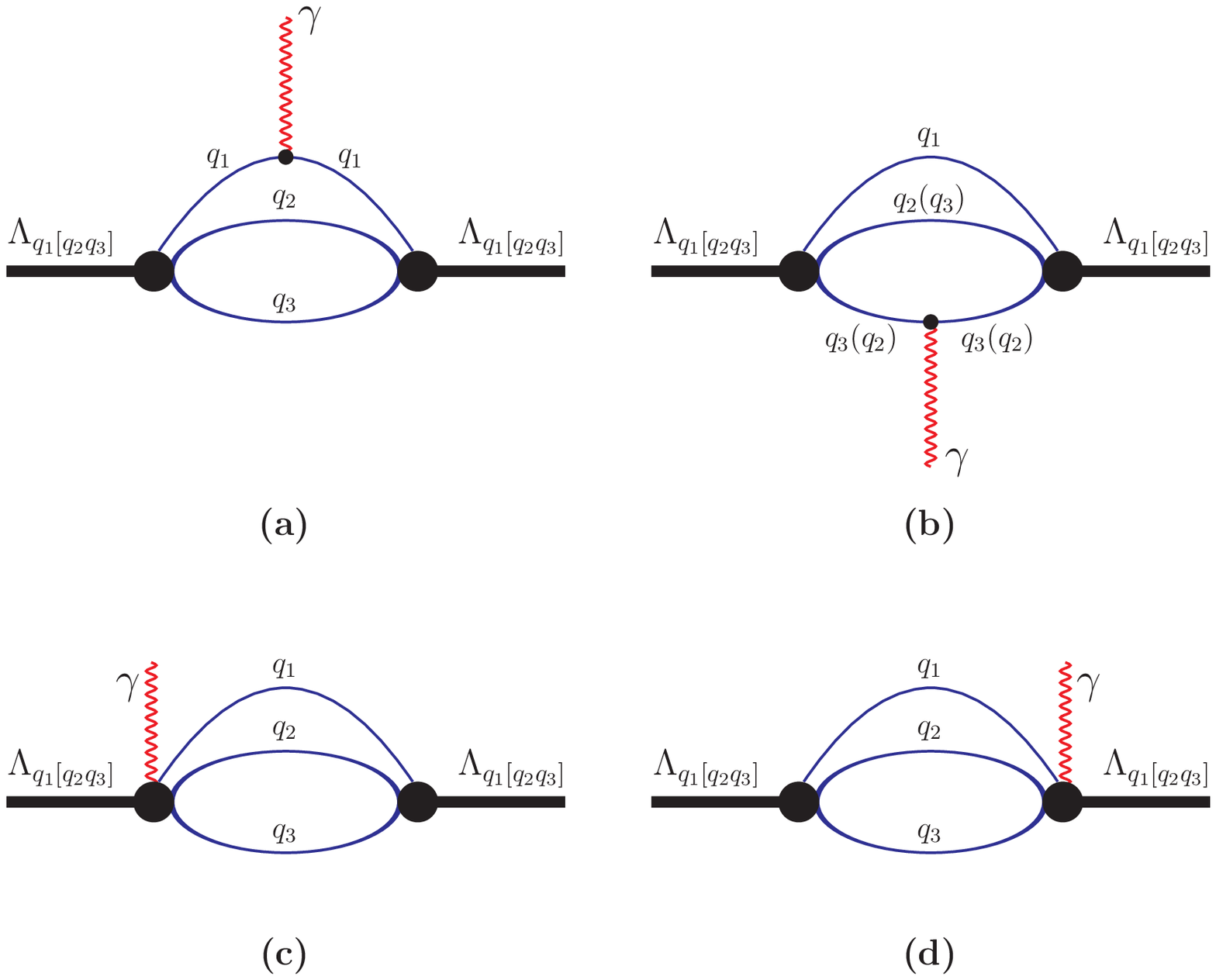,scale=.8} 
\end{center}
\vspace*{-.75cm}
\caption{Electromagnetic vertex function of the $\Lambda_{q_1[q_2q_3]}$ 
baryon: 
(a) triangle diagram with the (off-shell) photon attached to the quark $q_1$; 
(b) triangle diagram with the (off-shell) photon attached to 
quarks $q_2$ or $q_3$; 
(c) bubble diagram with the (off-shell) photon attached to the vertex
of the ingoing baryon; 
(d) bubble diagram with the (off-shell) photon attached to the vertex
of the outgoing baryon.}

\begin{center}
\epsfig{figure=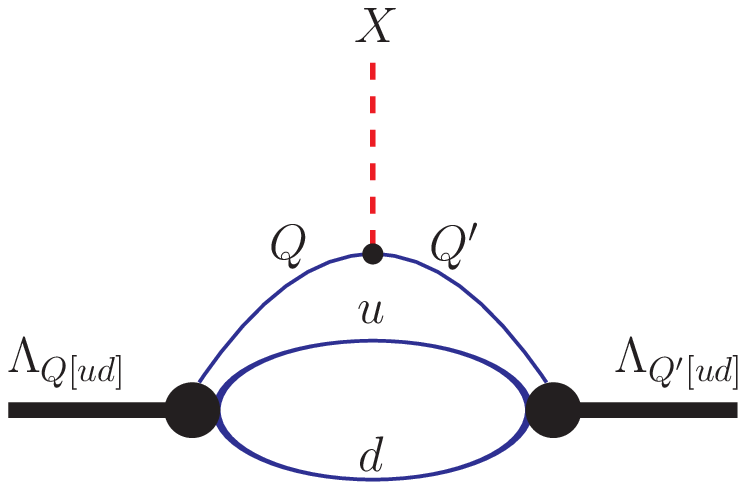,scale=.8} 
\end{center}
\vspace*{-.75cm}
\caption{Diagrams contributing to the flavor-changing transition 
$\Lambda_{Q[ud]} \to \Lambda_{Q'[ud]} + X$ , 
where $X=\ell^-\bar\nu_\ell, \ell^+\ell^-$ or $\gamma$.}  
\end{figure} 

\newpage 

\begin{figure}[ht]
\begin{center}
\hspace*{-0.5cm}
\begin{tabular}{lr}
\includegraphics[width=0.35\textwidth]{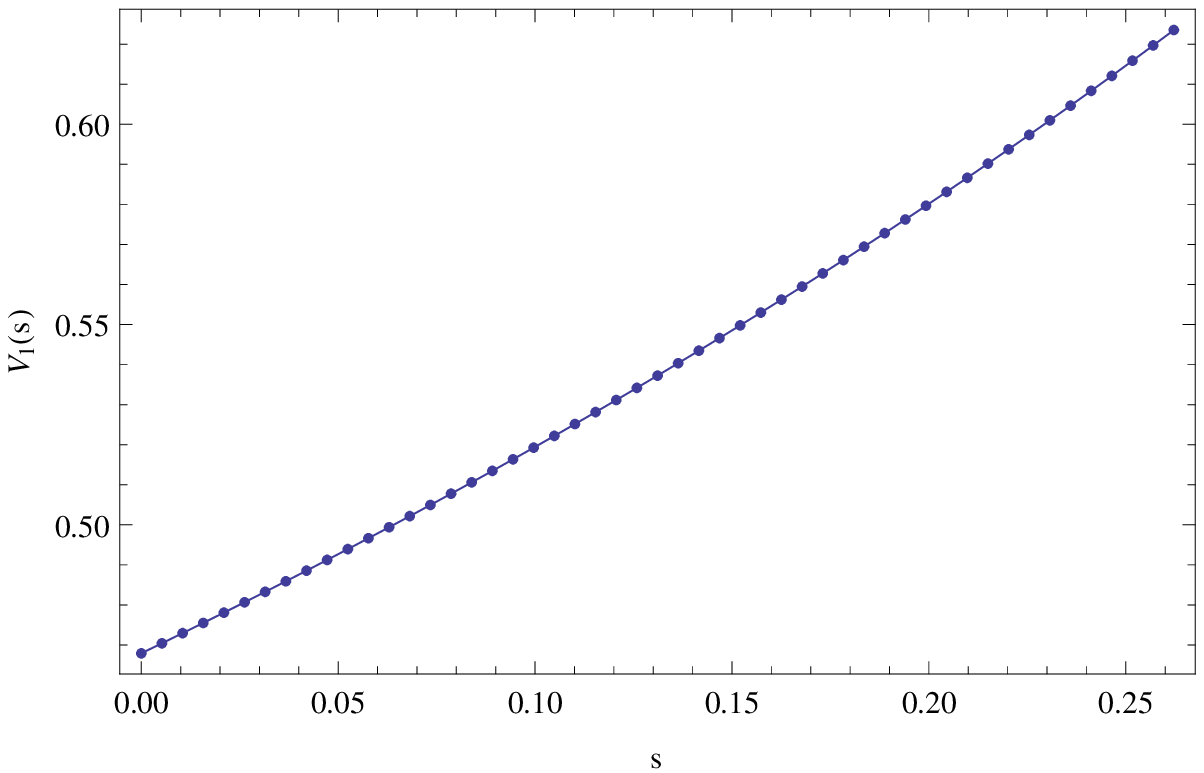}   & \hspace*{.5cm}
\includegraphics[width=0.35\textwidth]{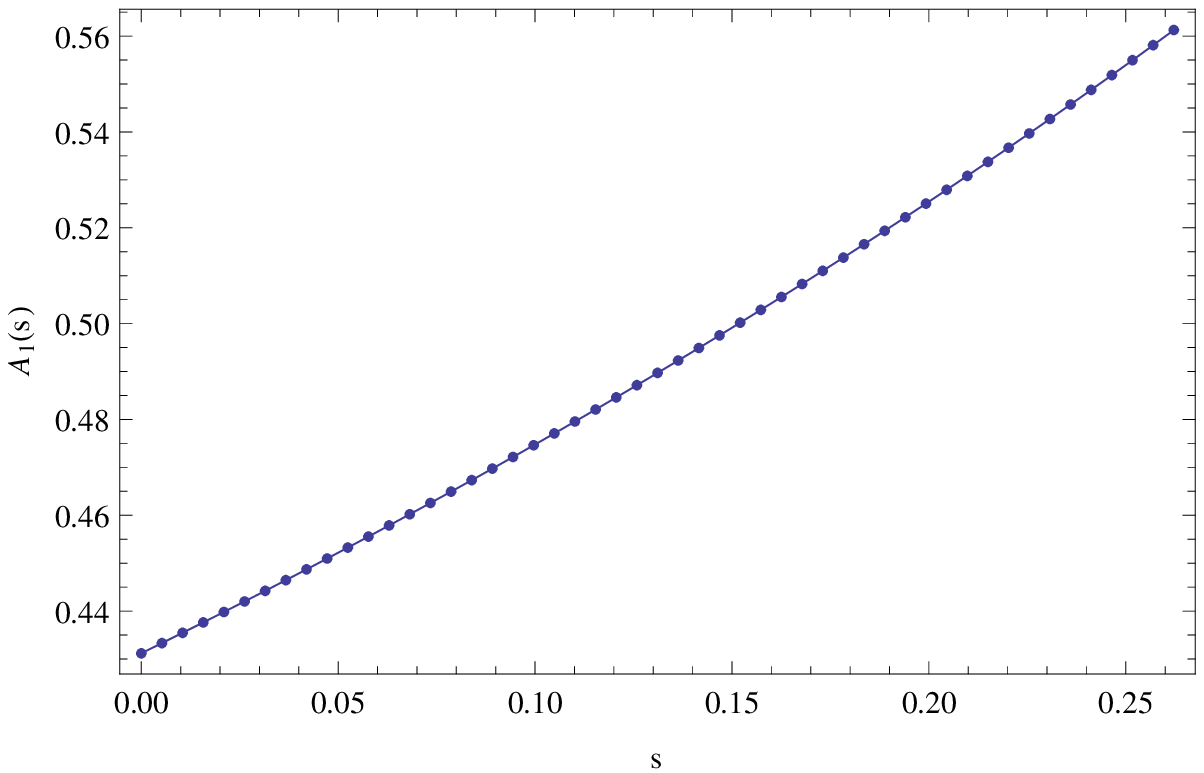}      \\[2ex]

\includegraphics[width=0.35\textwidth]{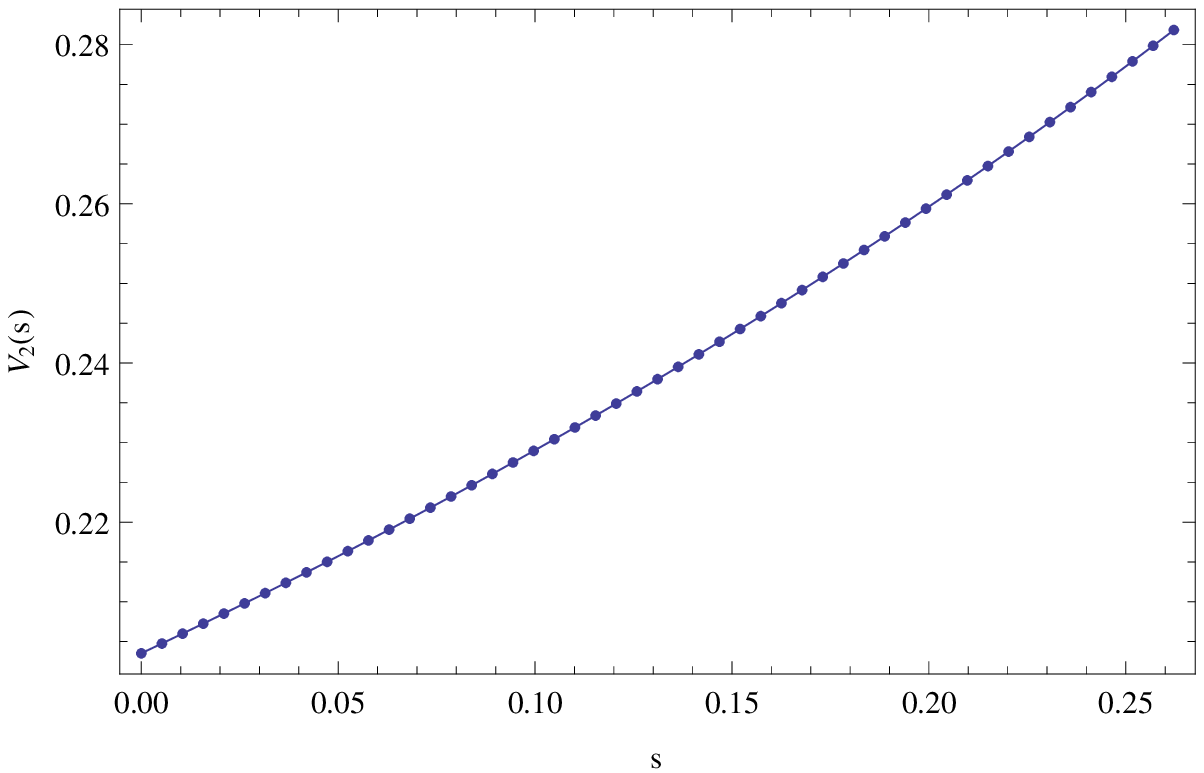}  & \hspace*{.5cm} 
\includegraphics[width=0.35\textwidth]{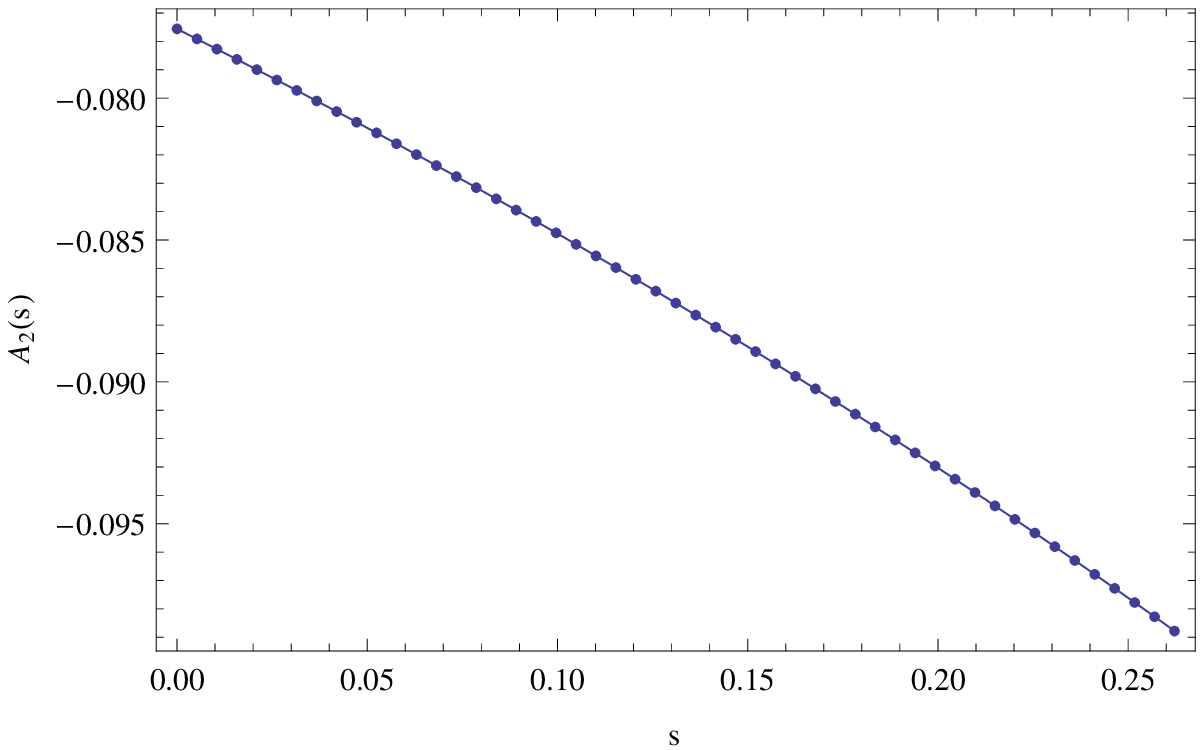}    \\[2ex]

\includegraphics[width=0.35\textwidth]{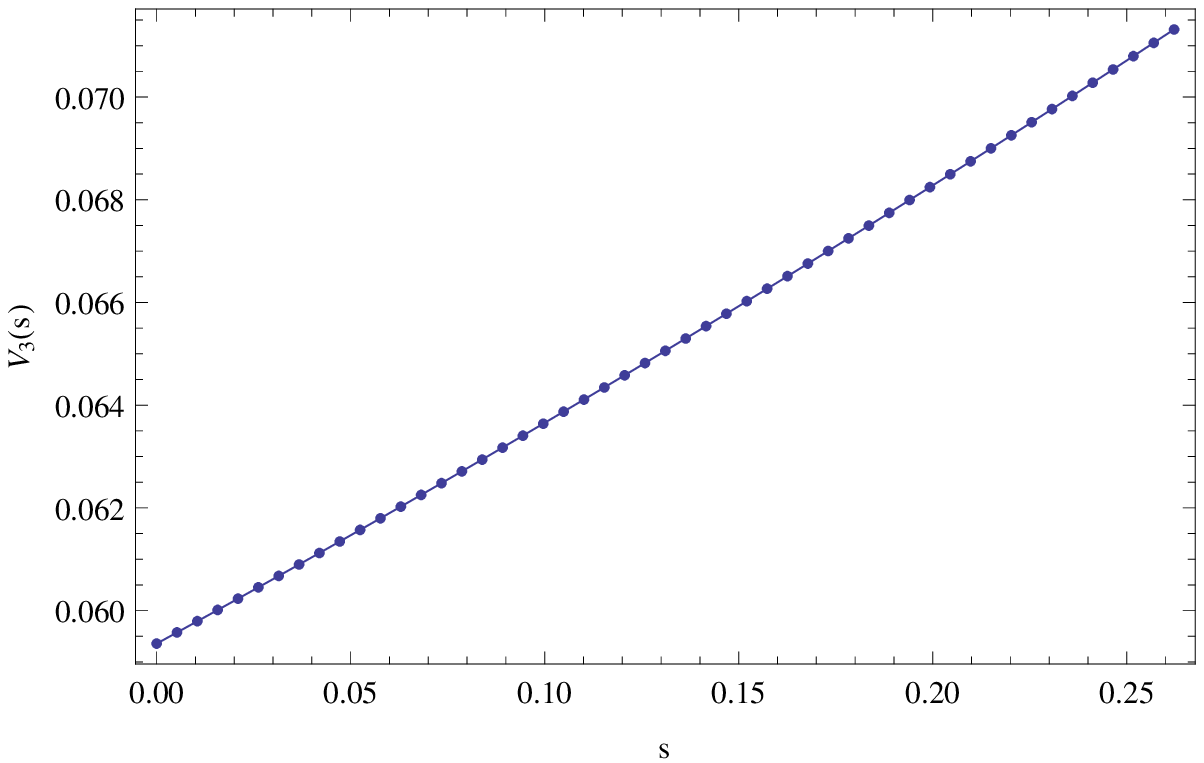}  & \hspace*{.5cm} 
\includegraphics[width=0.35\textwidth]{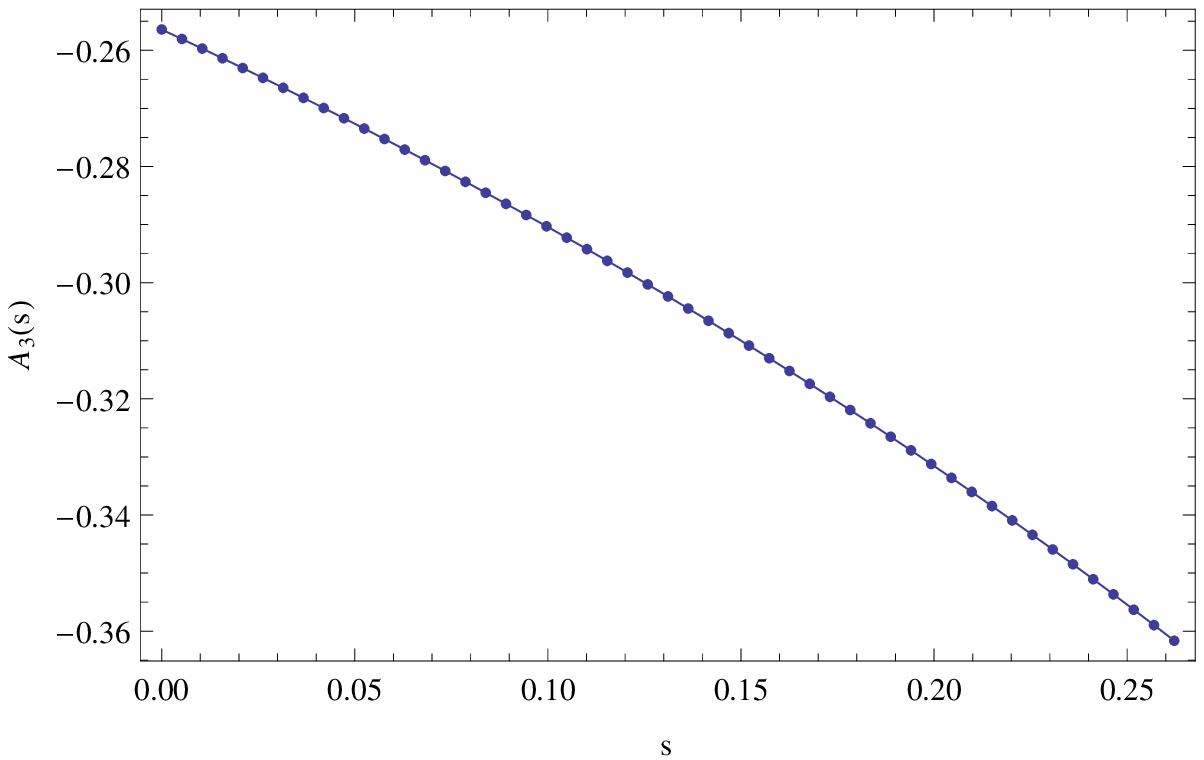}   
\end{tabular}
\end{center}
\caption{\label{fig:ff_cs} 
Form factors defining the transition $\Lambda_c\to\Lambda$: 
approximated results (solid line), exact result(dotted line).
}
\end{figure} 

\newpage 

\begin{figure}[ht]
\begin{center}
\hspace*{-0.5cm}
\begin{tabular}{lr}
\includegraphics[width=0.35\textwidth]{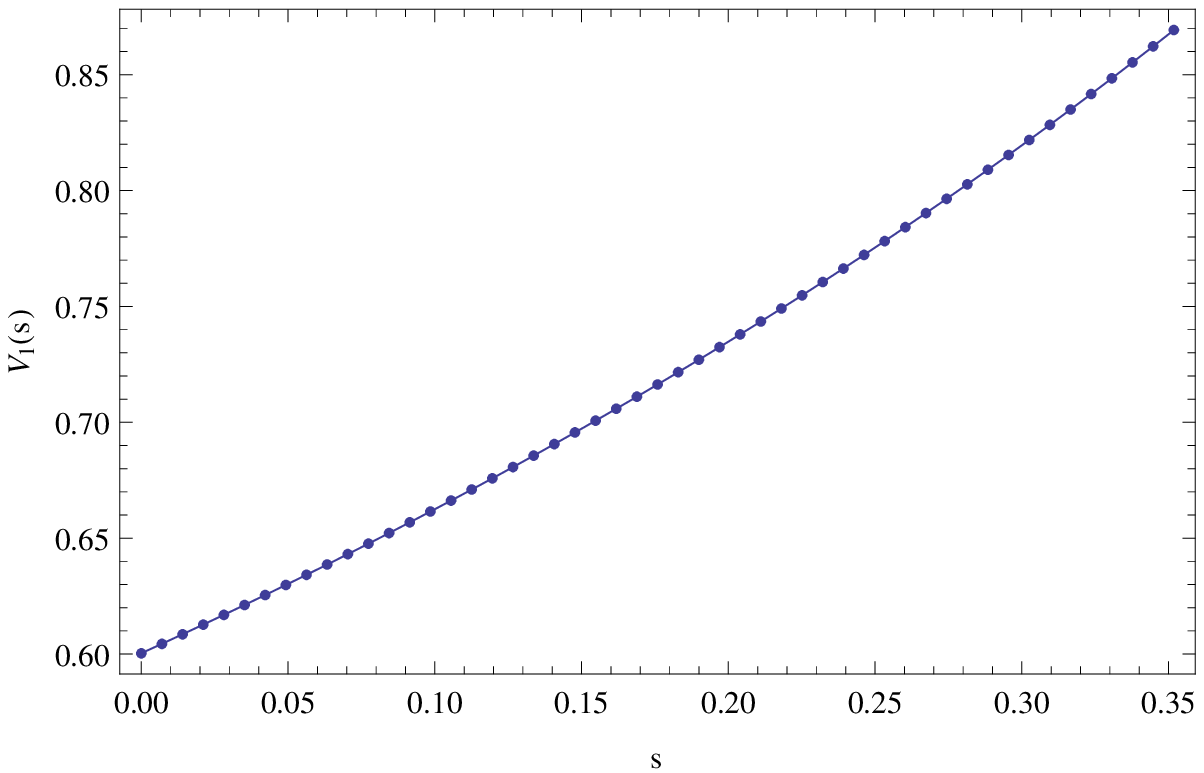}   & \hspace*{.5cm}
\includegraphics[width=0.35\textwidth]{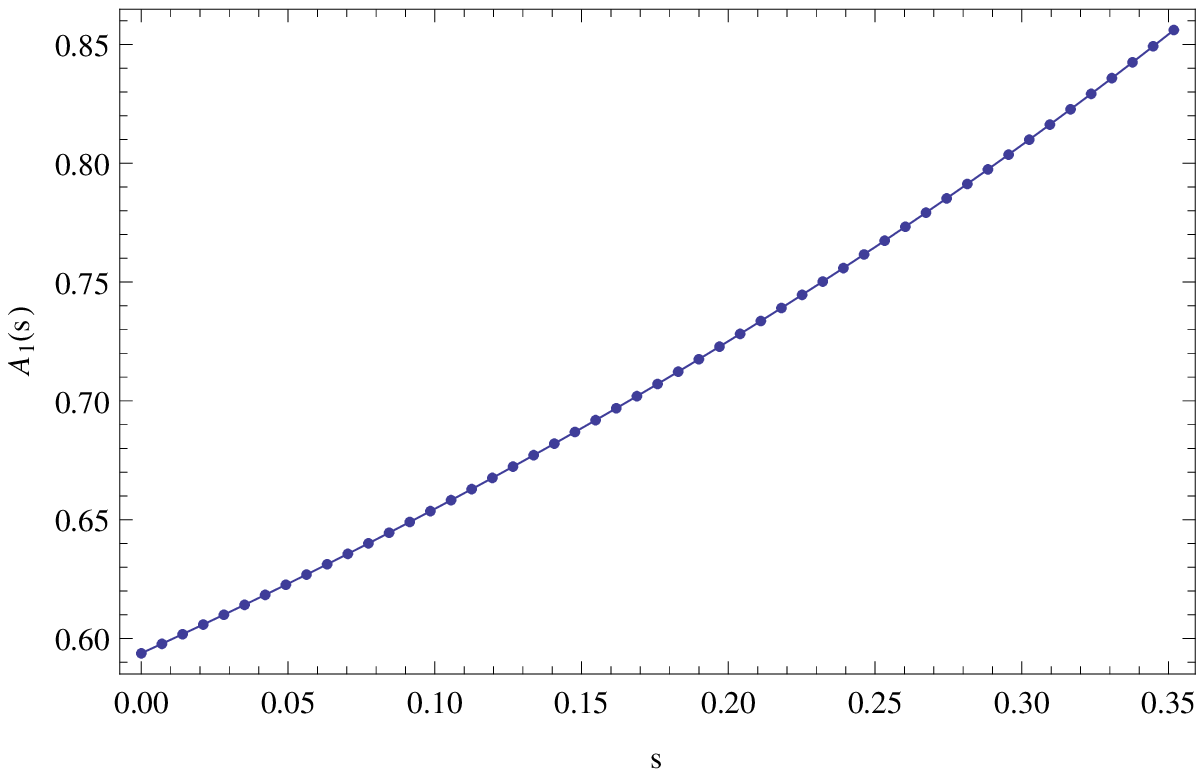}      \\[2ex]

\includegraphics[width=0.35\textwidth]{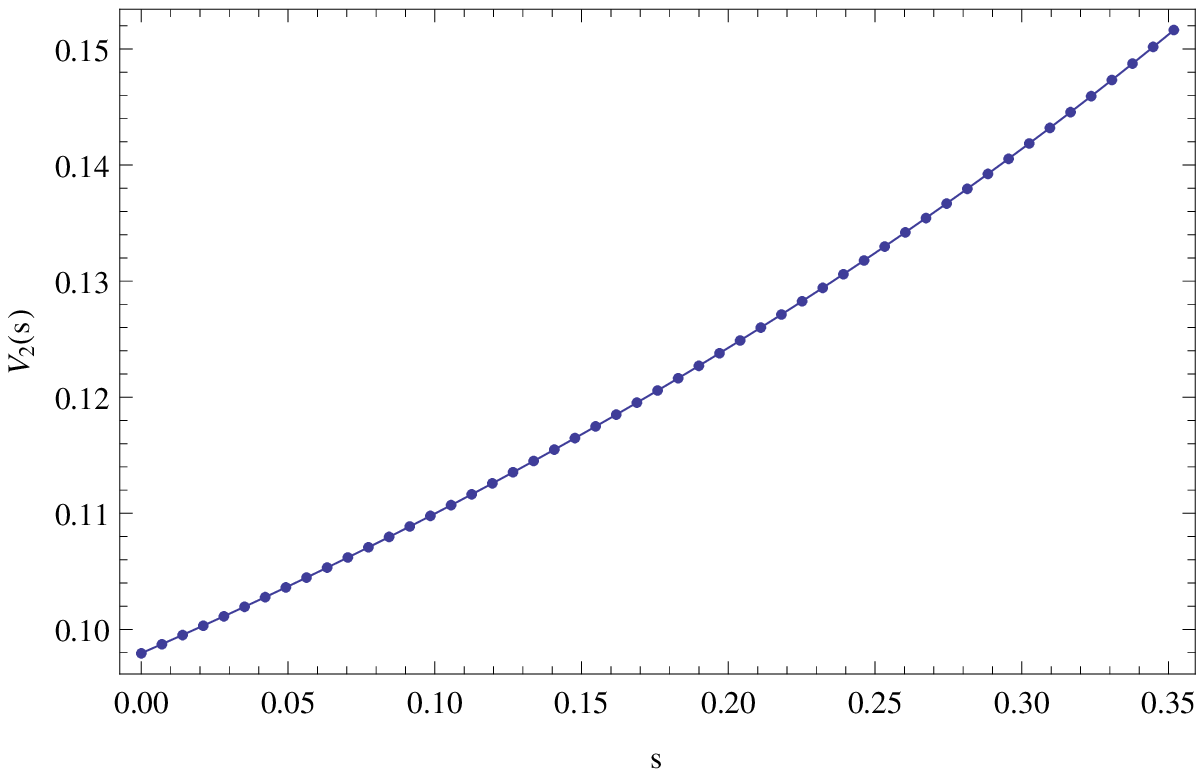}  & \hspace*{.5cm} 
\includegraphics[width=0.35\textwidth]{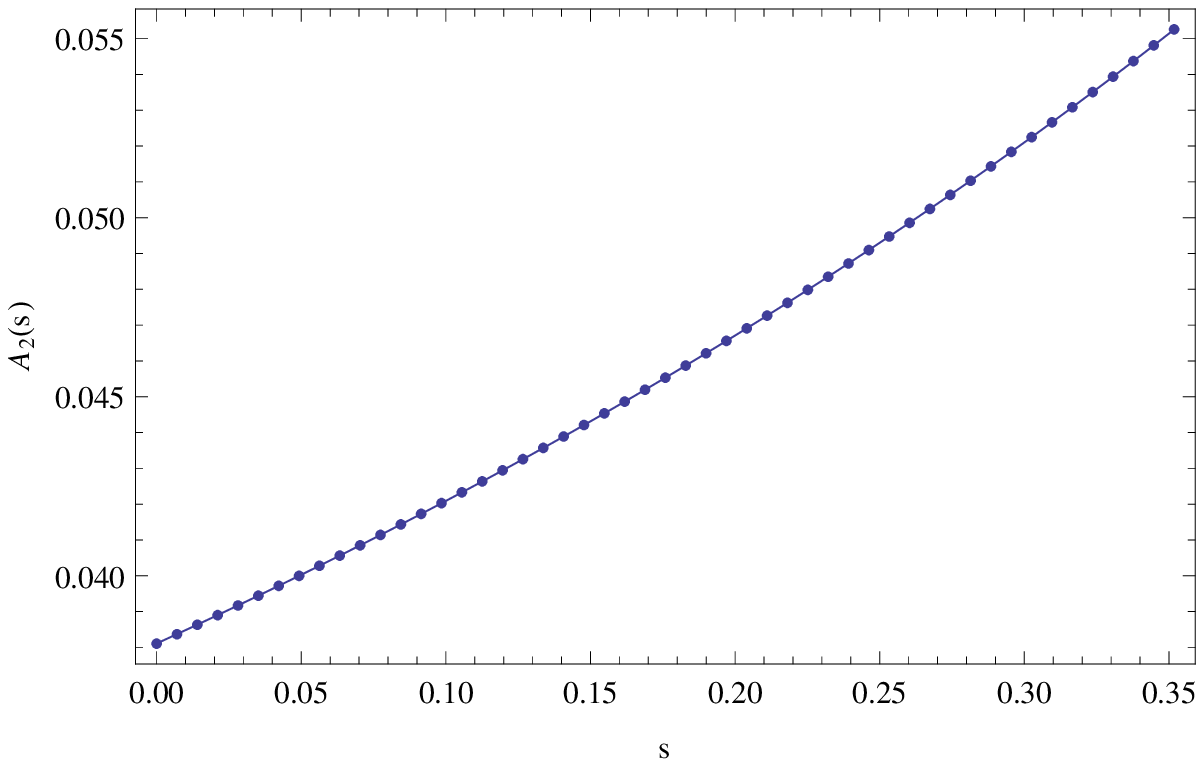}    \\[2ex]

\includegraphics[width=0.35\textwidth]{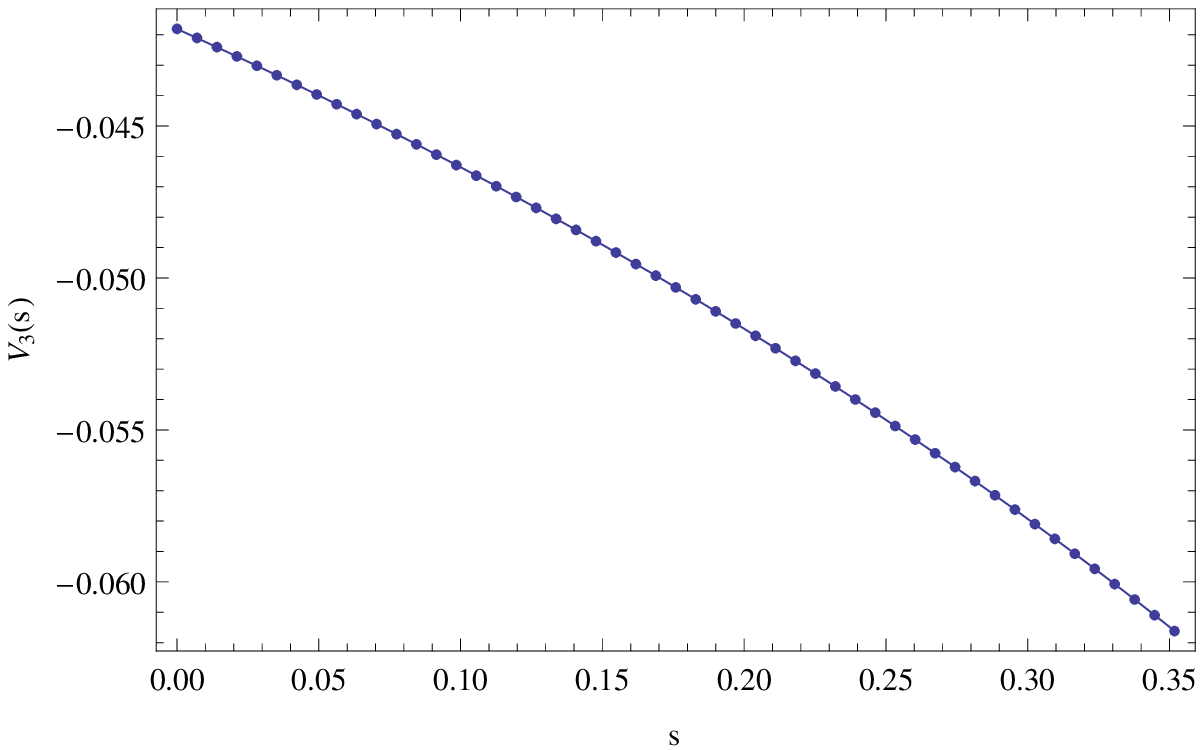}  & \hspace*{.5cm} 
\includegraphics[width=0.35\textwidth]{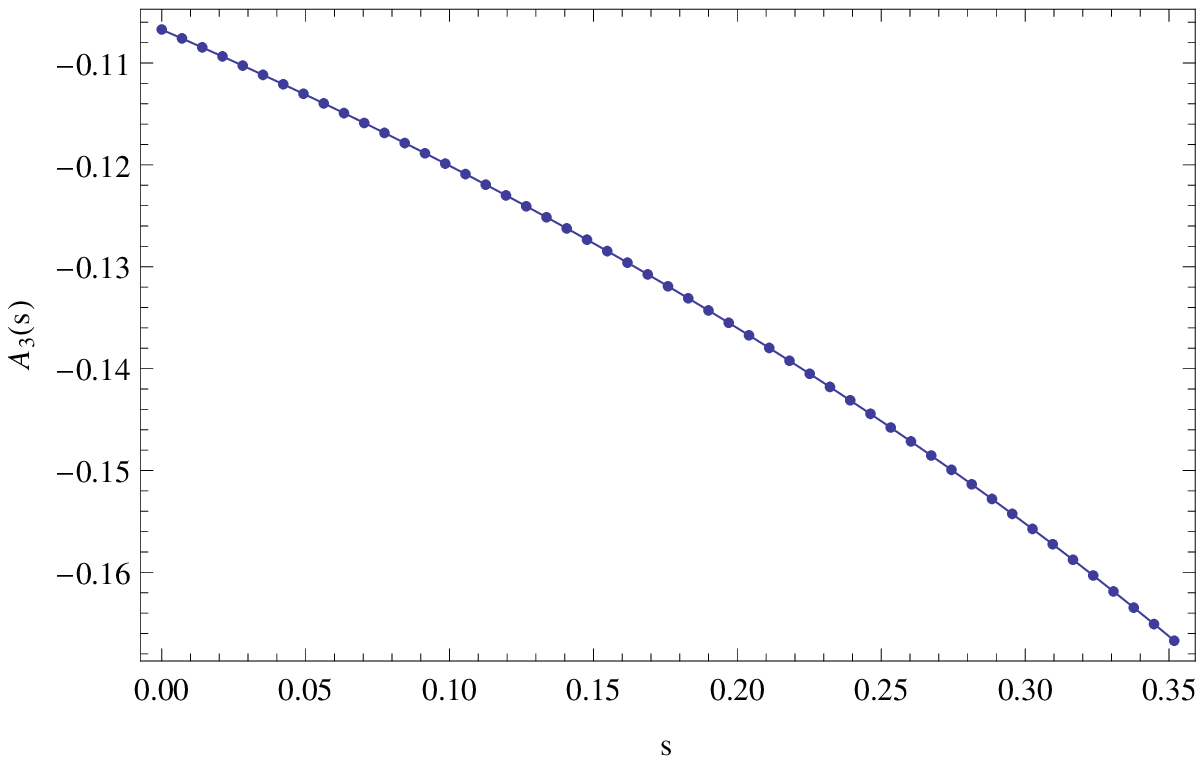}    
\end{tabular}
\end{center}
\caption{\label{fig:ff_bc}
Form factors defining the transition $\Lambda_b\to\Lambda_c$: 
approximated results (solid line), exact result (dotted line).
}
\end{figure}

\newpage 

\begin{figure}[ht]
\begin{center}
\hspace*{-0.5cm}
\begin{tabular}{lr}
\includegraphics[width=0.35\textwidth]{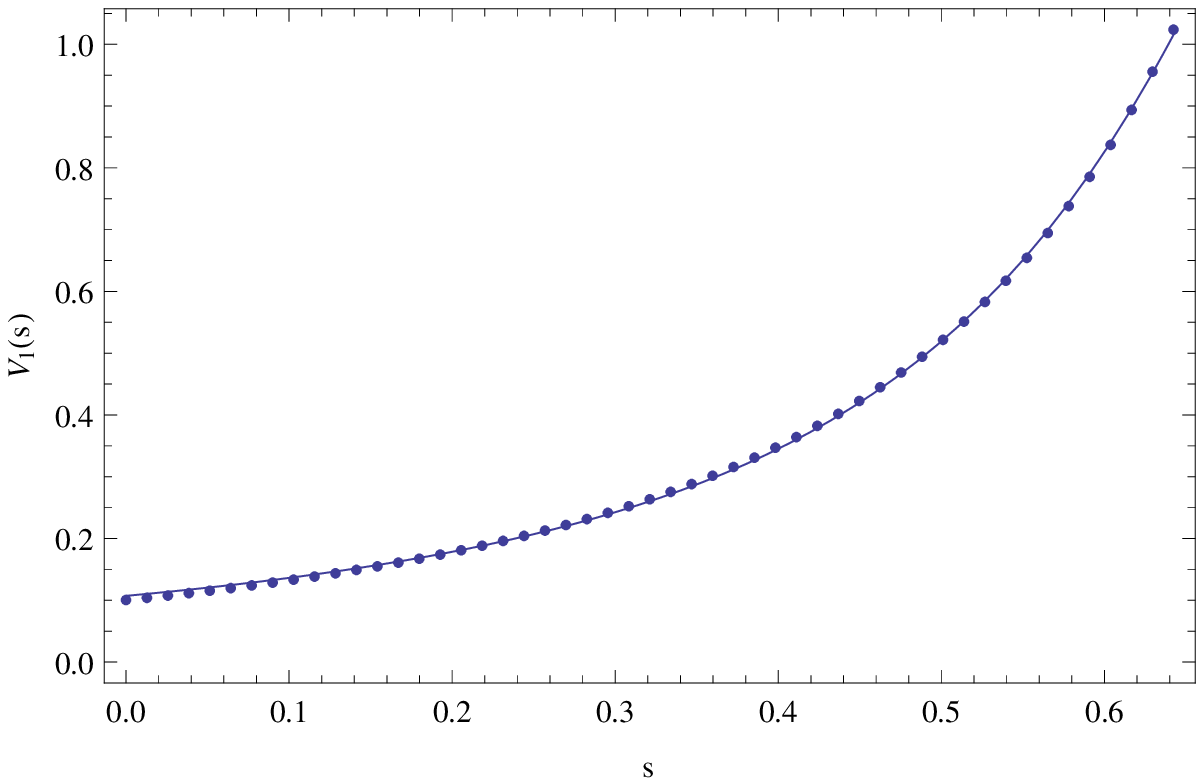}   & \hspace*{.5cm}
\includegraphics[width=0.35\textwidth]{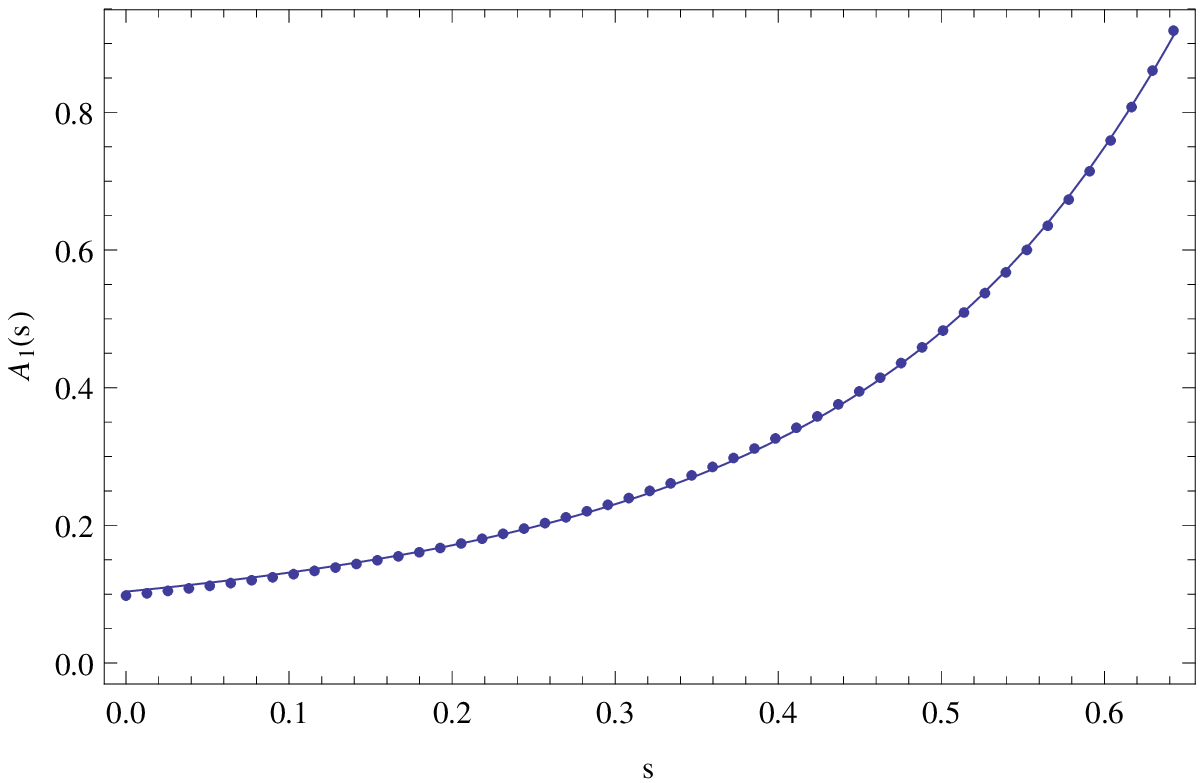}      \\[2ex]

\includegraphics[width=0.35\textwidth]{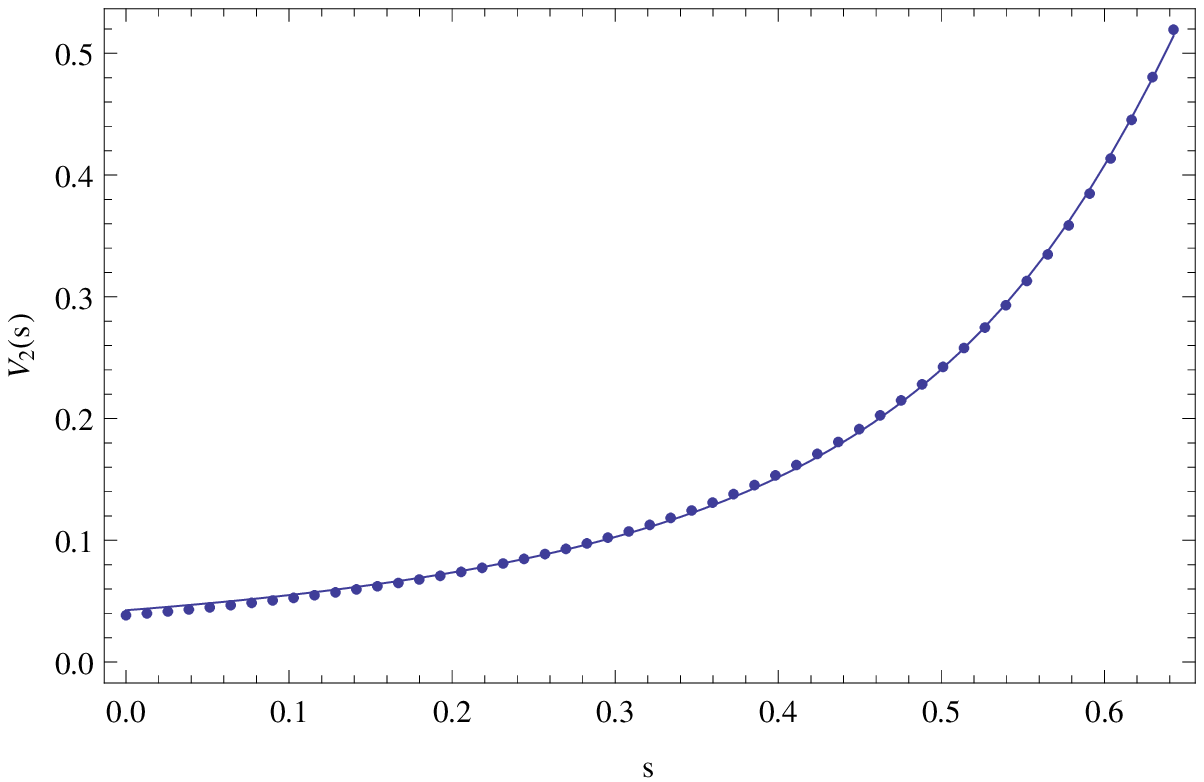}  & \hspace*{.5cm} 
\includegraphics[width=0.35\textwidth]{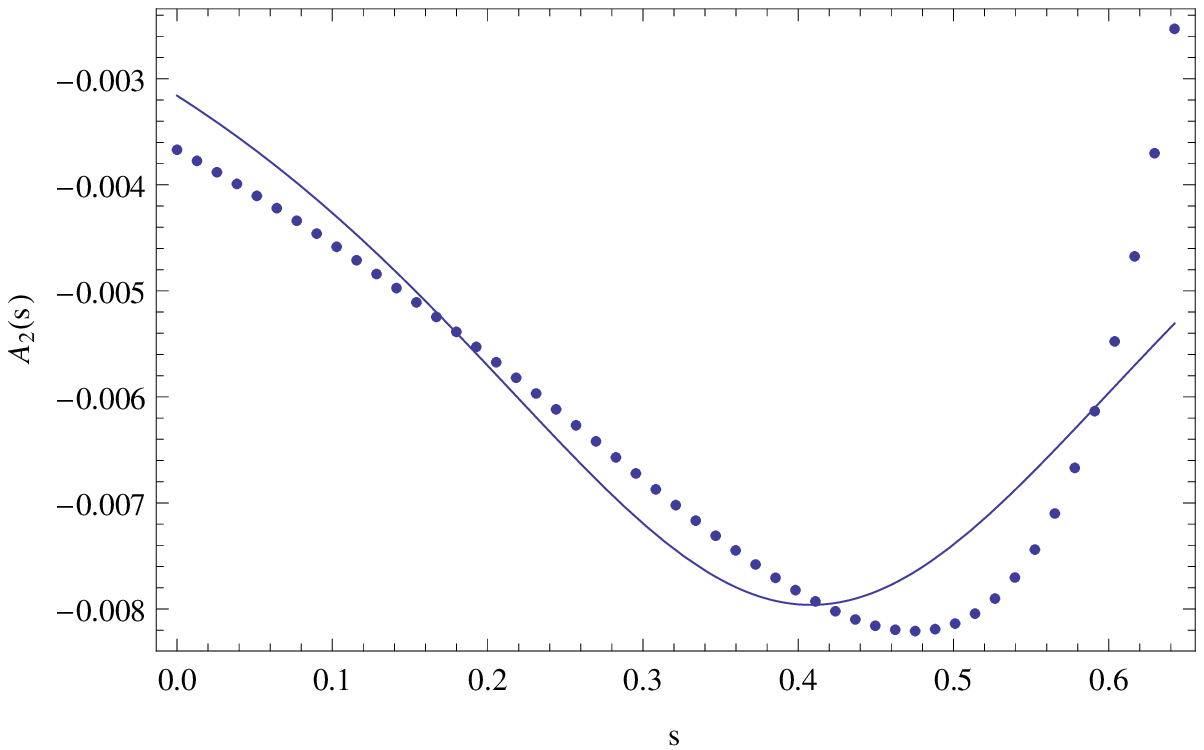}    \\[2ex]

\includegraphics[width=0.35\textwidth]{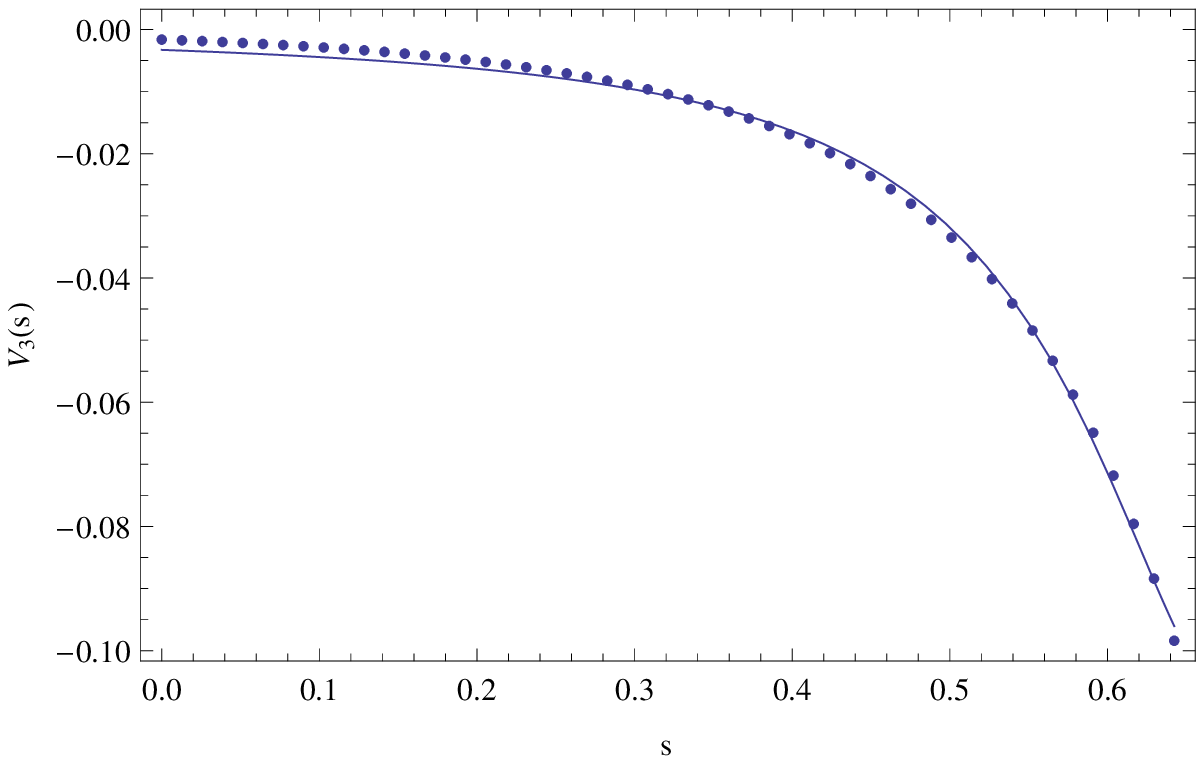}  & \hspace*{.5cm} 
\includegraphics[width=0.35\textwidth]{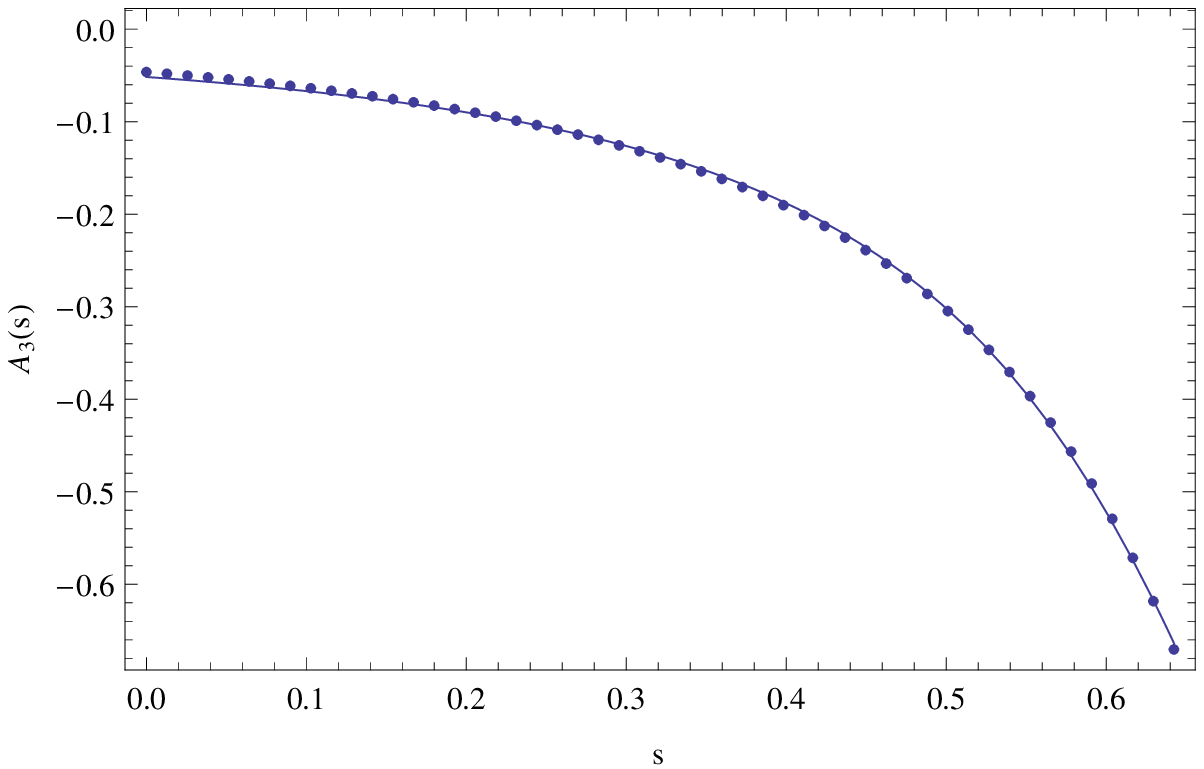}    \\[2ex]

\includegraphics[width=0.35\textwidth]{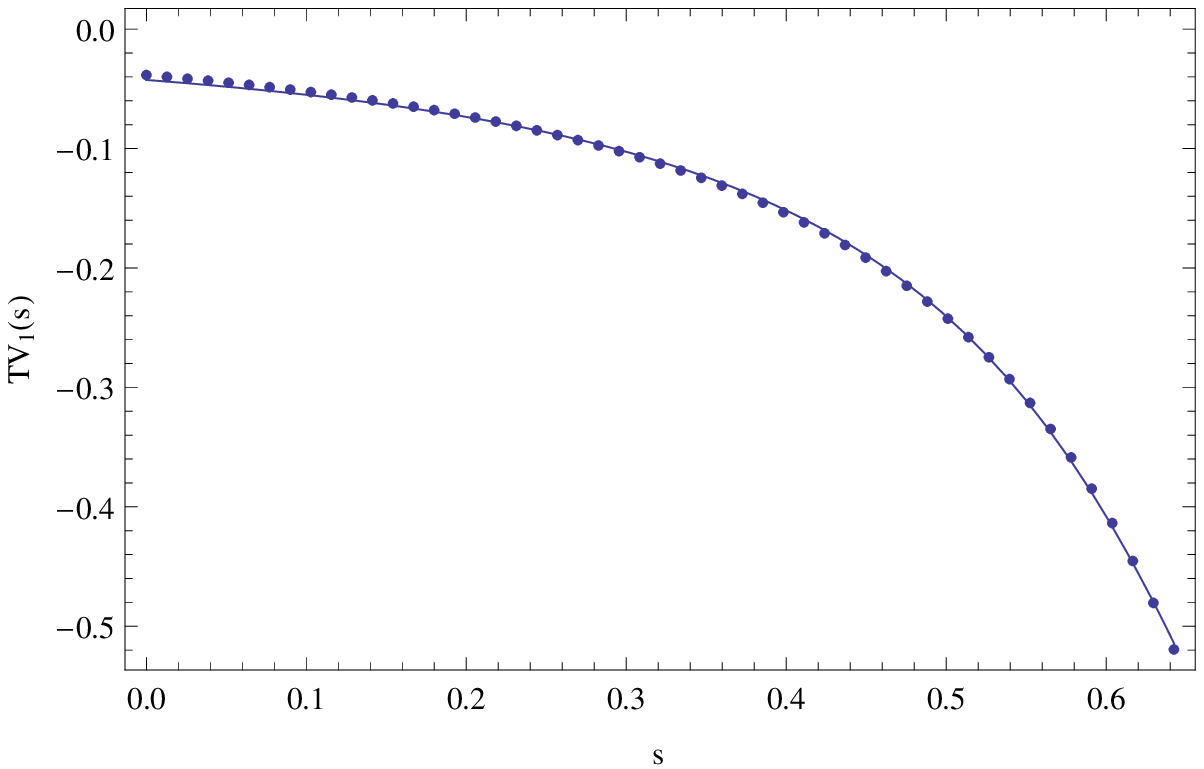}  & \hspace*{.5cm} 
\includegraphics[width=0.35\textwidth]{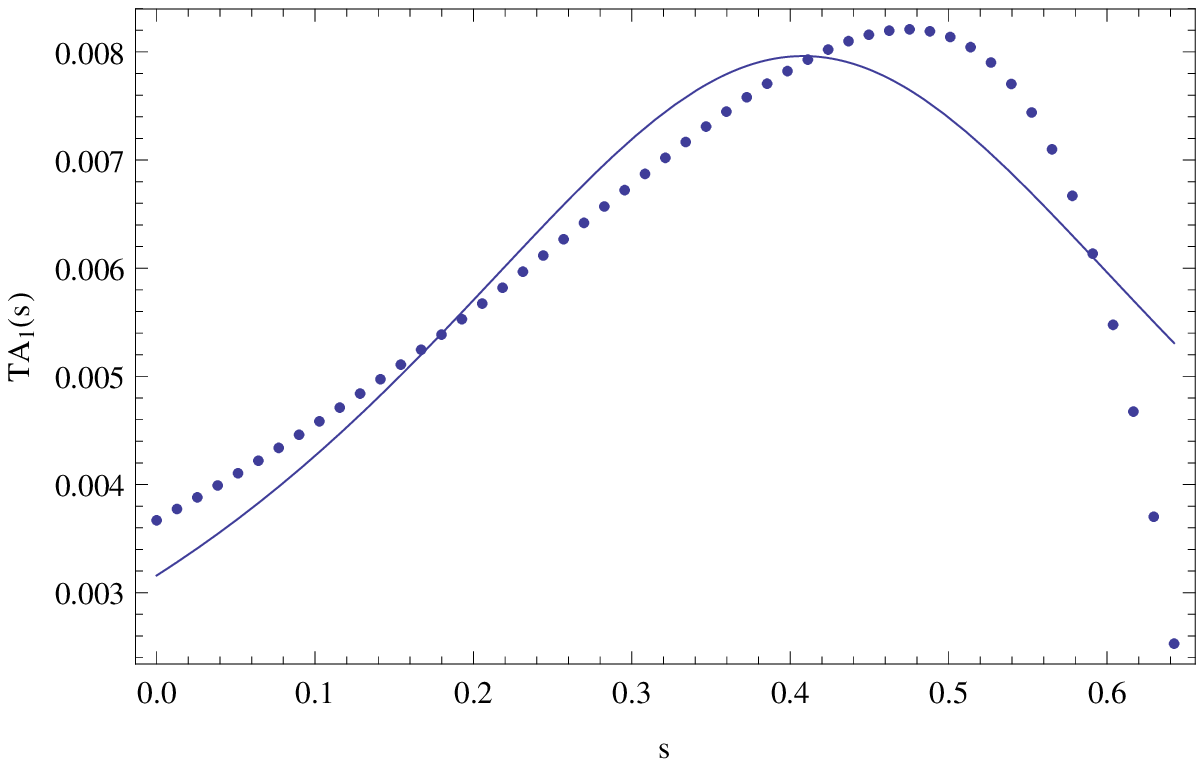}    \\[2ex]

\includegraphics[width=0.35\textwidth]{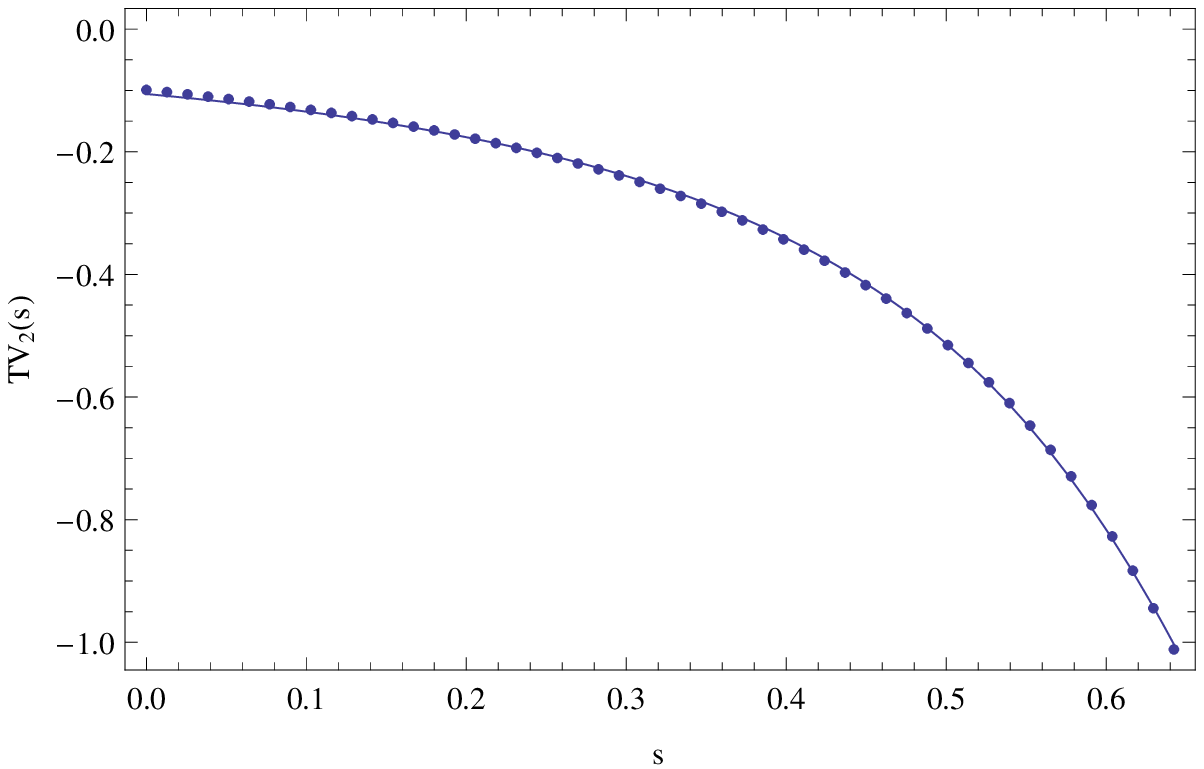}  & \hspace*{.5cm} 
\includegraphics[width=0.35\textwidth]{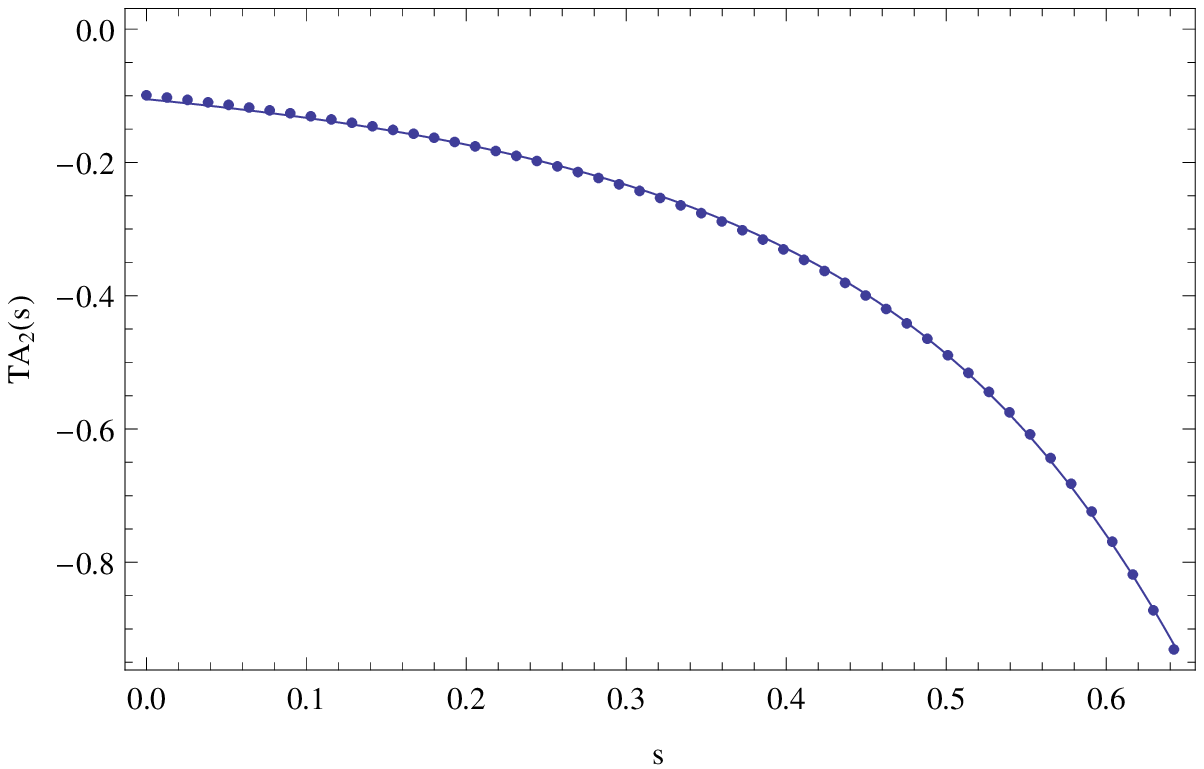}    
\end{tabular}
\end{center}
\caption{\label{fig:ff_bs}
Form factors defining the transition $\Lambda_b\to\Lambda$: 
approximated results (solid line), exact result (dotted line).
}
\end{figure}

\newpage 

\begin{figure}[ht]
\begin{center}
\includegraphics[scale=0.75]{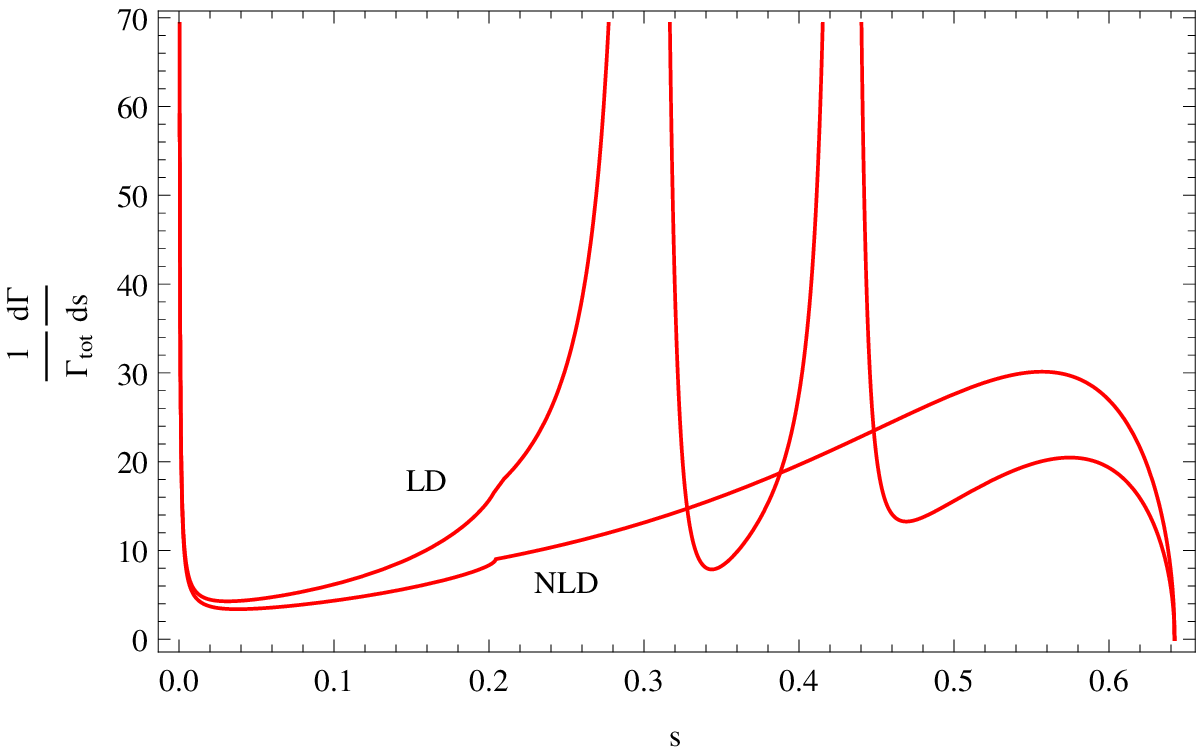} 
\caption{Differential rate 
$\frac{1}{\Gamma_{\rm tot}} \, 
\frac{d\Gamma(\Lambda_{b}\to \Lambda \,e^{+}e^{-})}{ds}$ 
in units of $10^{-7}$ GeV$^{-2}$\,.\label{fig:Distee}} 

\vspace*{.5cm}
\includegraphics[scale=0.75]{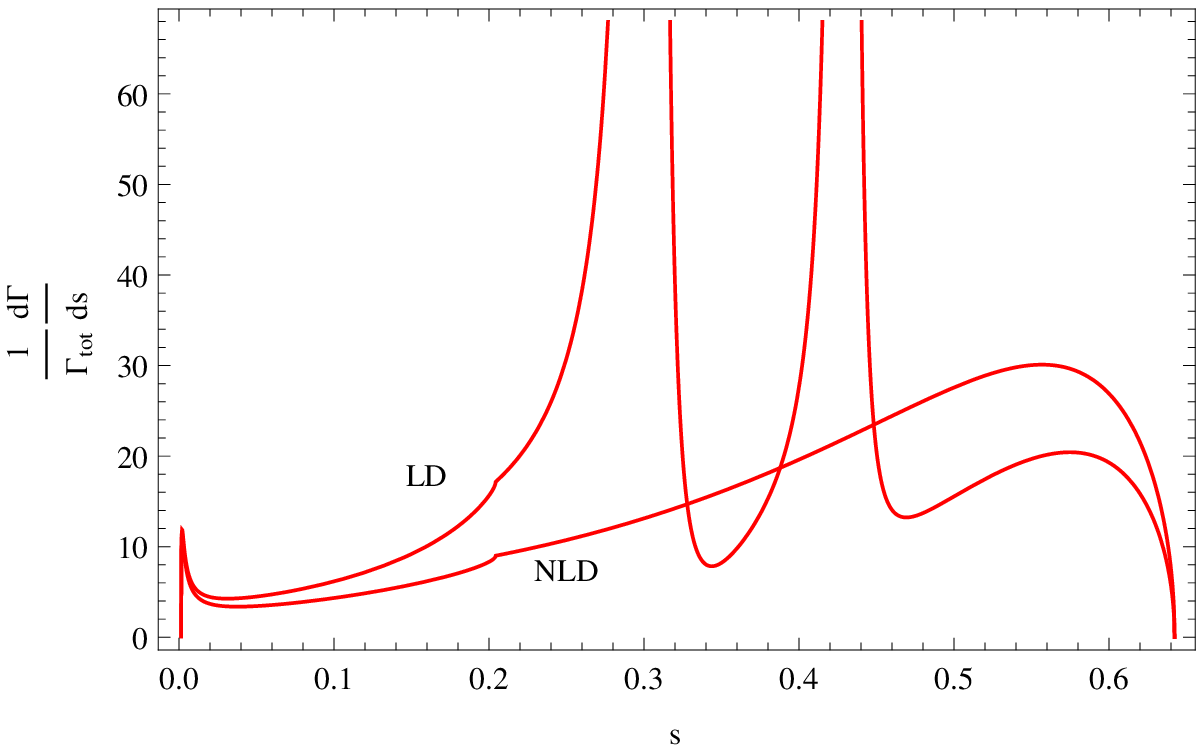} 
\caption{Differential rate 
$\frac{1}{\Gamma_{\rm tot}} \, 
\frac{d\Gamma(\Lambda_{b}\to \Lambda \,\mu^{+}\mu^{-})}{ds}$ 
in units of $10^{-7}$ GeV$^{-2}$\,.\label{fig:Distmm}}   

\vspace*{.5cm}
\includegraphics[scale=0.75]{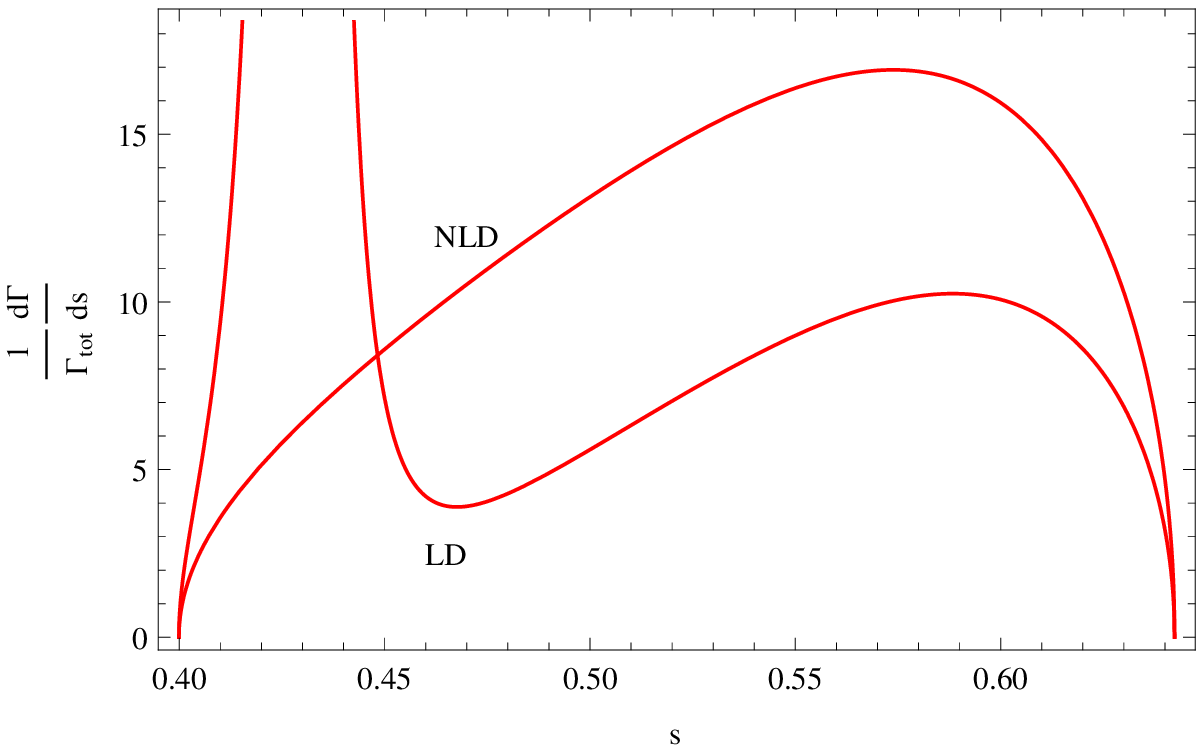} 
\caption{Differential rate 
$\frac{1}{\Gamma_{\rm tot}} \, 
\frac{d\Gamma(\Lambda_{b}\to \Lambda \,\tau^{+}\tau^{-})}{ds}$ 
in units of $10^{-7}$ GeV$^{-2}$\,.\label{fig:Disttt}}  
\end{center}
\end{figure}

\newpage 

\begin{figure}[ht]
\begin{center}
\includegraphics[scale=0.75]{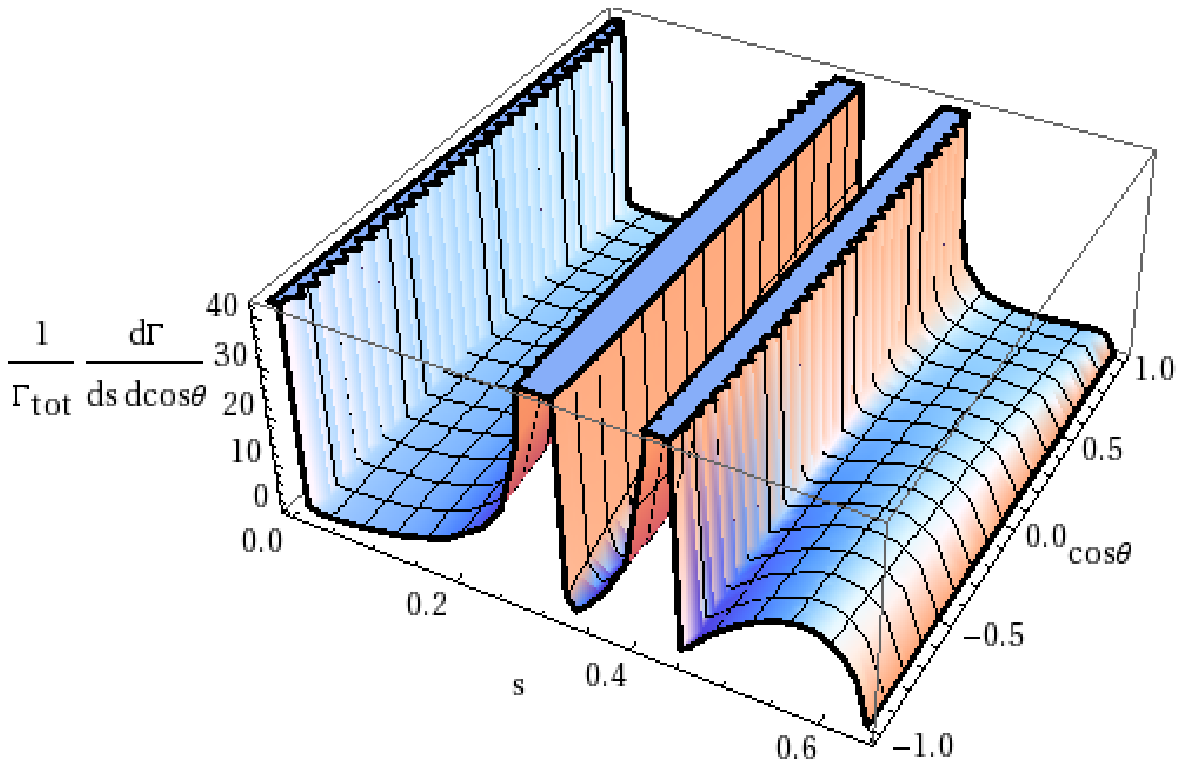} 
\caption{Lepton-side angular decay distribution 
$\frac{1}{\Gamma_{\rm tot}} \, 
\frac{d\Gamma(\Lambda_{b}\to \Lambda \,e^{+}e^{-})}{ds d\cos\theta}$ 
in units of $10^{-7}$ GeV$^{-2}$\,.\label{fig:LSDee}} 

\vspace*{.5cm}
\includegraphics[scale=0.75]{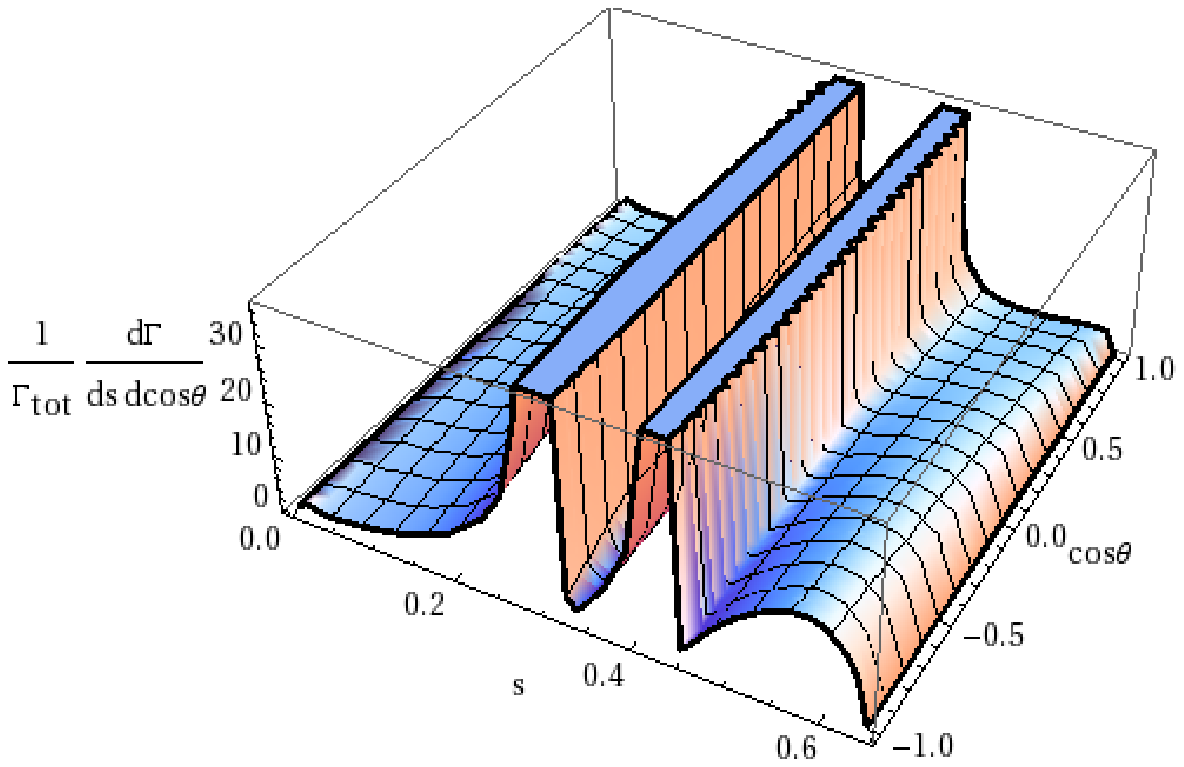} 
\caption{Lepton-side angular decay distribution 
$\frac{1}{\Gamma_{\rm tot}} \, 
\frac{d\Gamma(\Lambda_{b}\to \Lambda \,\mu^{+}\mu^{-})}{ds d\cos\theta}$ 
in units of $10^{-7}$ GeV$^{-2}$\,.\label{fig:LSDmm}} 

\vspace*{.5cm}
\includegraphics[scale=0.75]{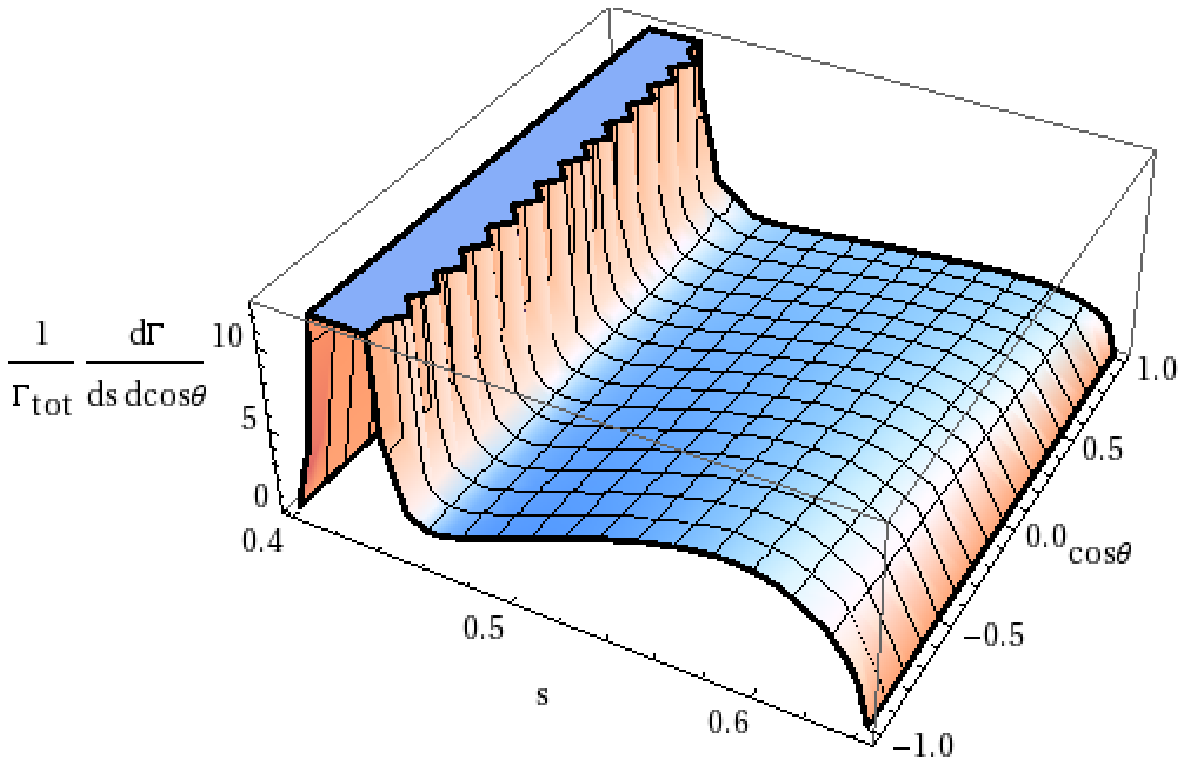} 
\caption{Lepton-side angular decay distribution 
$\frac{1}{\Gamma_{\rm tot}} \, 
\frac{d\Gamma(\Lambda_{b}\to \Lambda \,\tau^{+}\tau^{-})}{ds d\cos\theta}$ 
in units of $10^{-7}$ GeV$^{-2}$\,.\label{fig:LSDtt}} 
\end{center}
\end{figure}

\newpage 

\begin{figure}[ht]
\begin{center}
\includegraphics[scale=0.75]{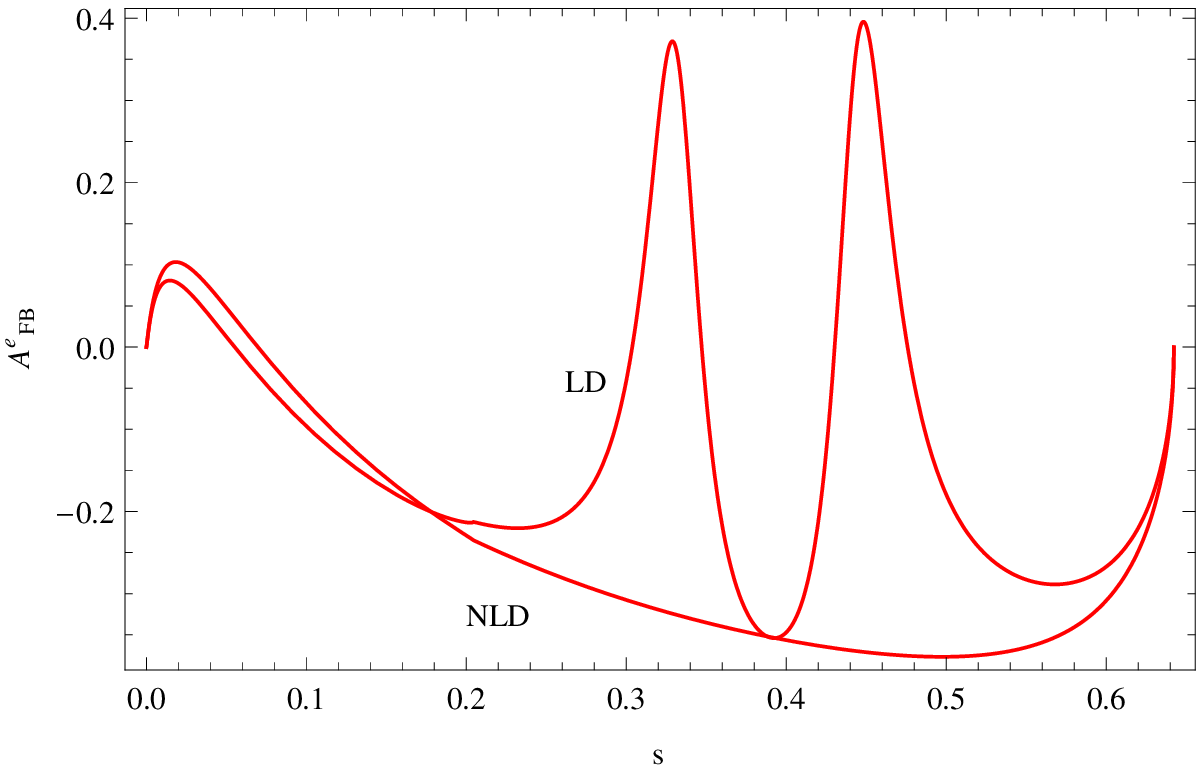} 
\caption{Lepton--side forward--backward asymmetry 
$A^l_{FB}$ 
in the decay $\Lambda_{b}\to \Lambda \,e^{+}e^{-}$\,. \label{fig:AeFB}}  

\vspace*{.5cm}
\includegraphics[scale=0.75]{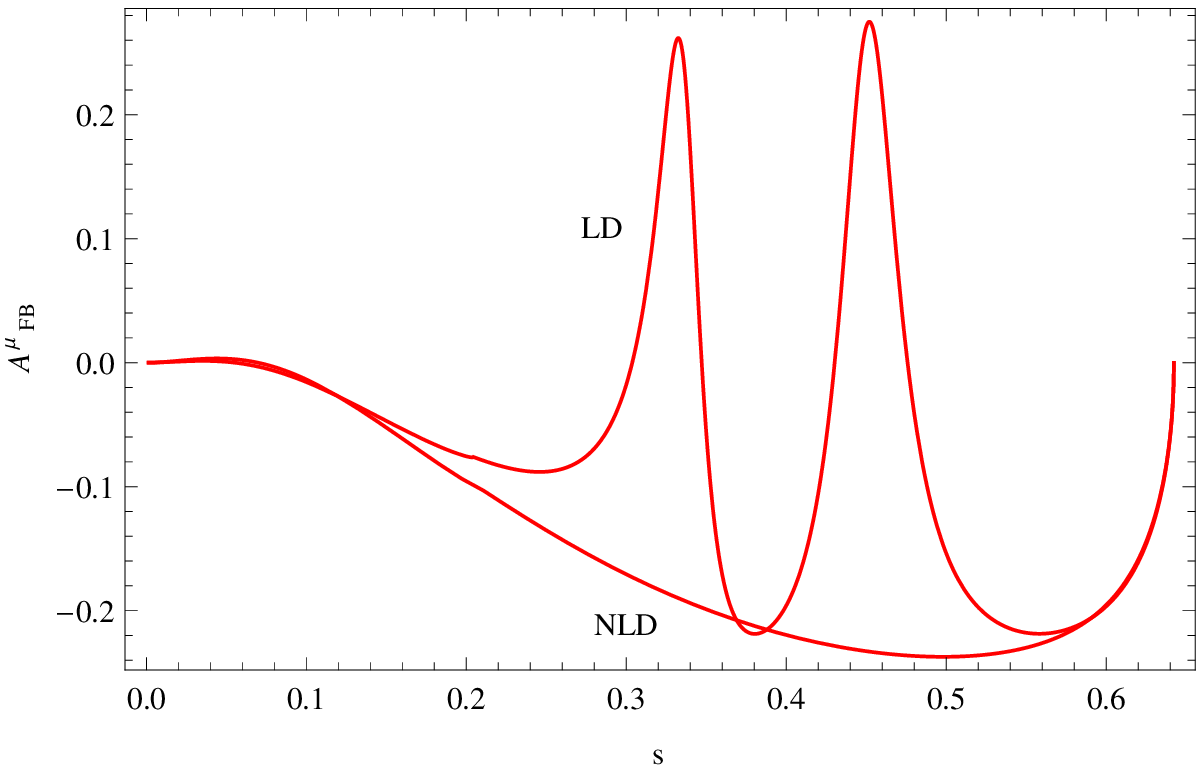} 
\caption{Lepton--side forward--backward asymmetry 
$A^l_{FB}$ 
in the decay $\Lambda_{b}\to \Lambda \,\mu^{+}\mu^{-}$\,.\label{fig:AmFB}}  

\vspace*{.5cm}
\includegraphics[scale=0.75]{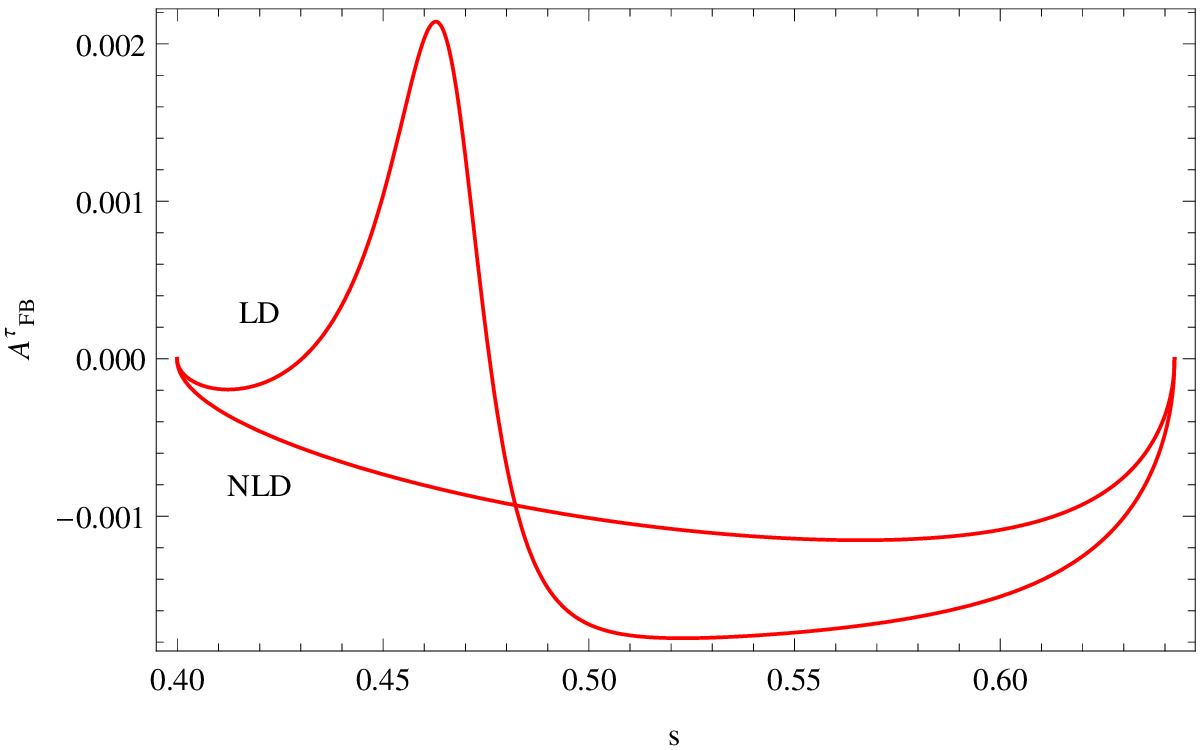} 
\caption{Lepton--side forward--backward asymmetry 
$A^l_{FB}$ 
in the decay $\Lambda_{b}\to \Lambda \,\tau^{+}\tau^{-}$\,.\label{fig:AtFB}}  
\end{center}
\end{figure}

\newpage 

\begin{figure}[ht]
\begin{center}
\includegraphics[scale=0.75]{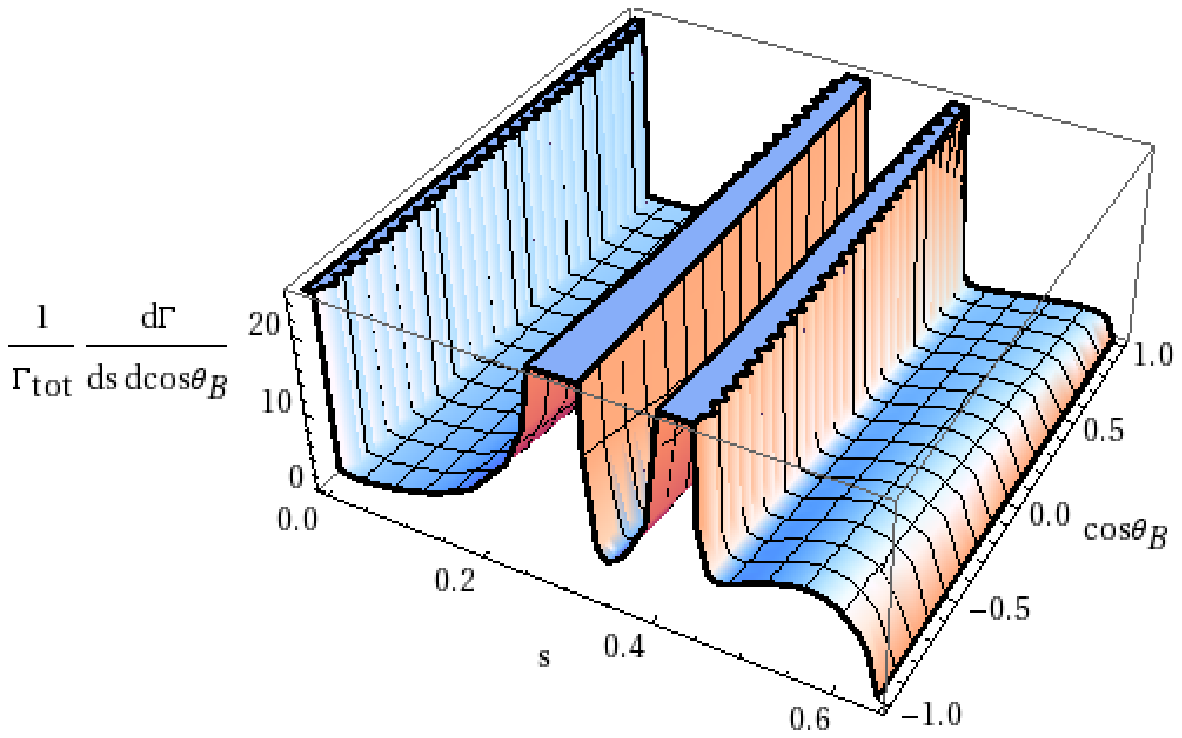} 
\caption{Hadron-side angular decay distribution 
$\frac{1}{\Gamma_{\rm tot}} \, 
\frac{d\Gamma(\Lambda_{b}\to \Lambda \,e^{+}e^{-})}{ds d\cos\theta_B}$ 
in units of $10^{-7}$ GeV$^{-2}$\,.\label{fig:HSDee}}    

\vspace*{.5cm}
\includegraphics[scale=0.75]{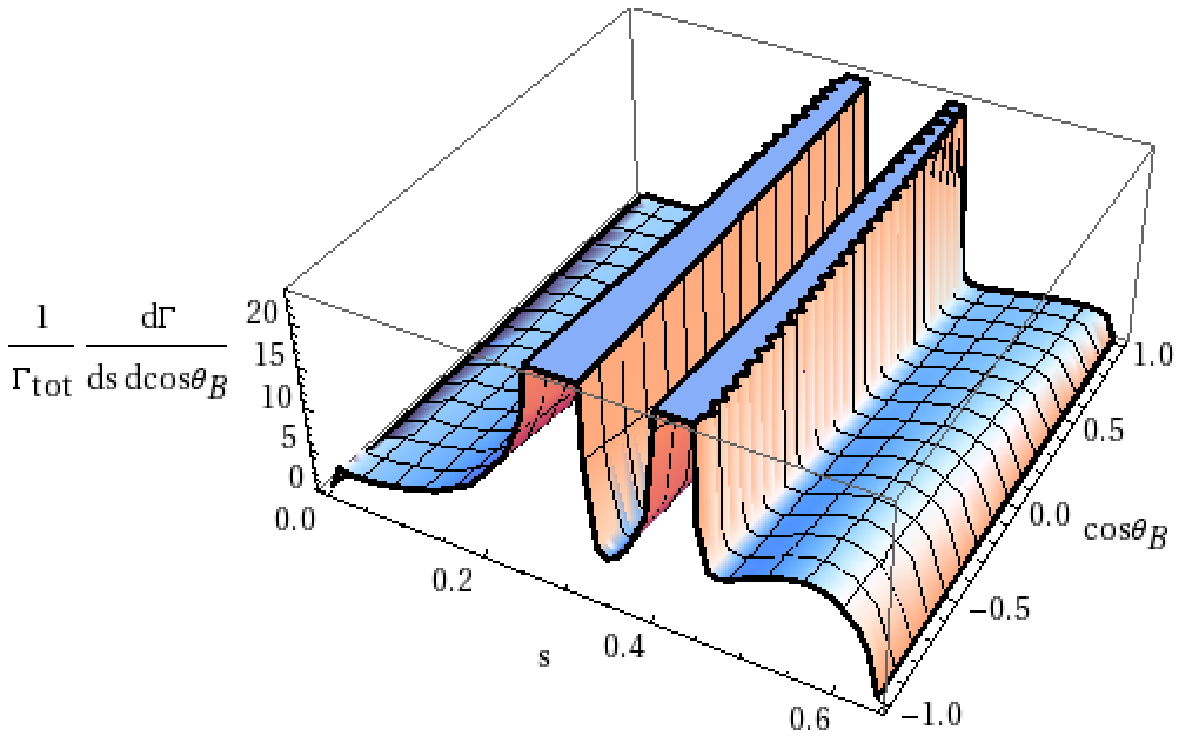} 
\caption{Hadron-side angular decay distribution 
$\frac{1}{\Gamma_{\rm tot}} \, 
\frac{d\Gamma(\Lambda_{b}\to \Lambda \,\mu^{+}\mu^{-})}{ds d\cos\theta_B}$ 
in units of $10^{-7}$ GeV$^{-2}$\,.\label{fig:HSDmm}}    

\vspace*{.5cm}
\includegraphics[scale=0.75]{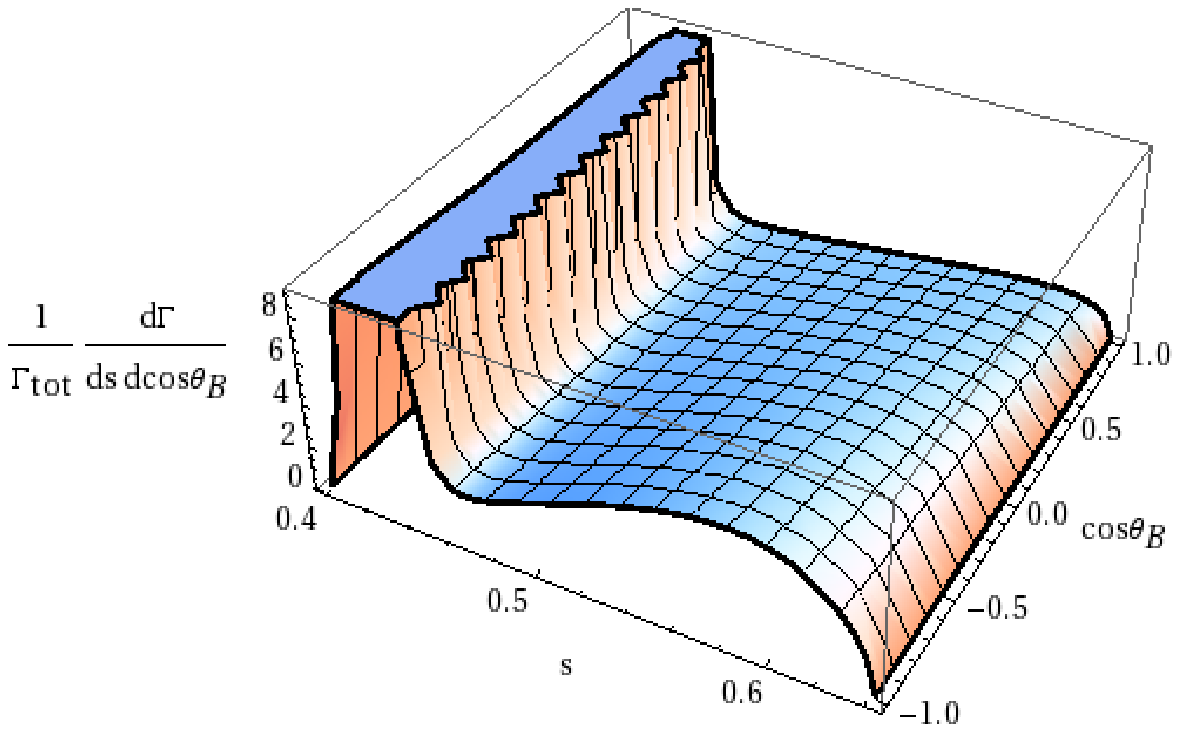} 
\caption{Hadron-side angular decay distribution 
$\frac{1}{\Gamma_{\rm tot}} \, 
\frac{d\Gamma(\Lambda_{b}\to \Lambda \,\tau^{+}\tau^{-})}{ds d\cos\theta_B}$ 
in units of $10^{-7}$ GeV$^{-2}$\,.\label{fig:HSDtt}}    
\end{center}
\end{figure}

\newpage 

\begin{figure}[ht]
\begin{center}
\includegraphics[scale=0.75]{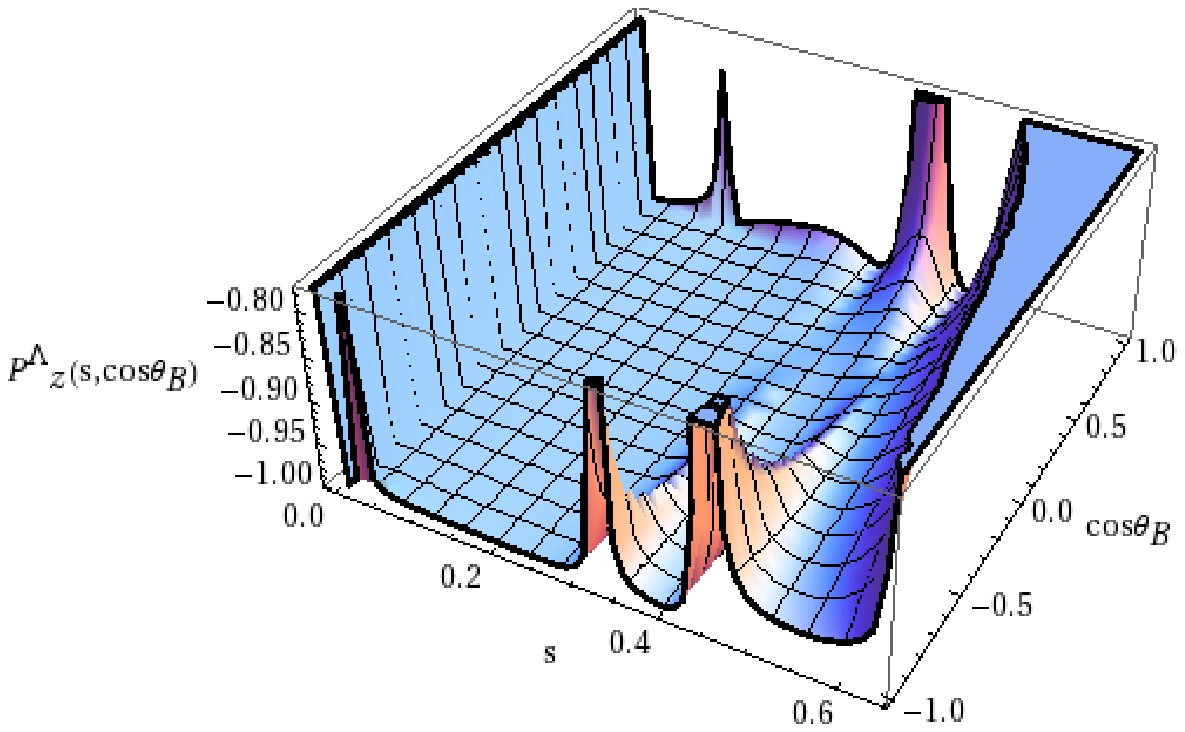} 
\caption{Polarization $P_z^\Lambda(s,\cos\theta_B)$ 
for the decay $\Lambda_{b}\to \Lambda \,e^{+}e^{-}$\,. \label{fig:HpSDee}}     

\vspace*{.5cm}
\includegraphics[scale=0.75]{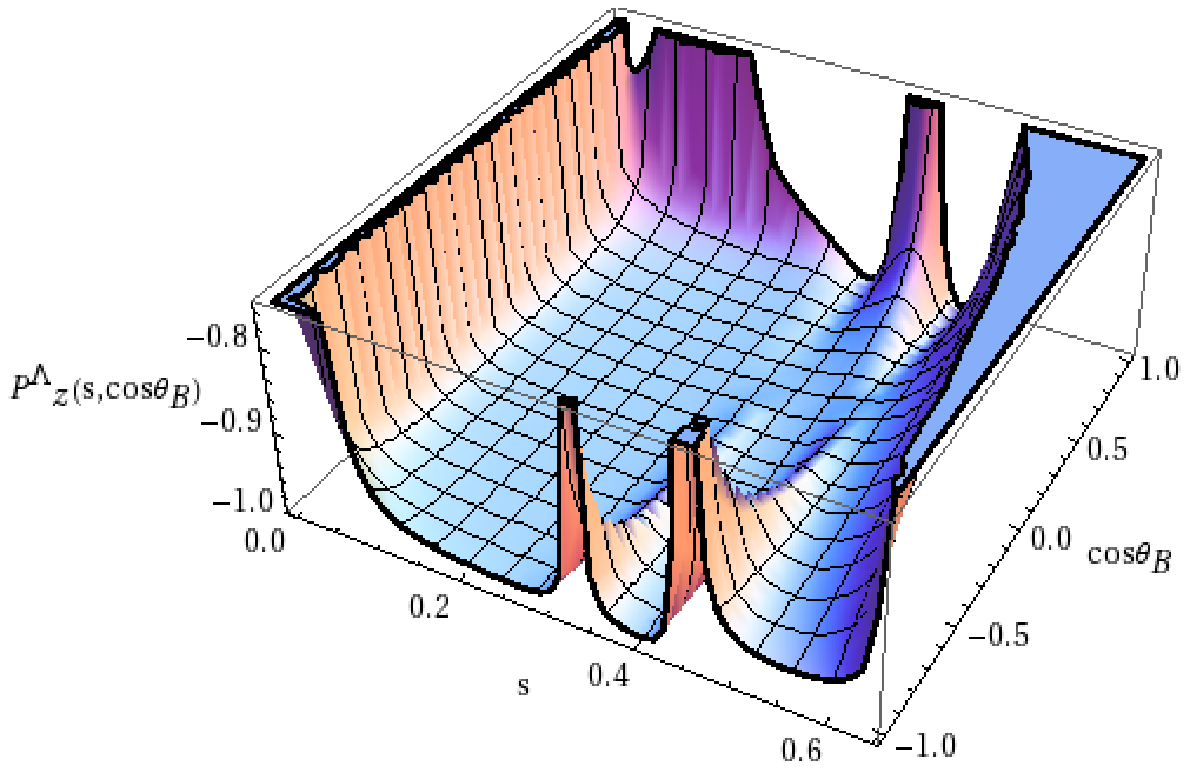} 
\caption{Polarization $P_z^\Lambda(s,\cos\theta_B)$ 
in the decay $\Lambda_{b}\to \Lambda \,\mu^{+}\mu^{-}$\,. 
\label{fig:HpSDmm}}     

\vspace*{.5cm}
\includegraphics[scale=0.75]{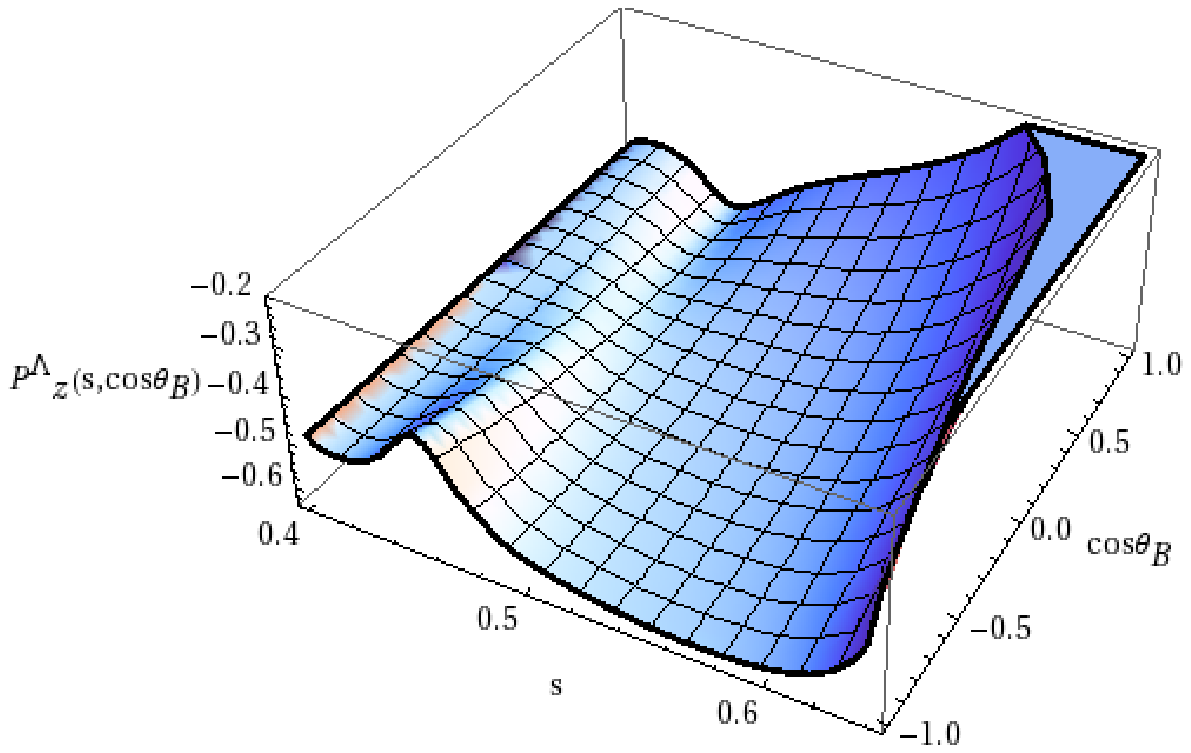} 
\caption{Polarization $P_z^\Lambda(s,\cos\theta_B)$ 
in the decay $\Lambda_{b}\to \Lambda \,\tau^{+}\tau^{-}$\,.
\label{fig:HpSDtt}}     
\end{center}
\end{figure}

\newpage 

\begin{figure}[ht]
\begin{center}
\includegraphics[scale=0.75]{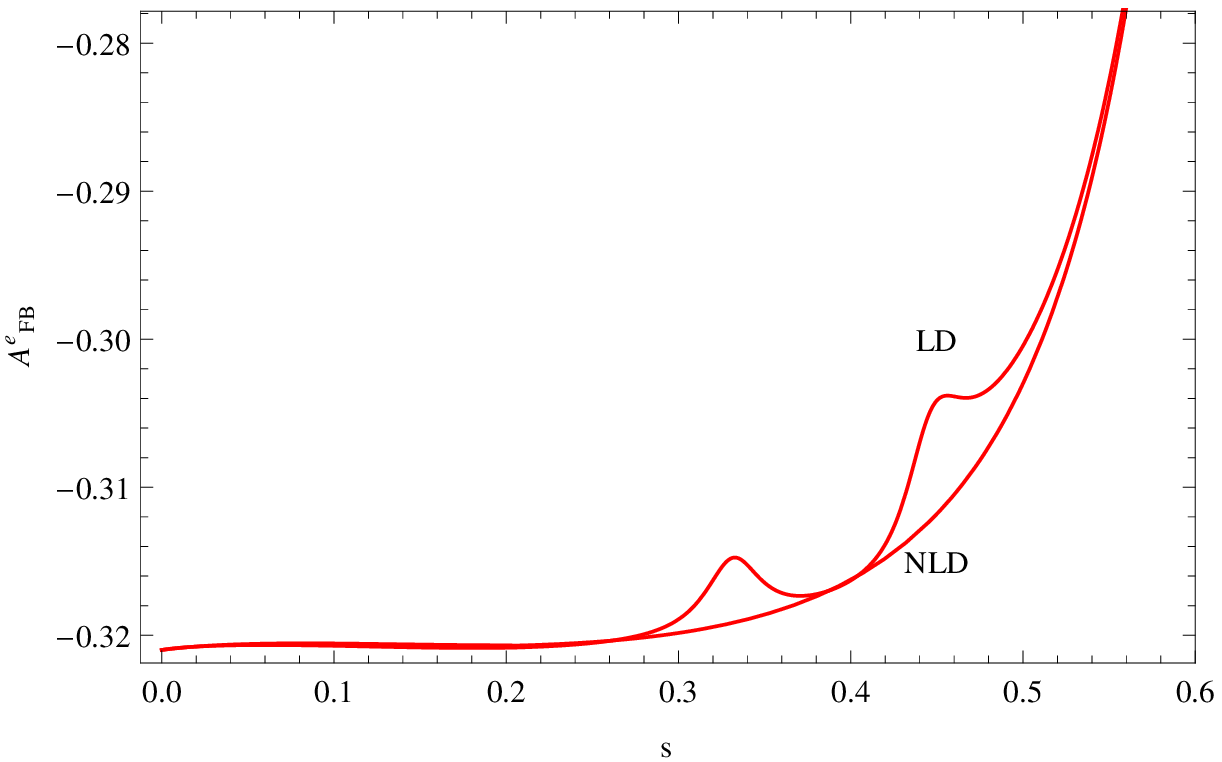} 
\caption{Hadron--side forward--backward asymmetry 
$A^h_{FB}$ 
in the decay $\Lambda_{b}\to \Lambda \,e^{+}e^{-}$\,. 
\label{fig:AheFB}}     

\vspace*{.5cm}
\includegraphics[scale=0.75]{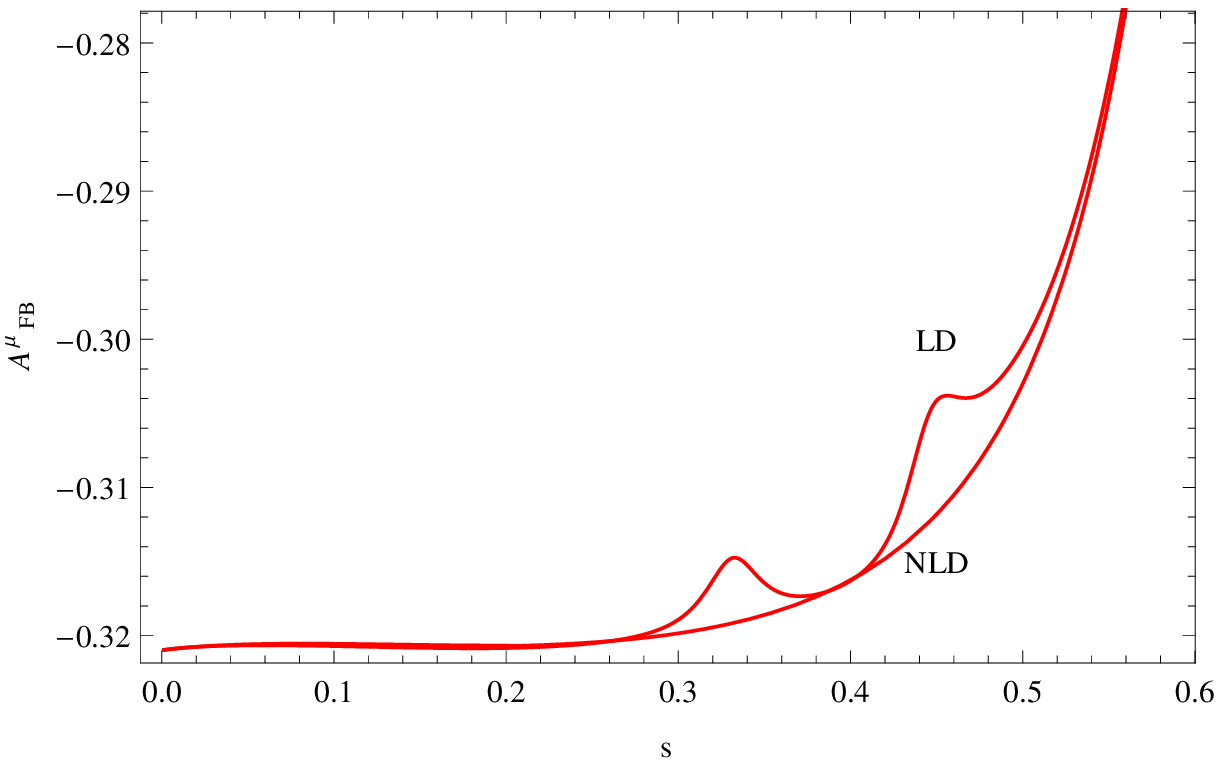} 
\caption{Hadron--side forward--backward asymmetry 
$A^h_{FB}$ 
in the decay $\Lambda_{b}\to \Lambda \,\mu^{+}\mu^{-}$\,. 
 \label{fig:AhmFB}}     

\vspace*{.5cm}
\includegraphics[scale=0.75]{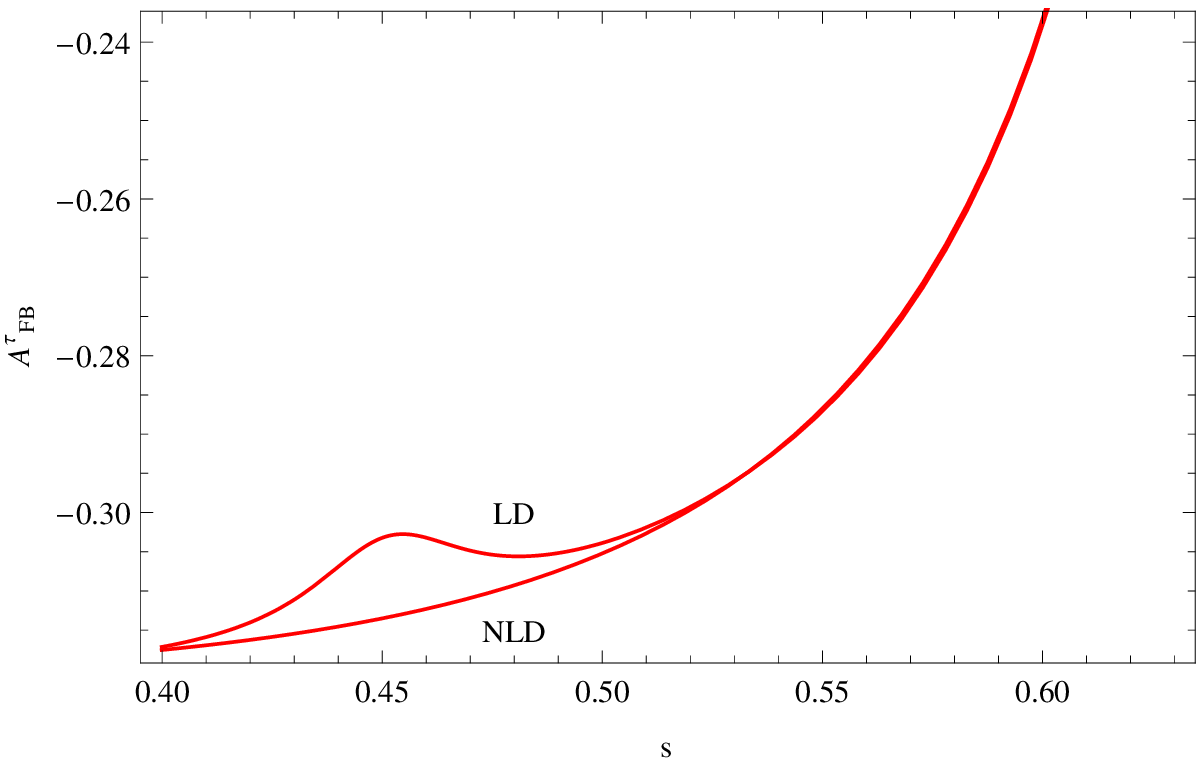} 
\caption{Hadron--side forward--backward asymmetry 
$A^h_{FB}$ 
in the decay $\Lambda_{b}\to \Lambda \,\tau^{+}\tau^{-}$\,. 
\label{fig:AhtFB}}     
\end{center}
\end{figure}

\newpage

\begin{table} 
\caption{Parameters for the approximated form factors
$f(s)=f(0)/(1 - a s + b s^2)\,,$ \
$s=q^2/m_{\Lambda_b}^2$ in the $\Lambda_c \to \Lambda$ transition.}
\begin{center}
\def\arraystretch{.9}\begin{tabular}{|l|l|l|l|l|l|l|}
\hline
& $f_1^V$ & $f_2^V$ & $f_3^V$ 
& $f_1^A$ & $f_2^A$ & $f_3^A$
\\
\hline
$f(0)$ & 0.468 & 0.204 & 0.059 & 0.431 & -0.078 & -0.256\\ 
$a$    & 1.017 & 1.148 & 0.698 & 0.939 & 0.870 &  1.208\\ 
$b$    & 0.249 & 0.337 & 0.221 & 0.211 & 0.195 &  0.377\\ 
\hline
\end{tabular}
\label{tab:fflcs}
\end{center}

\caption{Parameters for the approximated form factors
$f(s)=f(0)/(1 - a s + b s^2)\,,$ \
$s=q^2/m_{\Lambda_b}^2$ in the $\Lambda_b \to \Lambda_c$ transition.}
\begin{center}
\def\arraystretch{.9}
\begin{tabular}{|l|l|l|l|l|l|l|}
\hline
& $f_1^V$ & $f_2^V$ & $f_3^V$ 
& $f_1^A$ & $f_2^A$ & $f_3^A$
\\
\hline
$f(0)$ & 0.600 & 0.098 & 0.042 & 0.594 & 0.038 & -0.107\\ 
$a$    & 0.961 & 1.127 & 1.008 & 0.951 & 0.971 &  1.148\\ 
$b$    & 0.233 & 0.344 & 0.267 & 0.228 & 0.254 &  0.357\\ 
\hline
\end{tabular}
\label{tab:fflbc}
\end{center}

\caption{Parameters for the approximated form factors
$f(s)=f(0)/(1 - a s + b s^2)\,,$ \
$s=q^2/m_{\Lambda_b}^2$ in the $\Lambda_b \to \Lambda$ transition.} 
\begin{center}
\def\arraystretch{.9}
\begin{tabular}{|l|l|l|l|l|l|l|l|l|l|l|}
\hline
& $f_1^V$ & $f_2^V$ & $f_3^V$ 
& $f_1^A$ & $f_2^A$ & $f_3^A$
& $f_1^{TV}$ & $f_2^{TV}$ 
& $f_1^{TA}$ & $f_2^{TA}$ 
\\
\hline
$f(0)$ & 0.107 & 0.043 & 0.003 & 0.104 & 0.003 & -0.052 & -0.043 & -0.105 & 0.003 & -0.105\\
$a$    & 2.271 & 2.411 & 2.815 & 2.232 & 2.955 & 2.437  & 2.411  & 0.072  & 2.955  & 2.233\\
$b$    & 1.367 & 1.531 & 2.041 & 1.328 & 3.620 & 1.559  & 1.531  & 0.001  & 3.620  & 1.328\\
\hline
\end{tabular}
\label{tab:fflbs}
\end{center}

\begin{center}
\caption{Branching ratios of semileptonic decays of heavy baryons 
(in \%).} 

\vspace*{.15cm}

\def\arraystretch{1}
    \begin{tabular}{|c|c|c|}
      \hline
Mode  & Our results & Data~\cite{Beringer:1900zz} \\
\hline
$\Lambda_c \to \Lambda e^+\nu_e$     
& $2.0$ & $2.1 \pm 0.6$ \\
\hline
$\Lambda_c \to \Lambda \mu^+\nu_\mu$     
& $2.0$ & $2.0 \pm 0.7$ \\
\hline
$\Lambda_b \to \Lambda_c e^-\bar\nu_e$     
& $6.6$ & $6.5^{+3.2}_{-2.5}$ \\ 
\hline
$\Lambda_b \to \Lambda_c \mu^-\bar\nu_\mu$     
& $6.6$ &  \\ 
\hline
$\Lambda_b \to \Lambda_c \tau^-\bar\nu_\tau$     
& $1.8$ &  \\ 
\hline
\end{tabular}
\label{tab:rates}
\end{center}

\begin{center}
\caption{Asymmetry parameter $\alpha$ 
in the semileptonic decays of heavy baryons.} 

\vspace*{.15cm}

\def\arraystretch{1}
    \begin{tabular}{|c|c|c|}
      \hline
Mode  & Our results & Data~\cite{Beringer:1900zz}  \\
\hline
$\Lambda_c \to \Lambda e^+\nu_e$     
& $0.828$ & $0.86 \pm 0.04$ \\
\hline
$\Lambda_c \to \Lambda \mu^+\nu_\mu$     
& $0.825$ & \\ 
\hline
$\Lambda_b \to \Lambda_c e^-\bar\nu_e$     
& $0.831$ & \\ 
\hline
$\Lambda_b \to \Lambda_c \mu^-\bar\nu_\mu$     
& $0.831$ &  \\ 
\hline
$\Lambda_b \to \Lambda_c \tau^-\bar\nu_\tau$     
& $0.731$ &  \\ 
\hline
\end{tabular}
\label{tab:asym}
\end{center}
\end{table}

\newpage 

\begin{figure}
\begin{center}
\includegraphics[scale=0.75]{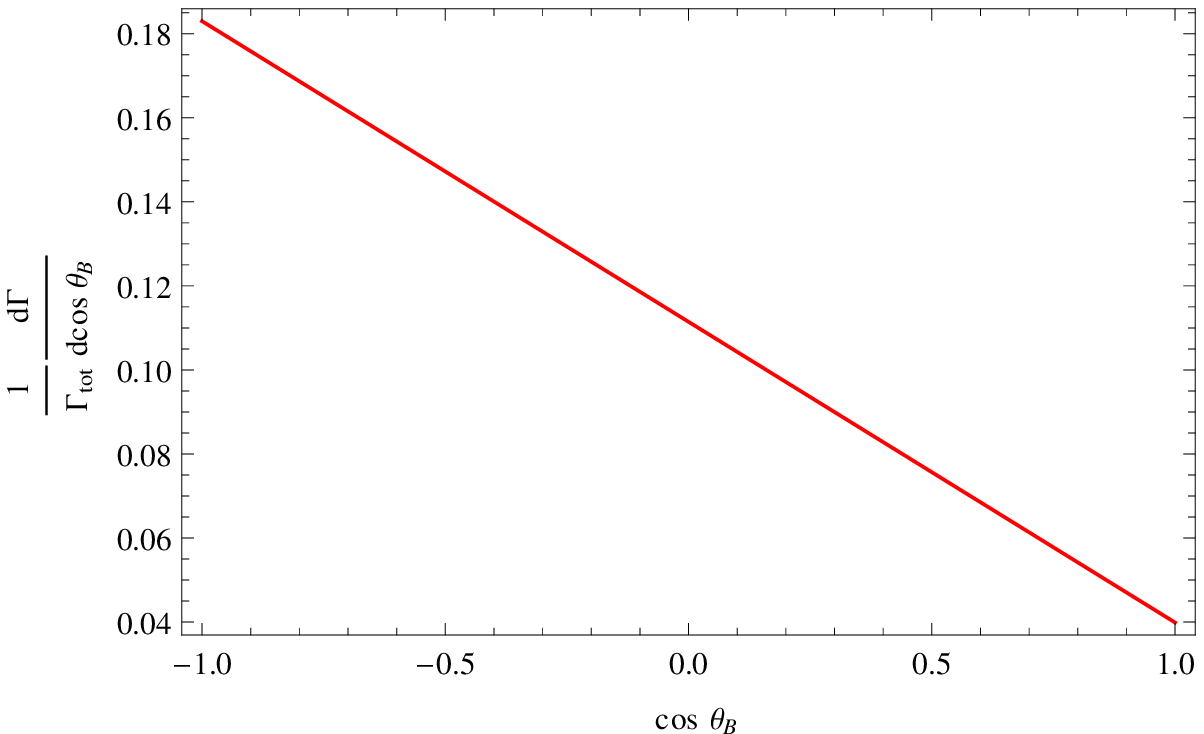}
\caption{Angular decay distribution for 
the decay $\Lambda_{b}\to \Lambda(\to p \pi^{-})\gamma$ 
in units of $10^{-5}$\,. \label{fig:ang_radiative}}
\end{center}
\end{figure}

\begin{table} 
\begin{center}
\caption{Branching ratios of semileptonic decays 
$\Lambda_b \to \Lambda \ell^+\ell^-$ with (without) 
long--distance contributions (in units of $10^{-6}$)\,. 
\label{tab:sl_leptons}}   

\vspace*{.15cm}

\hspace*{-.6cm}
\def\arraystretch{1}
    \begin{tabular}{|c|c|c|c|}
      \hline
Mode & Our results  & Theoretical predictions & Data \\
\hline
$\Lambda_b \to \Lambda e^+e^-$
& 1.0 (1.0)
& 2.79 $\pm$ 0.56~\cite{Chen:2001zc};
  4.6 $\pm$ 1.6~\cite{Aliev:2010uy};
  53 (2.3) ~\cite{Chen:2001ki} & \\
\hline
$\Lambda_b \to \Lambda \mu^+\mu^-$
& 1.0 (1.0)
& 26.5 $\pm$ 5.5 (0.8 $\pm$ 0.2)~\cite{Mott:2011cx};
& 1.73 $\pm$ 0.42 $\pm$ 0.55~\cite{Aaltonen:2011qs}\\
& & 53 (2.1)~\cite{Chen:2001ki};
        2.64 $\pm$ 0.56~\cite{Chen:2001zc};
& \\
& &46 (6.1)~\cite{Wang:2008sm};
         39 (5.9)~\cite{Aslam:2008hp};
         4.0 $\pm$ 1.2~\cite{Aliev:2010uy};
& \\
& &3.96$^{+0.38}_{-0.08}$~\cite{Lb_QCDSR};
   2.03$^{+0.26}_{-0.09}$~\cite{Lb_QCDSR}
& \\
\hline
$\Lambda_b \to \Lambda \tau^+\tau^-$
& 0.2 (0.3)
& 0.63 $\pm$ 0.13 (0.30 $\pm$ 0.08)~\cite{Mott:2011cx};
  0.23 $\pm$ 0.05~\cite{Chen:2001zc};
& \\
& &4.3 (2.4)~\cite{Wang:2008sm};
       4.0 (2.1)~\cite{Aslam:2008hp}; & \\
& &0.8 $\pm$ 0.3~\cite{Aliev:2010uy};
       11 (0.18)~\cite{Chen:2001ki}
& \\
\hline
\end{tabular}
\end{center}

\begin{center}
\caption{Branching ratio of the radiative decay  
$\Lambda_b \to  \Lambda \gamma$  (in units of $10^{-5}$)\,. 
\label{tab:sl_gamma}}   

\vspace*{.15cm}

\def\arraystretch{1}
    \begin{tabular}{|c|c|c|}
      \hline
Our result & Theoretical predictions & 
Data~\cite{Beringer:1900zz} \\
\hline
0.4
& 2.75 $\pm$ 1.75~\cite{Mannel:1997xy}; 3.7 $\pm$ 0.5~\cite{Huang:1998ek};
& $<$ 130 \\
&3.1 $\pm$ 0.6~\cite{Chen:2001ki}; 0.68 $\pm$ 0.05~\cite{Wang:2008ni};
& \\
&1.99$^{+0.34}_{-0.31}$~\cite{Lb_QCDSR};
 0.61$^{+0.14}_{-0.13}$~\cite{Lb_QCDSR}; & \\
&1.55$\pm$0.35~\cite{Cheng:1994kp}; 0.6~\cite{Cheng:1994kp}; & \\
&0.23~\cite{Mohanta:1999id}; 5.55$\pm$1.25~\cite{He:2006ud}   & \\
\hline
\end{tabular}
\end{center}

\begin{center}
\caption{Asymmetries $\bar A^l_{FB}$ and $\bar A^h_{FB}$ 
with (without) long--distance contributions\,. \label{tab:FB_int}}   

\vspace*{.15cm}

\def\arraystretch{1}
\begin{tabular}{|c|c|c|}
      \hline
Mode & $\bar A^l_{FB}$ & $\bar A^h_{FB}$ \\     
\hline
$\Lambda_b \to \Lambda e^+e^-$     
& $3.2 \times 10^{-10}$ 
 ($1.2\times 10^{-8}$)
& -0.321 (-0.321) \\
\hline
$\Lambda_b \to \Lambda \mu^+\mu^-$ 
& $1.7 \times 10^{-4}$ 
 ($8.0 \times 10^{-4}$)
& -0.300 (-0.294) \\
\hline
$\Lambda_b \to \Lambda \tau^+\tau^-$ 
& $5.9 \times 10^{-4}$ 
 ($9.6 \times 10^{-4}$)
& -0.265 (-0.259) \\ 
\hline
\end{tabular} 
\end{center}

\begin{center} 
\caption{Values of Wilson coefficients.} 

\vspace*{.25cm} 
\def\arraystretch{1.5}
\begin{tabular}{|l|r||l|r||l|r|}
\hline 
 $C_1$  &   -0.248 & $C_4$ &  -0.026 & $C_7^{\rm eff}$ & -0.313 \\
\hline
 $C_2$  &    1.107 & $C_5$ &   0.007 & $C_9$           &  4.344 \\           
\hline 
 $C_3$  &    0.011 & $C_6$ &  -0.031 & $C_{10}$          & -4.669 \\
\hline 
\end{tabular}
\label{tab:Wilson}
\end{center}
\end{table}

\end{document}